\documentclass[galaxies,review,accept,moreauthors,pdftex]{Definitions/mdpi} 

\firstpage{1} 
\pubvolume{1}
\issuenum{1}
\articlenumber{0}
\pubyear{2021}
\copyrightyear{2019}
\datereceived{} 
\dateaccepted{} 
\datepublished{} 
\hreflink{https://doi.org/} 

\newcommand{\be}{\begin{equation}}
\newcommand{\ee}{\end{equation}}
\newcommand{\ba}{\begin{eqnarray}}
\newcommand{\ea}{\end{eqnarray}}

\newcommand{\fracb}[2]{\left(\frac{#1}{#2}\right)}

\newcommand{\mean}[1]{\langle{#1}\rangle}

\usepackage{enumitem}
\usepackage{aas_macros}
\usepackage{amssymb}
\usepackage{amsmath}
\usepackage{subfig}

\definecolor{blazeorange}{rgb}{1.0, 0.4, 0.0}
\definecolor{seagreen}{rgb}{0.18, 0.55, 0.34}
\definecolor{rufous}{rgb}{0.66, 0.11, 0.03}
\definecolor{royalfuchsia}{rgb}{0.79, 0.17, 0.57}
\definecolor{scarlet}{rgb}{1.0, 0.13, 0.0}
\definecolor{royalpurple}{rgb}{0.47, 0.32, 0.66}

\Title{GRB Polarization: A Unique Probe of GRB Physics}

\TitleCitation{Title}

\Author{Ramandeep Gill $^{1,2,3}$\orcidA{}, Merlin Kole $^{4}$ and Jonathan Granot $^{3,2,1}$}

\AuthorNames{Ramandeep Gill, Merlin Kole and Jonathan Granot}

\AuthorCitation{Gill, R.; Kole, M.; Granot, J.}

\address{%
$^{1}$ \quad Department of Physics, The George Washington University, Washington, DC 20052, USA\\
$^{2}$ \quad Astrophysics Research Center of the Open university (ARCO), The Open University of Israel, P.O Box 808, Ra'anana 43537, Israel\\
$^{3}$ \quad Department of Natural Sciences, The Open University of Israel, P.O Box 808, Ra'anana 43537, Israel\\
$^{4}$ \quad Department of Nuclear and Particle Physics, University of Geneva, 24 Quai Ernest-Ansermet, 1205 Geneva, Switzerland%
}

\corres{Correspondence: rsgill.rg@gmail.com (R.G.);  merlin.kole@unige.ch (M.K.); granot@openu.ac.il (J.G.)}

\abstract{
Over half a century from the discovery of gamma-ray bursts (GRBs), the dominant radiation mechanism responsible for 
their bright and highly variable prompt emission  remains poorly understood. Spectral information alone 
has proven insufficient for understanding the composition and main energy dissipation mechanism 
in GRB jets. High-sensitivity polarimetric observations from upcoming instruments in this decade may help answer 
such key questions in GRB physics. This article reviews the current status of prompt GRB polarization measurements 
and provides comprehensive predictions from theoretical models. A concise overview of the fundamental questions in 
prompt GRB physics is provided. Important developments in gamma-ray polarimetry including a critical overview of
different past instruments are presented. Theoretical predictions for different radiation mechanisms and jet structures 
are confronted with time-integrated and time-resolved measurements. The current status and capabilities of upcoming instruments regarding the prompt emission are presented. The very complimentary information that can be 
obtained from polarimetry of X-ray flares as well as reverse-shock and early to late forward-shock (afterglow) 
emission is highlighted. Finally, promising directions for overcoming the inherent difficulties in obtaining statistically 
significant prompt-GRB polarization measurements are discussed, along with prospects for improvements in the theoretical 
modeling, which may lead to significant advances in the field.
}


\begin{document}

\section{Introduction} 

Gamma-ray bursts (GRBs) are one of the most energetic, and electromagnetically the brightest, transient phenomena in the Universe. They are the ideal test 
beds for understanding nature at its extreme that involves an explosive release of energy over a short timescale, producing a burst of $\gamma$-rays with 
isotropic-equivalent luminosity of $L_{\gamma,\rm iso}\sim10^{51}-10^{54}\,{\rm erg\,s}^{-1}$. It is now well established that most 
GRBs are cosmological sources and that they are powered by ultrarelativistic (with bulk Lorentz factors $\Gamma\gtrsim100$) bipolar beamed outflows driven by a central engine -- a 
compact object. The identity of the central 
engine, which could be either a black hole (BH) or a millisecond magnetar, is not entirely clear as the highly variable emission is produced far away from it at a 
radial distance of $R\sim10^{12}-10^{16}$cm. The most luminous phase of the burst, referred to as the ``prompt'' phase is short lived with a bimodal duration 
distribution, where the short GRBs have typical durations of $t_{\rm GRB}\sim10^{-1}\,$s and the long GRBs typically last for $t_{\rm GRB}\sim30\,$s while the 
dividing line sits at $t\sim 2\,$s \citep{Kouveliotou+93}. These two classes of GRBs are also distinct spectrally, with the short GRBs being spectrally harder 
as compared to the long GRBs that produce softer $\gamma$-rays. Other clues, e.g. the association of long-soft GRBs with star-forming regions \citep{Fruchter+06} 
and type-Ib/c supernovae \citep{Galama+98,Hjorth+03,Stanek+03} and that of the short-hard GRBs with early type galaxies \citep{Gehrels+05,Barthelmy+05} lead to the 
identification of two distinct progenitors. The long-soft GRBs are associated with the core-collapse of massive ($\gtrsim(20-30)M_\odot$) Wolf-Rayet stars 
\citep{Woosley-93} whereas the short-hard GRBs were theorized to originate in compact object mergers, namely that of two neutron stars (NSs) or a NS-BH pair 
\citep{Eichler+89,Narayan+92}. The unequivocal proof of the 
latter association had to wait until the gravitational wave (GW) detectors, LIGO and Virgo, became operational, which lead to the coincident detection of GWs from 
the merger of two NSs and a short-hard GRB by \textit{Fermi}-GBM and the INTEGRAL-ACS from GW 170817/GRB 170817A \citep{Abbott+17-GW170817-Ligo-Detection,Abbott+17-GW170817-GRB170817A}.

Although the global picture is fairly clear, the details of the energy dissipation process, the exact radiation mechanism, and the transfer of radiation in 
the highly dynamical flow remain poorly understood. All of these different processes combine to produce a non-thermal spectrum that is often well described 
by the Band-function \citep{Band+93}, an empirical fit to the spectrum featuring a smoothly broken power law. In $\nu F_\nu$ space, 
which indicates the observed energy flux around the frequency $\nu$ with $F_\nu$ being the spectral flux density this break manifests as 
a peak at the mean photon energy $\langle E_{\rm br}\rangle\simeq250\,$keV, which also represents the energy at which most of the energy of the burst 
is released, and the asymptotic power-law photon indices below and above the break energy have mean values of $\langle\alpha_{\rm Band}\rangle\simeq-1$ and 
$\langle\beta_{\rm Band}\rangle\simeq-2.3$, respectively \citep{Preece+00,Kaneko+06}. After decades of spectral modeling of the prompt emission the basic questions 
of GRB physics remain unanswered and it is becoming challenging to advance our understanding with spectral modeling alone.

An exciting opportunity was presented by the claimed detection of high level of linear polarization, with $\Pi=80\%\pm20\%$, in GRB 021206 \citep{Coburn-Boggs-03}. 
Although this result had a detection significance of $5.7\sigma$, further scrutiny by other works \citep{Rutledge,Wigger} cast irrevocable doubts and ultimately 
refuted the final 
result. Nevertheless, this one result initiated vigorous theoretical effort to understand the polarization of prompt GRB emission with the expectation that 
highly sensitive measurements will be able to resolve many of the outstanding questions of GRB physics. Over the past several years the number of prompt GRB 
polarization measurements (in some cases time-resolved) have grown, however the main results remain inconclusive due to inherent difficulties in obtaining 
highly statistically significant measurements. Therefore, it is hoped that the next generation of $\gamma$-ray polarimeters that will be launched in this 
decade will provide further important clues.

The main objectives of this review are to provide a concise yet comprehensive overview of the current status of theoretical developments as well as observations 
in the field of prompt GRB polarization, and also to highlight the need for developing more sensitive instruments and better analysis tools which are hoped 
to yield statistically significant measurements in the coming decade. Many points presented here have also been covered in earlier reviews on the topic  
\citep[e.g.,][]{Lazzati-06,Toma+09,Toma-13,Covino-Gotz-16,McConnell-17,Gill+20}. This review begins with a summary of the fundamental questions in GRB physics 
(\S\ref{sec:key-questions}) that can be addressed with measurements of linear polarization along with insights gained from prompt GRB spectral modeling. 
These include the outflow composition and dynamics (\S\ref{sec:composition-dynamics}), energy dissipation mechanisms (\S\ref{sec:energy-dissipation}), 
radiation mechanisms (\S\ref{sec:Rad_mech}), and angular structure of the outflow (\S\ref{sec:structure}). An overview of $\gamma$-ray polarimetry is presented 
in \S\ref{sec:gamma-ray-polarimetry} that includes the fundamental principles of $\gamma$-ray polarization measurement (\S\ref{sec:measurement-principles}) and a 
summary of the different detectors that have been used for GRB polarimetry (\S\ref{sec:polarimeters}). The theory of GRB polarization is presented next in 
\S\ref{sec:theory} which covers several topics, such as polarization from uniform (\S\ref{sec:Pol-Uniform-Jets}) and structured (\S\ref{sec:Pol-Structured-Jets}) 
jets with different radiation mechanisms, temporal evolution of polarization (\S\ref{sec:Pol-temp-evol}), polarization arising from multiple overlapping pulses 
(\S\ref{sec:muliple-pulses}), the most likely polarization for a given radiation mechanism (\S\ref{sec:most-likely-pol}), and the energy dependence of polarization 
(\S\ref{sec:energy-dependence}). The current status of prompt GRB polarization measurements is presented next that includes time-integrated 
(\S\ref{sec:time-integrated-measurements}), time-resolved (\S\ref{sec:time-resolved-measurements}), and energy resolved (\S\ref{sec:energy-resolved-measurements}) 
measurements. The importance of polarization measurements from the other phases of the burst, namely X-ray flares, reverse-shock emission (optical flash and 
radio flare), and forward-shock emission, that also probe the properties of the GRB outflow is emphasized in \S\ref{sec:other-pol}. Finally, \S\ref{sec:outlook} 
touches upon the outlook for this decade that will see the launch of more sensitive instruments (\S\ref{sec:future-instruments}) the predicted performance of some 
is compared in \S\ref{sec:instrument-comparison}. This review concludes by offering some suggestions for improvements in the polarization data analysis 
(\S\ref{sec:improvements-analysis}) and its theoretical modeling (\S\ref{sec:improvements-theory}).

\section{Key questions that can be addressed with GRB polarization}
\label{sec:key-questions}
Measurements of the prompt GRB polarization may help shed light on many critical aspects of the relativistic outflow whose knowledge has 
evaded us so far. Below we summarize key open questions in GRB physics which can be probed with spectro-polarimetric observations. More 
detailed reviews and discussions of these topics can be found in other
review articles (e.g. \citep{Piran-04,Zhang-Meszaros-04,Meszaros-06,Granot-Ramirez-Ruiz-12,Kumar-Zhang-15}). Theoretical modeling of prompt 
GRB polarization and comprehensive results are provided in \S\ref{sec:theory}.

\subsection{\textbf{\textit{What are the outflow composition and dynamics?}}}
\label{sec:composition-dynamics}
The main dissipation and radiation mechanisms that produce the GRB prompt emission are dictated by the composition of the outflow. 
The two most widely discussed scenarios invoke an outflow that is either kinetic-energy-dominated (KED) \citep{Rees-Meszaros-94} or 
Poynting-flux-dominated (PFD) \citep{Thompson-94,Lyutikov-Blandford-03}. In the former most of the energy is initially thermal (fireball) 
and is eventually transferred to the kinetic energy of the cold baryons, while in the latter the main energy reservoir is the (likely 
ordered) magnetic field that drives the expansion and acceleration of the flow. 
If the radiation mechanism is indeed synchrotron (see \S\ref{sec:Rad_mech}), then the level of polarization in both types of flows depends on the 
structure of the magnetic field that is either generated in situ, e.g. in internal shocks in a KED flow, or survives at large distances from the 
central engine, which could happen in both types of flows. Our theoretical understanding of the B-field structure in the emission region in a given 
type of flow is still limited and rather speculative. Any measurement of polarization will put strong constraints on the B-field structure. 
Therefore, in combination with polarization measurements, spectral and temporal (pulse profiles) modeling will allow us to constrain the composition.

The distinction between a KED and PFD flow can be conveniently parameterized using the magnetization parameter,
\begin{equation}
    \sigma \equiv \frac{w_B'}{w_m'} = \frac{B'^2}{4\pi\left(\rho' c^2 + \frac{\hat\gamma}{\hat\gamma-1}P'\right)} \xrightarrow[cold]{} 
    \frac{B'^2}{4\pi\rho'c^2}\,,
\end{equation}
which is defined as the ratio of the comoving (all quantities measured in the comoving/fluid frame are primed) magnetic field 
enthalpy density $w_B' = B'^2/4\pi$, where $B'$ is the magnetic field strength, to that of matter, 
$w_m' = \rho' c^2+\frac{\hat{\gamma}}{\hat{\gamma}-1}P'$ or $w_m' = \rho' c^2$ when it is cold ($P'\ll\rho'c^2$), 
where\footnote{We assume here for simplicity that the baryons dominate the total rest mass.} 
$\rho' = m_p n'$, $n'$ is the particle number density, $m_p$ is the proton mass, and $c$ is the speed of light. The baryons are 
assumed to be cold with adiabatic index $\hat\gamma=5/3$ ($\hat\gamma=4/3$ for a relativistic fluid) and negligible pressure $P'$ 
when compared with the particle inertia. A KED flow will have $\sigma<1$ and magnetic fields, if present, are weak and randomly 
oriented with short coherence length scales and are unimportant in governing the dynamics of the outflow. On the other hand, a PFD flow 
will have $\sigma>1$ and the magnetic field is much more ordered where it's responsible for accelerating the flow.

A prime example of a KED flow is the standard `fireball' scenario \citep[][]{Goodman-86,Paczynski-86}, in which total energy $E$ is 
released close to the central engine, launching a radiation dominated and optically thick outflow, with Thomson optical depth 
$\tau_T\gg1$. The temperature at the base of the flow is typically $k_BT\gtrsim\,$MeV, which leads to copious production of $e^\pm$-pairs 
via $\gamma\gamma$-annihilation that further enhances the optical depth. The enormous radiation pressure causes the flow to expand 
adiabatically thereby converting the radiation field energy to the kinetic energy of baryons, that are inefficient radiators due to 
their large Thomson cross-sections. The bulk Lorentz factor (LF) of the fireball grows linearly with radius, $\Gamma(R_0<R<R_s)\approx R/R_0$,
where $R=R_0$ is the launching radius, while its comoving temperature declines as $T'(R)\propto R^{-1}$. The amount 
of baryon loading, i.e. the amount of baryons with total mass $M_b$ entrained in the flow of given energy $E$,  
determines the terminal LF, $\Gamma_\infty = E/M_bc^2$, which is attained at the saturation radius 
$R=R_s\sim\Gamma_\infty R_0$ at which point the growth in the bulk $\Gamma$ saturates and the flow simply coasts at $\Gamma=\Gamma_\infty$. 
The kinetic energy of the baryons is tapped at a large distance ($R>R_s$) from the central engine via internal shocks (see below).

In a PFD flow, large-scale magnetic fields propagate outwards from the central engine with an angular coherence scale $\theta_B>1/\Gamma$, 
where $1/\Gamma$ represents the characteristic angular scale over which the flow is causally connected and, as discussed later, also 
the angular scale into which the emitted radiation is beamed towards the observer from a relativistic flow. While the fireball scenario 
is well agreed upon and has enjoyed many successes since it is fairly robust, no such \textit{standard model} exists for a magnetized outflow 
to explain GRB properties. In several works \citep[e.g.][]{Li+92,Vlahakis-Konigl-03,Beskin-Nokhrina-06,Lyubarsky-10a,Komissarov+09}, 
ideal-MHD models for a steady-state, axisymmetric, and non-dissipative outflow have been developed in which the flow expands adiabatically 
due to magnetic stresses. The flow is launched highly magnetized near the light cylinder radius, $R_L$, with $\sigma(R_L)=\sigma_0\gg1$ and 
bulk LF $\Gamma(R_L)=\Gamma_0\sim1$. As the flow expands, its magnetization declines with radius and in the case of a radial wind (i.e. 
unconfined, with a negligible external pressure) the flow is limited to a terminal LF of $\Gamma_\infty\sim\sigma_0^{1/3}$ 
where the corresponding magnetization of the flow is $\sigma\sim\sigma_0^{2/3}$ \citep{Goldreich-Julian-70}. For weak external confinement 
(an external pressure profile $p_{\rm ext}\propto z^{-\kappa}$ with $\kappa>2$, where $z\approx R=(z^2+r_{\rm cyl}^2)^{1/2}$ is the distance 
from the central source along the jet's symmetry axis and $r_{\rm cyl}$ is the cylindrical radius) the acceleration saturates at a terminal 
LF of $\Gamma_\infty\sim\sigma_0^{1/3}\theta_j^{-2/3}$ and magnetization $\sigma_\infty\sim(\sigma_0\theta_j)^{2/3}\sim(\Gamma_\infty\theta_j)^2\gg1$ 
where $\theta_j$ is the jet's asymptotic half-opening angle \citep{Lyubarsky-09}. For strong external confinement ($p_{\rm ext}\propto z^{-\kappa}$ 
with $\kappa<2$) the jet maintains lateral causal contact and equilibrium, leading $\Gamma\sim r_{\rm cyl}/R_L\sim(z/R_L)^{\kappa/4}$, which 
saturates at $\Gamma_\infty\sim\sigma_0$, $\sigma_\infty\sim1$, and $\Gamma_\infty\theta_j\sim1$. Since prompt GRB observations demand the dissipation 
region to be expanding ultrarelativistically with $\Gamma_\infty\gtrsim100$, to avoid the \textit{compactness problem} \citep{Ruderman-75,Piran-04}, 
and afterglow observations suggest that typically $\theta_j\gtrsim0.05-0.1$, this implies $\Gamma_\infty\theta_j\gtrsim10$ in GRBs. This suggests 
that the weakly confined regime is most relevant for GRBs, however it implies $\sigma_\infty\sim(\Gamma_\infty\theta_j)^2\gg1$ which suppresses 
internal shocks. It has been pointed out \citep{Tchekhovskoy+10,Komissarov+10} that the sharp drop in the surrounding (lateral) pressure as the jet 
exits the progenitor star in long GRBs can lead to $\Gamma_\infty\theta_j\gg1$ along with a more modest asymptotic magnetization $\sigma_\infty\gtrsim 1$, 
but even then internal shocks remain inefficient.

When the steady-state assumption is relaxed, alternative models that consider an impulsive and highly variable flow yield a much larger 
terminal LF with $\Gamma_\infty\sim\sigma_0$ and may achieve $\sigma_\infty<1$ or even $\sigma_\infty\ll1$ under certain conditions 
\citep{Granot+11,Granot12b}. In this scenario, a thin shell of initial width $\ell_0$ is accelerated due 
to magnetic pressure gradients that causes its bulk LF to grow as $\Gamma\sim(\sigma_0R/R_0)^{1/3}$, where $R_0\approx\ell_0$, while its 
magnetization drops as $\sigma\sim\sigma_0^{2/3}(R/R_0)^{-1/3}$. The bulk LF of the shell saturates at $R_s\sim\sigma_0^2R_0$ at which 
point its $\Gamma\sim\Gamma_\infty\sim\sigma_0$ and $\sigma\sim1$. For $R>R_s$, the magnetization continues to drop further as $\sigma\sim (R/R_s)^{-1}$ as the 
shell starts to spread radially. For a large number of shells initially separated by $\ell_{\rm gap}$ the radial expansion is limited as 
neighboring shells collide and one expects an asymptotic mean magnetization of $\sigma_\infty\sim\ell_0/\ell_{\rm gap}$. This scenario offers 
the dual possibility of magnetic energy dissipation via MHD instabilities when $\sigma>1$ at $R<R_s$ as well as kinetic energy dissipation 
via internal shocks when $\sigma<1$ at $R>R_s$.

Similar outflow dynamics are obtained in a popular model that makes the magnetohydrodynamic (MHD) approximation and features a 
striped-wind magnetic field structure \citep{Lyubarsky-Kirk-01,Spruit+01,Drenkhahn-02,Drenkhahn-Spruit-02,Begue+17}, in which magnetic field lines 
reverse polarity over a characteristic length scale $\lambda\sim\pi R_L=\pi c/\Omega = cP/2 = 1.5\times10^7P_{-3}\,$cm. Here $\Omega = 2\pi/P$ is 
the central engine's angular frequency with $P$ being its rotational period. Close to the central engine the flow may be accelerated by 
magneto-centrifugal, and to some extent, thermal acceleration. At distances larger than the Alfv\'{e}n radius, where $R_A \gtrsim R_L$, these 
effects are negligible and when collimation-induced acceleration is ineffective then the properties of the flow can be described using radial dynamics. 
If a reasonable fraction (the usual assumption is approximately half) of the dissipated energy in the flow goes towards its acceleration, conservation 
of the total specific energy, while ignoring any radiative losses, yields the relation $\Gamma(R)[1+\sigma(R)]=\Gamma_0[1+\sigma_0]$ 
for a cold flow, which simplifies to $\Gamma(R)\sigma(R)\approx\Gamma_0\sigma_0$ for $\sigma(R)\gg1$. At the Alfv\'{e}n radius, the four velocity 
of the flow is $u_A=\Gamma_A\beta_A=\sigma_A^{1/2}\approx\Gamma_A\approx\Gamma_0\sigma_0/\sigma_A\approx\sigma_0/\sigma_A$ that implies that $\sigma_A\approx\sigma_0^{2/3}$ and $\Gamma_A\approx\sigma_0^{1/3}$. The terminal LF is achieved at the saturation radius 
$R_s$ when $\sigma(R_s)\sim1$, at which point $\Gamma_\infty\approx\Gamma_0\sigma_0\approx\sigma_0=\sigma_A^{3/2}$. 
In this scenario, the saturation radius is given by $R_s = \Gamma_\infty^2\lambda/6\epsilon=1.7\times10^{13}\Gamma_{\infty,3}^2(\lambda/\epsilon)_8\,$cm, 
where $\epsilon = v_{\rm in}/v_A \sim0.1$ is a measure of the reconnection rate where it quantifies the plasma inflow velocity $v_{\rm in}$ into the 
reconnection layer in terms of the Alfv\'{e}n speed. For magnetized flows, $v_A=c\sqrt{\sigma/(1+\sigma)}$ which approaches the speed of light for $\sigma\gg1$.
Beyond the Alfv\'{e}n radius the bulk LF grows as a power law in radius, with $\Gamma(R)=\Gamma_\infty(R/R_s)^{1/3}$, while the magnetization declines as 
$\sigma(R)=(R/R_s)^{-1/3}$.

In the regime of high magnetization ($\sigma\gg1$), an alternative model that does not make the MHD approximation is considered by \citet{Lyutikov-Blandford-03} 
and \citet{Lyutikov-06}.

\subsection{\textbf{\textit{How and where is the energy dissipated?}}}
\label{sec:energy-dissipation}
The composition of the outflow has a strong impact on the dominant energy dissipation channel. To produce the prompt GRB emission, the baryonic 
electrons as well as any $e^\pm$-pairs, which are the primary radiators, cannot be cold and they need to be accelerated or heated to 
raise their internal energy. The observed photon energy spectrum is not only shaped by the underlying radiation mechanism but also the radial location 
in the flow where energy is dissipated. If most of the energy is dissipated much below the photospheric radius, at $R\ll R_{\rm ph}$, 
where the Thomson optical depth of the flow is $\tau_T\gg1$ and where the radiation field and particles are tightly coupled via Compton 
scattering (baryons are coupled with the leptons via Coulombic interactions) and assume a thermal distribution, the final outcome is a 
quasi-thermal spectrum \citep{Goodman-86,Paczynski-86,Thompson-94}. The observed spectrum in this case is not a 
perfect blackbody due to the observer seeing different parts of the jet with different Doppler boosts, but close to one with low-energy 
(below the spectral peak energy) photon index $\alpha_{\rm ph}=d\ln N_\gamma/d\ln E\approx 0.4$ which is softer from 
$\alpha_{\rm ph} = 1$ expected for a Rayleigh-Jeans thermal spectrum \citep{Peer-08,Beloborodov-10}. If  instead, most of the energy is dissipated 
in the optically thin ($\tau_T<1$) parts of the flow, then a non-thermal spectrum emerges.  When the flow is continuously heated across 
the photosphere, the final spectrum is a combination of two components: quasi-thermal and non-thermal. 

If the flow is uniform (i.e. quasi-spherical with negligible angular dependence within angles of $\lesssim1/\Gamma$ around the line of sight) 
then any thermal component will show negligible polarization as there is no preferred direction for the polarization vector to align with. 
Even if different parts of the flow may be significantly polarized at the photosphere \citep{Beloborodov-11}, the net polarization 
averages out to zero after integrating over the GRB image on the sky. However, angular structure in the flow properties can lead to modest 
($\Pi\lesssim20\%$) polarization \citep{Ito+14,Lundman+14,Gill+20,Parsotan+20}. The polarization of the non-thermal spectral component ultimately 
depends on the radiation mechanism, discussed in section \ref{sec:Rad_mech}.

In a KED flow, after an initial phase of rapid acceleration of the fireball when the bulk LF saturates, the particles are cold in the 
comoving frame with negligible pressure ($P'\ll\rho'c^2$). The energy of the flow is dominated by the kinetic energy of the baryons which is very 
\textit{ordered}. To produce any radiation particle motion needs to be \textit{randomized}. A simple and robust way to achieve that 
is via shocks. The canonical model of internal shocks \citep{Rees-Meszaros-94,Papathanassiou-Meszaros-96,Sari-Piran-97,Daigne-Mochkovitch-98} 
posits that the central engine accretes intermittently and ejects shells of matter 
that are initially separated by a typical length scale $\sim ct_v/(1+z)$ and have fluctuations in their bulk LFs of order $\Delta\Gamma\sim\Gamma$, 
with $\Gamma$ being the mean bulk LF. Here $t_v$ is the observed variability of the prompt emission lightcurve and $z$ is the redshift of 
the source. Typically $R_0\sim10^7\;$cm and $\Gamma_\infty\sim10^2-10^3$ so that the acceleration saturates at $R_s\sim\Gamma_\infty R_0\sim10^{9}-10^{10}\;$cm. 
For $R>R_s$, faster moving shells catch up from behind with slower ones and collide to dissipate their kinetic energy at internal shocks occurring 
at the dissipation radius of $R_{\rm dis}=2\Gamma_\infty^2ct_v/(1+z)=6\times10^{13}(1+z)^{-1}\Gamma_{\infty,2}^2t_{v,-1}\,$cm.

\end{paracol}
\begin{figure}
    \widefigure
    \centering
    \includegraphics[width=0.6\textwidth]{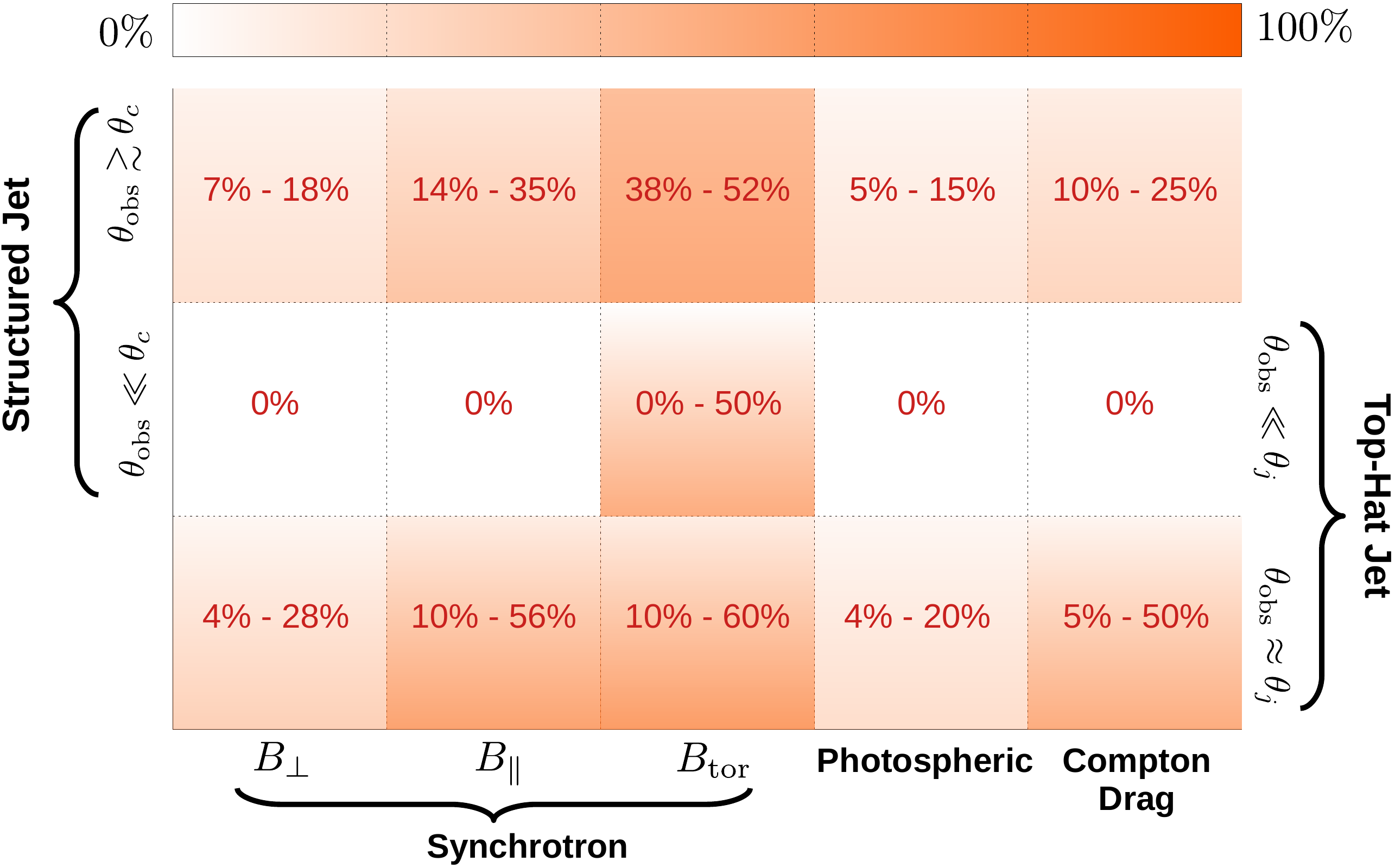}
    \caption{Approximate degree of polarization for different radiation mechanisms and jet structures \citep{Gill+20}. 
    If the emission is synchrotron then polarization for different B-field configurations is given (assuming $\Pi_{\max}=70\%$). 
    For each jet structure distinction is made between two cases: (i) when the observer's viewing angle 
    ($\theta_{\rm obs}$) is much smaller than the half-opening angle ($\theta_j$) of a top-hat jet 
    or, in the case of a structured jet, if it is much smaller than the core angle ($\theta_c$), 
    (ii) when $\theta_{\rm obs}$ is close to $\theta_j$, i.e. the edge of the jet. For a structured jet 
    $\theta_{\rm obs}$ can exceed $\theta_c$ by an order unity factor before the fluence starts to drop 
    significantly. When $\theta_{\rm obs}\approx\theta_j$ the minimum value of polarization can be zero in 
    all cases, except for $B_{\rm tor}$, for a pulse with a given $\xi_j=(\Gamma\theta_j)^2$, where $\xi_j^{1/2}$ 
    is the ratio of the angular sizes of the jet and of the beaming cone. Different pulses may have 
    slightly different $\xi_j$ (typically with a similar $\theta_j$ but different $\Gamma$), which on 
    average would yield a finite polarization. The quoted lower range reflects this mean value (see \citep{Gill+20} 
    for more details). For the $B_{\rm tor}$ case, $\Pi=0\%$ when $\theta_{\rm obs}=0$ due to symmetry and 
    $\vert\Pi\vert>0\%$ otherwise while $\Pi\approx50\%$ at $1/\Gamma<\theta_{\rm obs}<\theta_j$. 
    }
    \label{fig:Pol-Schematic}
\end{figure}
\begin{paracol}{2}
\switchcolumn

When the shells collide a double shock structure forms with a forward shock going into the slower shell and accelerating it while a reverse 
shock goes into the faster shell and decelerates it. These shocks heat a fraction $\xi_e$ of the electrons into a power-law energy distribution, 
with $dN_e/d\gamma_e\propto\gamma_e^{-p}$ for $\gamma_e>\gamma_m$, where these electrons hold a fraction $\epsilon_e$ of the total internal energy 
density behind the shock. The LF of the minimal energy electrons, $\gamma_m=[(p-2)/(p-1)](\epsilon_e/\xi_e)(m_p/m_e)(\Gamma_{\rm ud}-1)$ (for $p>2$), 
depends on the relative bulk LF, $\Gamma_{\rm ud}$, of the upstream to downstream matter across the relevant shock. 
A fraction $\epsilon_B$ of the internal energy density behind the shock is held by the shock-generated magnetic field of strength $B'\sim10^2-10^3\,$G. 
More generally one can express the comoving magnetic field in terms of the radius and outflow Lorentz factor and magnetization at that radius, 
as well as the observed isotropic equivalent $\gamma$-ray luminosity, $L_{\gamma,{\rm iso}}$, and the $\gamma$-ray emission efficiency, 
$\epsilon_\gamma$ (i.e. fraction of the total outflow energy channeled into gamma-rays), 
$B' = 1.8\times10^5\Gamma_2^{-1}R_{14}^{-1}(\frac{\sigma}{1+\sigma})^{1/2}L_{\gamma,{\rm iso},52}^{1/2}\epsilon_{\gamma,-1}^{-1/2}\;$G. 
The exact structure of the magnetic field is still an open 
question, but it has been argued that streaming instabilities \citep{Weibel-59,Gruzinov-Waxman-99,Medvedev-Loeb-99,Bret-09,Keshet+09}, 
e.g. the relativistic two-stream and/or Weibel (filamentation) instability, are responsible for generating a small-scale field with coherence 
scale on the order of the electron and/or proton skin depth, $c/\omega_p' = c(4\pi n'e^2/m)^{-1/2}$ where $\omega_p'$ is the plasma frequency, 
which depends on the particle number density $n'$, mass $m$ is the particle mass, and $e$ is the elementary charge. Since the coherence 
length of the shock-generated field is much smaller than angular size of the beaming cone ($\theta_B\ll1/\Gamma$), the net polarization is 
limited to $\Pi\lesssim30\%$. 

Alternatively, interaction of the shock with density inhomogeneities in the upstream can cause macroscopic turbulence in the downstream (e.g. 
excited by the Richtmyer-Meshkov instability), which can in turn amplify a shock-compressed large-scale upstream magnetic field to near-equipartition 
with the downstream turbulent motions \citep{Sironi-Goodman-07,Zhang+09,Inoue+11,Mizuno+11,Mizuno+14,del-Valle+16}. The dynamo-amplified magnetic 
field is expected to form multiple mutually incoherent patches (with angular sizes up to a fraction of the visible region) in which the field 
is largely ordered. The expected polarization, after averaging over such patches in the observed region, is typically expected to be small, with 
$\Pi\lesssim2\%$ \citep{Inoue+11}.

As mentioned earlier, in a PFD flow, the main dissipation channel is magnetic reconnection and /or MHD instabilities. Both of these are non-ideal 
effects that cannot be treated in an ideal MHD formalism. Magnetic field energy is dissipated when opposite polarity field lines reconnect, which 
leads to acceleration of electrons that then cool by either emitting synchrotron radiation outside of the reconnection sites or inverse-Compton 
scattering either synchrotron photons or a pre-existing radiation field advected with the flow. Exactly where dissipation commences depends on the 
initial magnetic field geometry in the flow as the field lines expand outward from the central engine to larger distances \citep{Romanova-Lovelace-92}. 
If the flow is axisymmetric and is not permeated by polarity switching field lines, magnetic energy can still be dissipated due to current-driven 
instabilities, e.g. the kink instability \citep{Lyubarsky-92,Eichler-93,Begelman-98,Giannios-Spruit-06}. Such an instability may also occur at the 
interface between the jet and the confining medium, e.g. the stellar interior of a Wolf-Rayet star in long-soft GRBs \citep{Levinson-Begelman-13} 
and the dynamically ejected wind during a binary neutron star merger in short-hard GRBs. 
Magnetic field lines that reverse polarity on some characteristic length scale $\lambda$ can be embedded into the outflow in a variety of ways 
\cite{Mckinney-Uzdensky-12}. These can indeed be injected at the base of the flow where field polarity reversals are obtained in the accretion 
disk due to the magnetorotational instability, as demonstrated in several shearing-box numerical MHD simulations \cite{Davis+10} as 
well global simulations of black hole accretion \citep{ONiell+11}. Depending on how particles are heated/accelerated when magnetic energy is dissipated 
as the flow becomes optically-thin, as discussed in the next section, the polarization will be energy dependent and can be $\Pi\lesssim60\%$ 
if synchrotron emission dominates and the B-field angular coherence length near the line of sight is $\theta_B\gtrsim1/\Gamma$.

In the striped-wind model \citep{Drenkhahn-02,Drenkhahn-Spruit-02,Giannios-Uzdensky-19}, magnetic dissipation commences beyond the Alfv\'{e}n radius 
and becomes the dominant contributor towards flow acceleration. Below the Alfv\'{e}n radius the flow is accelerated due to magneto-centrifugual effects 
as well as collimation provided by the confining medium \citep{Tchekhovskoy+08,Komissarov+09}.  If the confining medium has a sharp outer boundary 
(e.g. the edge of the massive star progenitor for a long GRB) then as the jet breaks out of the confining medium, the flow becomes conical and expands 
ballistically. The sudden loss of pressure also leads to some further acceleration via the mechanism of \textit{rarefaction acceleration} 
\citep{Komissarov+10} that operates in PFD relativistic jets. While these ideal MHD processes may continue to operate at length scales relevant for 
prompt GRB emission, magnetic reconnection in a striped wind provides a source for gradual acceleration out to the saturation radius $R_s$. 
Beyond this radius magnetic reconnection subsides, and therefore acceleration ceases and the flow starts to coast. When the prompt emission 
is produced in an accelerating flow, the degree of polarization is not affected. Instead, the duration of the pulses becomes shorter in comparison to that obtained in a coasting flow \citep[see, e.g.,][]{Gill-Granot-21}.

Other variants of the PFD model, as presented above, include the internal-collision-induced magnetic reconnection and turbulent (ICMART) model 
\citep{Zhang-Yan-11}, in which high-$\sigma$ shells are intermittently ejected by the central engine that 
dissipate their energy at $R\sim10^{15}-10^{16}\,$cm, where collision-induced magnetic reconnection and turbulence radiates away the magnetic energy 
and reduces the initially high magnetization of the ejecta to order unity. The expected polarization from an ICMART event has been presented in 
\citet{Deng+16} using 3D numerical MHD simulations where they also find a $90^\circ$ change in polarization angle.

\subsection{\textbf{\textit{What radiation mechanism produces the Band-like GRB spectrum?}}\label{sec:Rad_mech}}

Few radiation mechanisms have been proposed to explain the Band-like spectrum of prompt emission, the most popular being synchrotron and inverse-Compton. Below we present a concise summary of the different proposed mechanisms and show the expected polarization in Fig.~\ref{fig:Pol-Schematic}.

\subsubsection{\textbf{Optically-Thin Synchrotron Emission}}\label{sec:Synchrotron}
Relativistic electrons gyrating around magnetic field lines cool by emitting synchrotron photons. When the energy distribution of these 
electrons is described by a power law, e.g., that obtained at collisionless internal shocks due to Fermi acceleration, the emerging 
synchrotron spectrum is described by multiple power-law segments that join at characteristic break energies \citep{Sari+98,Granot-Sari-02}. These 
correspond to the synchrotron frequency, $E_m=\Gamma(1+z)^{-1}h\nu_m' = \Gamma(1+z)^{-1}\gamma_m^2(\hbar eB'/m_ec)$, of minimal energy 
electron with LF $\gamma_m$ and the cooling frequency, 
$E_c=\Gamma(1+z)^{-1}h\nu_c' = 36\pi^2(1+z)^{-1}(\hbar em_ec^3/\sigma_T^2)(\Gamma^3\beta^2/B'^3R^2)$, of electrons that are cooling at the dynamical 
time, $t_{\rm cool}'=t_{\rm dyn}' = R/\Gamma\beta c$. Here $B'$ is the comoving magnetic field and $\sigma_T$ is the Thomson cross-section. 
The high radiative efficiency of prompt emission demands that the electrons be in the fast-cooling regime for which $E_c<E_m$ and the 
$\nu F_\nu$ spectrum peaks at $E_{\rm pk}=E_m$. In this case, the spectrum below the peak energy has photon index $\alpha_{\rm ph} = -2/3$ for 
$E<E_c$ and $\alpha_{\rm ph} = -3/2$ for $E_c<E<E_m$. Above the peak energy, the photon index is $\alpha_{\rm ph} = -(p+2)/2$ where $p$ is the power-law 
index of the electron distribution.

While synchrotron emission is still regarded as the default emission mechanism, the basic `vanilla' model has been argued to be not as 
robust as previously thought. First, its predictions have been challenged by a small fraction of GRBs that showed harder low-energy 
($E<E_{\rm pk}$) spectral slopes with $\alpha_{\rm Band} > -2/3$ \citep{Crider+97,Preece+98,Preece+02,Ghirlanda+03}, often identified as the 
synchrotron \textit{line of death}. Some possible alternatives that have been suggested to resolve this discrepancy include 
anisotropic electron pitch angle distribution and synchrotron self-absorption \citep{Lloyd-Petrosian-00}, jitter radiation \citep{Medvedev-00}, 
and photospheric emission \cite{Meszaros-Rees-00}. 
The line-of-death violation is generally derived by fitting the empirical Band-function to the observed spectrum. When synthetic synchrotron 
spectra (after having convolved with the energy response of a detector) are fit with the Band-function an even softer $\mean{\alpha_{\rm Band}}=-0.8$ 
is found due to the detector's limited energy range (e.g. \textit{Fermi}/GBM \citep{Burgess+15}) that doesn't quite probe the asymptotic value of 
$\alpha_{\rm ph}$. Since a significant fraction of GRBs show low-energy spectral indices that are harder than this value, it might indicate 
that another spectral component is possibly contributing at low energies and offsetting the spectral slope. Second, the spectral peak 
energy in the cosmological rest-frame of the source is given by $E_m(1+z)$ which depends on a combination of $\Gamma$, $\gamma_m$, and $B'$ to 
yield the measured peak energy in the range $200\,{\rm keV}\lesssim [E_{\rm pk,z} = E_{\rm pk}(1+z)]\lesssim1\,$MeV \citep{Poolakkil+21} with 
a possible peak around $E_{\rm pk,z}\sim m_ec^2$ \citep{Gruber+14}. 
Given that all of these quantities can vary substantially between different bursts, the synchrotron model doesn't offer any characteristic energy scale at 
which most of the energy is radiated in the event that the $E_{\rm pk,z}$ distribution indeed narrowly peaks around $\sim m_ec^2$ \citep{Beloborodov-Meszaros-17}.
Third, the synchrotron model predicts wider spectral peaks than that obtained by fitting the Band-function to observations 
\cite{Vurm-Beloborodov-16}. This issue has now been demonstrated for a large sample of GRBs where the spectral widths obtained with the 
simplest synchrotron model yielding the narrowest spectral peak, e.g. a slow-cooling Maxwellian distribution of electrons, is inconsistent 
with most of the GRBs \citep{Axelsson-Borgonovo-15,Yu+15}. Moreover, it is rather easy to get a wider spectral peak by having, e.g. 
fast-cooling particles, variable magnetic fields, etc., but much harder to obtain narrower peaks.

Several works that find the synchrotron model to be inconsistent with observations invariably use empirical models, e.g. the Band-function, 
a smoothly-broken power law, to determine low-energy spectral slopes and peak widths. This may become a problem in instances where such 
models are unable to capture the intrinsic complexity of the underlying data. Therefore, an arguably better approach is to directly fit 
physical models to the raw data to derive spectral parameters and remove any bias \citep{Vianello+18,Burgess-19,Yassine+20, Zhang_2016}. 
Such an approach has led to alleviating some of the issues encountered by the optically-thin 
synchrotron model, where it was shown that direct spectral fits (in count space rather than energy space) with synchrotron emission from cooling 
power-law electrons can explain the low-energy spectral slopes as well as the spectral width of the peak \citep{Burgess+20}. 
\\
\\
\noindent{\textbf{Magnetic Field Structure}} \\
\\
If the coherence length of the magnetic field is larger than the gyroradius of particles, the structure of the magnetic field in the dissipation 
region doesn't affect the spectrum or the pulse profile. However, it significantly affects the level of polarization. Therefore, spectro-polarimetric 
observations that strongly indicate synchrotron emission as the underlying radiation mechanism can be used to also determine the magnetic field 
structure. At least four physically motivated axisymmetric magnetic field structures 
and the emergent synchrotron polarization, have been discussed in the literature \citep{Granot-03,Granot-Konigl-03,Lyutikov+03,Granot-Taylor-05}:
\begin{enumerate}
    \item $\boldsymbol{B_{\rm ord}}$: An ordered magnetic field with angular coherence length $1/\Gamma\lesssim\theta_B\ll\theta_j$, where $1/\Gamma$ 
    is the angular size of the beaming cone. It is envisioned that the jet surface is filled with several small radiating patches of angular size much 
    smaller than the jet aperture and that these are pervaded by mutually incoherent ordered magnetic fields. In this way, such a field configuration as a 
    whole remains axisymmetric in a statistical sense (despite having a local preferred direction for a given line of sight, namely the ordered field direction 
    at that line of sight) and also different from a globally ordered B-field. This type of field structure was motivated by the high-polarization claim of 
    $\Pi=80\%\pm20\%$ \citep{Coburn-Boggs-03} in 
    GRB 021206 and by the notion that the local synchrotron polarization can be very high with $\Pi_{\rm max}\sim75\%$. Magnetic fields with sufficiently 
    large coherence lengths that are not globally ordered can be advected with the flow from the central engine where their length scale is 
    altered en route to the emission site due to hydromagnetic effects.
    \item $\boldsymbol{B_\perp}$: A random magnetic field (i.e. with $\Gamma\theta_B\ll1$) confined to the plane transverse to the local velocity vector of the 
    fluid element in the flow. In many cases, the flow is assumed to be expanding radially which is a good approximation when the prompt emission is generated 
    since no significant lateral motion is expected at that time. This field structure is motivated by the theoretical predictions of small-scale magnetic fields 
    generated by streaming instabilities at collisionless shocks \citep{Weibel-59,Gruzinov-Waxman-99,Medvedev-Loeb-99,Bret-09,Keshet+09}.
    \item $\boldsymbol{B_\parallel}$: An ordered field aligned along the local velocity vector of the outflow. This field structure presents the opposite extreme 
    of $B_\perp$, and in reality the shock-generated field may likely be (at least its emissivity-weighted mean value over the emitting region downstream of the 
    shock) more isotropic than anisotropic whereby it would be a distribution in the 
    $B_\perp-B_\parallel$ parameter space (see, e.g., \citep{Granot-Konigl-03,Gill-Granot-20} in the context of afterglow collisionless shocks). 
    \item $\boldsymbol{B_{\rm tor}}$: A globally ordered toroidal field symmetric around the jet symmetry axis. Such a field configuration naturally arises in a 
    high magnetization flow in which the dynamically dominant field is anchored either to the rotating central engine or in the accretion disk. The 
    azimuthal motion of the magnetic footpoints tightly winds up the field around the axis of rotation, which is also the direction along which 
    the relativistic jet is launched. Due to magnetic flux conservation, the poloidal component declines ($B_p\propto R^{-2}$) more rapidly 
    as compared to the toroidal component ($B_\phi\propto R^{-1}$) as the flow expands. Therefore, at large distances from the central engine the 
    toroidal field component dominates.
\end{enumerate}

\end{paracol}
\begin{figure}
    \widefigure
    \centering
    \includegraphics[width=0.42\textwidth]{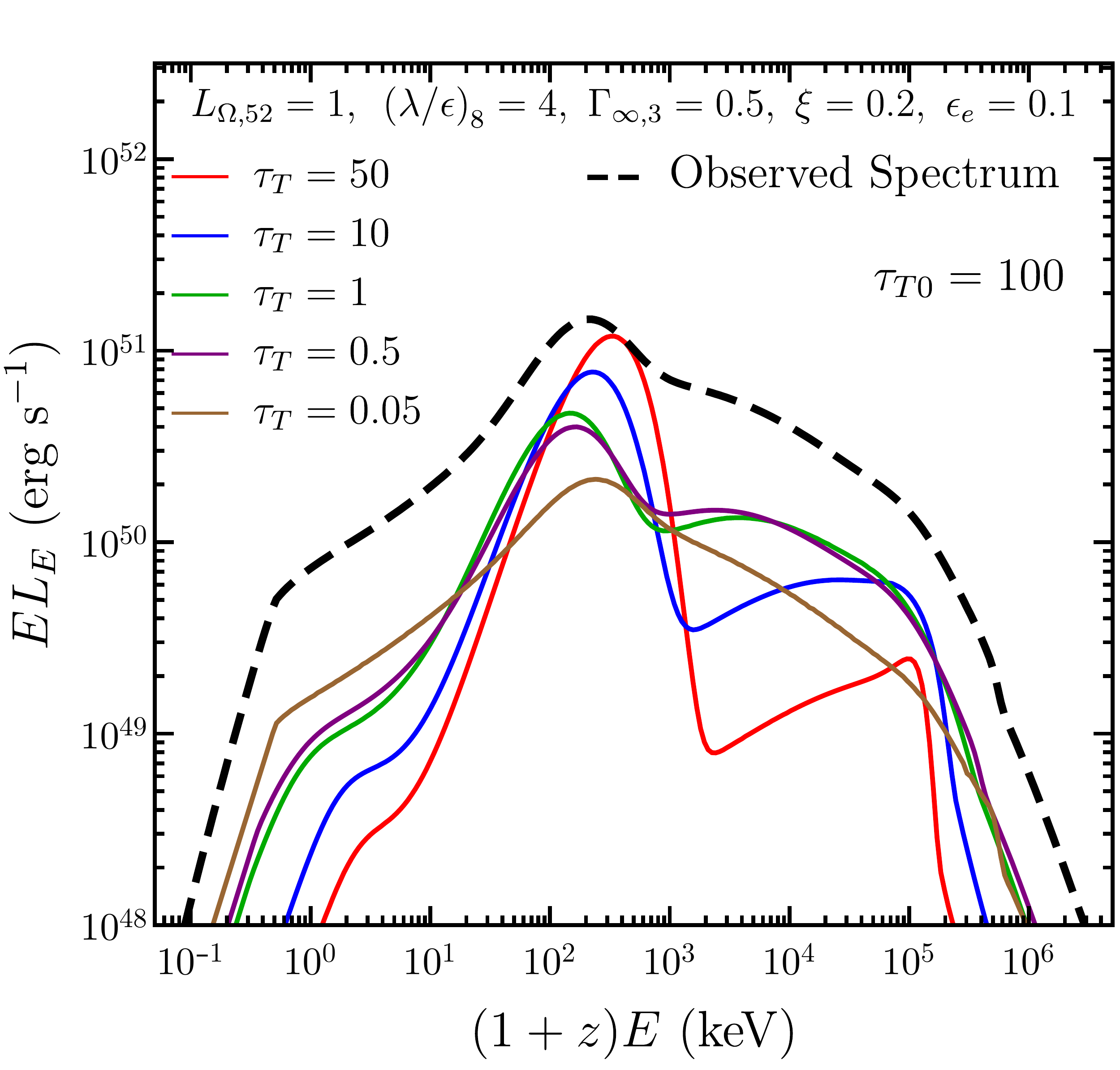}\hspace{3em}
    \includegraphics[width=0.42\textwidth]{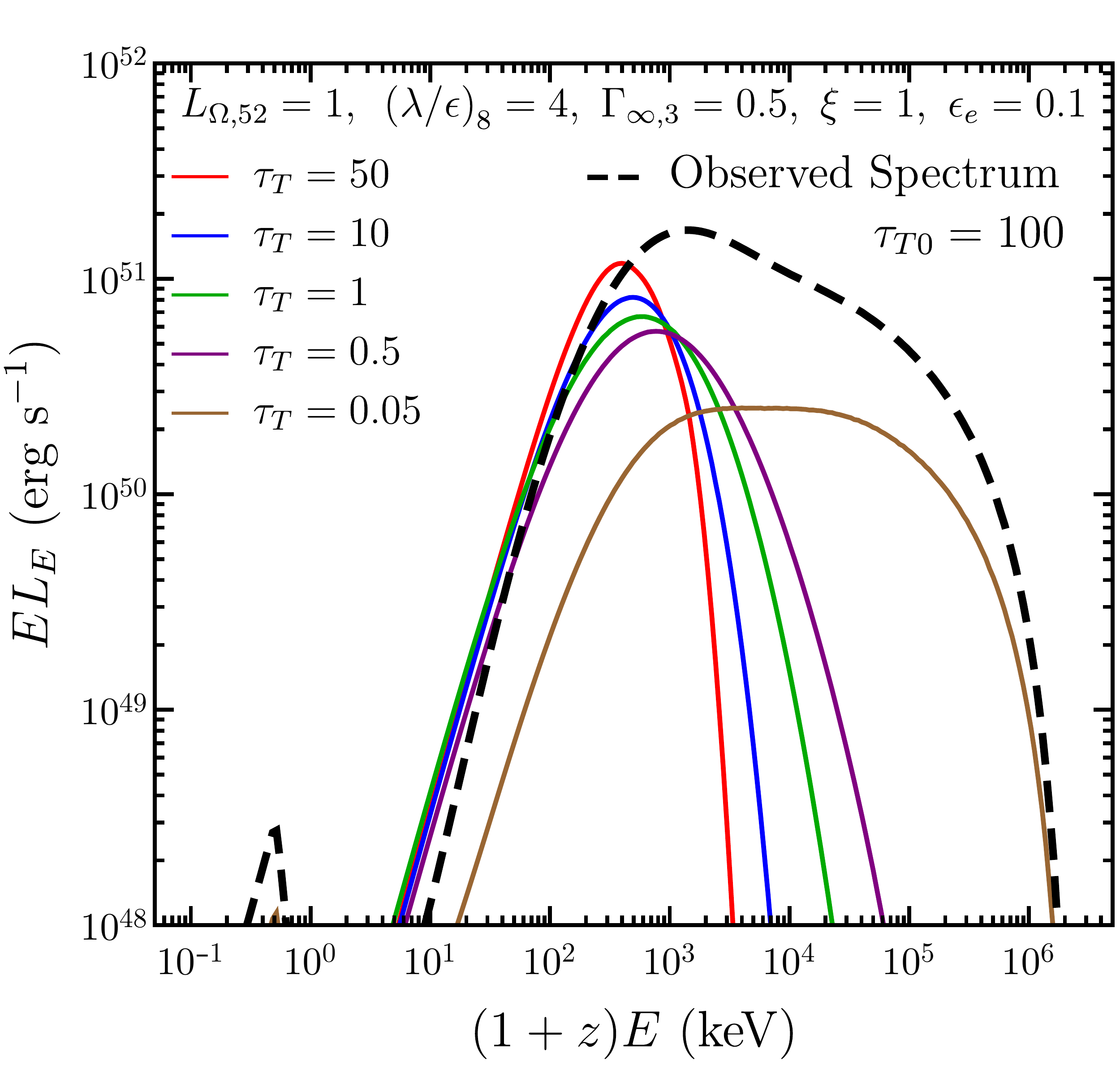}
    \caption{Spectral evolution in a dissipative steady PFD striped wind flow, shown as a function of the Thomson optical depth as the jet 
    is heated accross the photosphere. The spectra are shown for two different particle heating scenarios: \textbf{\textit{Left}} -- relativistic power-law 
    particles produced by magnetic reconnection, and \textbf{\textit{Right}} -- mildly relativistic particles forming an almost mono-energetic distribution 
    due to distributed heating and Compton cooling. The flow is evolved from initial $\tau_{T0}=100$ until the total optical depth of baryonic electrons 
    plus produced $e^\pm$-pairs is much less than unity. The observed spectrum is effectively a sum over the optically thin spectra. 
    See \citep{Gill+20b} for more details.}
    \label{fig:spectrum-evolution}
\end{figure}
\begin{paracol}{2}
\switchcolumn

\subsubsection{\textbf{Inverse-Compton Emission}}

If the energy density of the (isotropic) radiation field ($U_\gamma' = 3L_\gamma/16\pi R^2\Gamma^2c$, where $L_\gamma$ is the isotropic-equivalent 
luminosity) advected with the flow is much larger than that of the magnetic field ($U_B'=B'^2/8\pi$), relativistic particles with LF $\gamma_e$ cool predominantly by 
inverse-Compton upscattering softer seed photons, with energy $E_s'$, to higher energies with a mean value (for an isotropic seed photon field in the comoving frame),
$E'=(4/3)\gamma_e^2E_s'$. When the Thomson optical depth of the flow is $\tau_T >1$, 
these seed photons undergo multiple Compton scatterings, where the process is usually referred to as \textit{Comptonization}, until they are 
able to stream freely when $\tau_T<1$. Comptonization has been argued as a promising alternative to optically-thin synchrotron emission where it is able 
to explain a broader range of low-energy spectral slopes, provide a characteristic energy scale for the peak of the emission, and yield narrower 
spectral peaks \citep{Beloborodov-13,Beloborodov-Meszaros-17} 
It is the main radiation mechanism in a general class of models known as \emph{photospheric emission} 
models in which the outflow is heated across the photosphere due to some internal dissipation.

At the base of the flow, where $\tau_T\gg1$, the radiation field is thermalized and assumes a Planck spectrum. If the outflow remains non-dissipative 
the Planck spectrum is simply advected with the flow, cooled due to adiabatic expansion, and then released at the photosphere 
\citep{Goodman-86,Paczynski-86}. However, only a few GRBs show a clearly thermally-dominated narrow spectral peak \citep{Ryde-04} whereas most have a 
broadened non-thermal spectrum with low-energy photon index ($\alpha_{\rm ph} < 1$) softer than that obtained for the Planck spectrum ($\alpha_{\rm ph}=1$). In many 
cases, a sub-dominant thermal component in addition to the usual Band-function has been identified \citep{Guiriec+11,Guiriec+17}. These observations 
imply that photospheric emission plays an important role \citep{Ryde+11} but the pure thermal spectrum must be modified by dissipation across the photosphere 
\citep{Eichler-Levinson-00,Meszaros-Rees-00,Rees-Meszaros-05,Thompson+07,Beloborodov-10}. Several theoretical works tried to understand the thermalization 
efficiency of different radiative process, e.g. Bremmstrahlung, cyclo-synchrotron, and double Compton, below the photosphere to explain the location of 
the spectral peak and the origin of the low-energy spectral slope \citep[e.g.,][]{Vurm+13,Thompson-Gill-14,Begue-Peer-15}.

While sub-photospheric dissipation and Comptonization is able to yield the typical low-energy slope, further dissipation near and above the photosphere 
is needed to generate the high-energy spectrum above the thermal spectral peak. This can be achieved by inverse-Compton scattering of the thermal peak 
photons by mildly relativistic electrons \citep{Giannios-06,Peer+06,Giannios-08,Gill-Thompson-14,Vurm-Beloborodov-16,Gill+20}. If the flow is uniform, the net polarization of the Comptonized spectrum is negligible due to random orientations of the polarization vector at 
each point of the flow which, upon averaging over the visible part, adds up to zero polarization. Alternatively, if the flow has angular structure, particularly 
in the bulk-$\Gamma$ profile, then net polarization as large as $\Pi\lesssim20\%$ can be obtained \citep{Lundman+14,Gill+20}.

\subsubsection{\textbf{Dissipative Jet: Hybrid Spectrum}}\label{sec:hybrid-spectrum}

If the jet is dissipative across the photosphere, a hybrid spectrum can emerge where the spectral peak is dominated by a quasi-thermal component but the low and 
high-energy wings are dominated by non-thermal emission either from synchrotron or Comptonization. The final outcome depends on the nature of the 
dissipation and how that leads to particle acceleration/heating. \citet{Gill+20b}, who carried out numerical simulations and \citet{Beniamini-Giannios-17} 
who performed semi-analytic calculations, considered a steady PFD striped wind outflow, which is heated due to magnetic dissipation commencing at radii when 
the flow is optically thick to Thomson scattering with initial $\tau_{T0}=100$. At higher $\tau_T$, and equivalently lower radii, the flow maintains 
thermal equilibrium while it is being accelerated due to gradual magnetic dissipation. Localized reconnecting layers accelerate the baryonic electrons, 
as well as any produced $e^\pm$-pairs, into a relativistic power-law distribution. In this instance, since the flow is 
strongly magnetized with $\sigma>1$, the relativistic particles are predominantly cooled by synchrotron emission. The development of the spectrum as 
the flow expands is shown in the left panel of Fig.~\ref{fig:spectrum-evolution} as a function of the total $\tau_T$. The final observed spectrum 
is indeed Band-like but it is different from the optically-thin synchrotron spectrum even though by the end of the radially extended dissipation the total 
spectrum (energetically) is synchrotron dominated.

Alternatively, particle heating can occur in a distributed manner \citep{Thompson-94,Ghisellini-Celotti-99,Giannios-06,Giannios-Spruit-07,Giannios-08} 
throughout the whole causal region due to MHD instabilities. In this case, particles remain only mildly relativistic. Their mean energy is governed by a 
balance between (gradual and continuous) heating and Compton cooling, which leads to a mono-energetic distribution. The spectral evolution as the flow expands 
is shown in the right panel of Fig.~\ref{fig:spectrum-evolution}. In this case the high-energy spectrum is again Band-like but unlike the previous case 
it is completely formed through Comptonization \citep{Giannios-06,Giannios-Spruit-07,Giannios-08}. The mildly relativistic particles do produce some synchrotron 
emission, but only at energies $(1+z)E\lesssim1\,$keV.

Both particle energization mechanims can give rise to a Band-like spectra, however, they can produce completely different energy-dependent polarization. 
In both, if the jet is uniform and can be approximated as part of a spherical flow (i.e. away from the jet edge in a top-hat jet) then no polarization is expected 
near the spectral peak, as it is dominated by the quasi-thermal component. 
In such a scenario, away from the peak, where the spectrum is dominated by non-thermal emission, it is possible to measure high polarization ($\Pi\lesssim50\%$) 
if the emission is synchrotron and the flow has a large scale ordered magnetic field, e.g. a $B_{\rm tor}$ field. Other field configurations, 
namely $B_\perp$ and $B_\parallel$, will yield vanishingly small net polarization. Alternatively, if the non-thermal component is produced by Comptonization, then 
the expected polarization is again almost zero. On the other hand, if the LOS passes near the sharp edge of a top-hat jet or the edge of the almost uniform core 
in a structured jet, then the entire spectrum with non-thermal emission from Comptonization can produce $\Pi\lesssim20\%$. Similarly, the non-thermal wings coming 
from synchrotron emission can now yield significant polarization with $4\%\lesssim\Pi\lesssim28\%$ for $B_\perp$ and $10\%\lesssim\Pi\lesssim56\%$ for $B_\parallel$, 
while $B_{\rm tor}$ again yields higher levels with $10\%\lesssim\Pi\lesssim60\%$.

\subsubsection{\textbf{Other Proposed Mechanisms}}

\noindent (a) \textbf{\textit{Compton Drag}} \vspace{0.15cm}\\
\noindent
This model envisions the propagation of the relativistic outflow in a dense bath of seed photons with energy $E_{\rm seed}$ that provide the drag for 
the expanding outflow whereby cold electrons in the outflow Compton upscatter soft photons \citep{Lazzati+00,Ghisellini+00}. The seed photons can be 
provided either by radiation from the associated supernova remnant that exploded a time $\Delta t\simeq\,$few hours before the outflow is launched, or 
if $\Delta t$ is negligibly small then by the walls of the funnel that has been cleared in the massive star progenitor's envelope by the jet-driven bow 
shock post core-collapse. These requirements limit the applicability of this 
model to only long-soft GRBs and doesn't explain how such an emission would arise in short-hard GRBs. This scenario presents an entirely non-dissipative 
flow, which is insensitive to the magnetization but yields a high ($\lesssim50\%$) radiative efficiency . To produce the variability the flow is required 
to be unsteady. The required $\tau_T\gtrsim1$ in this model may make it difficult to produce prompt high-energy emission due to opacity to pair production.

When the prompt GRB emission originates inside the funnel, it is assumed that the funnel is pervaded by a blackbody radiation field emitted by the funnel 
walls. The spectral peak of the observed prompt emission is then simply the inverse-Compton scattered peak at energy $E_{IC}\sim\Gamma^2E_{\rm seed}$, 
where $\Gamma$ is the bulk LF of the outflow. Inhomogeneity in the funnel temperature and bulk-$\Gamma$ of subsequent shells, that could also collide 
to produce internal shocks, gives rise to a Band-like broadened spectrum. The local polarization, i.e. from a 
given point on the outflow surface, can be as high as $100\%$, however, the net observable polarization, e.g. in a top-hat jet, is reduced to $\Pi\lesssim50\%$ 
for a jet with $(\Gamma\theta_j)^2>10$ \citep{Gill+20}. If the jet is narrower than this with $(\Gamma\theta_j)^2<10$ then the net polarization can be much 
higher with $\Pi\lesssim95\%$ \citep{Lazzati+04}. However, such high polarization requires highly idealized assumptions that are hard to meet in reality.\\
\\
\noindent (b) \textbf{\textit{Jitter Radiation}} \vspace{0.15cm}\\
\noindent
If the magnetic field coherence length is much smaller than the gyroradius of particles, then synchrotron radiation is not the correct description of 
the radiative mechanism by which relativistic electrons cool, as it assumes homogeneous fields. In this case, the particles 
experience small pitch-angle scattering where their motion is deflected by magnetic field inhomogeneities by angles that are smaller than the beaming cone 
of the emitted radiation ($1/\gamma_e$). This scenario has been proposed as a viable alternative to synchrotron radiation \citep{Medvedev-00}, where 
it has been shown to yield harder spectral slopes that cannot be obtained in optically-thin synchrotron emission. In addition, it can produce sharper 
spectral peaks as compared to synchrotron radiation, which agrees better with observations. The small-scale magnetic fields needed in this scenario may 
potentially be produced in relativistic collisionless shocks via the Weibel instability (although this may not be easy to achieve in practice; see e.g. \citep{Sironi-Spitkovsky-09}). The polarization when this small-scale field is confined to a slab normal to the local fluid velocity is calculated in 
\citep{Mao-Wang-13,Mao-Wang-17}, where it is shown that the maximum degree of polarization is obtained only at large viewing angles when the slab is viewed 
almost edge-on. For small viewing angles that apply to distant GRBs, the polarization is indeed very weak. \\
\\
\noindent (c) \textbf{\textit{Synchrotron Self-Compton}} \vspace{0.15cm}\\
\noindent
Synchrotron self-Compton (SSC) emission has been considered in some works as a mechanism that could yield low-energy spectral slopes with photon indices 
as hard as $\alpha_{\rm ph}=0$, a change of $\Delta\alpha_{\rm ph}=2/3$ over the synchrotron line of death \citep{Panaitescu-Meszaros-00}. This is facilitated by the 
fact that for typical values of the model parameters in the internal shock scenario, optically-thin synchrotron emission peaks at much lower energies (at 
few eV when the SSC peak is at $\sim100\,$keV) and is mostly self-absorbed. One of the major drawbacks of this radiation mechanism is that 
it requires the synchrotron emission in the optical, which is the seed for the harder inverse-Compton emission, to be much (by a factor of $\gtrsim10^3$) 
brighter than observed (or upper limits \citep{Yost+07}). Otherwise, it requires the Compton-$y$ parameter to exceed unity by the same factor 
which is hard to accommodate while not strongly violating the total energy budget of the burst \cite{Derishev+01,Piran+09}. The energy-dependent local 
polarization for SSC in an ultrarelativistic spherical flow for two different ordered B-field configurations, one parallel and the other transverse to the 
local velocity vector, is calculated in \citep{Chang-Lin-14}, where they find that the local polarization can be as high as $\Pi\lesssim25\%$ under simplifying 
assumptions.

\subsection{\textbf{\textit{What's the angular structure of the outflow?}}}
\label{sec:structure}

The angular structure of the relativistic outflow in GRBs affects a number of important observables, such as prompt GRB pulse 
structure \citep{Gill-Granot-21}, polarization \citep{Gill+20}, afterglow lightcurve \citep{Gill-Granot-18a}, and the detectability 
of distant GRBs \citep{Lazzati+17a,Beniamini-Nakar-19}. Outflows in GRBs are collimated into narrowly beamed bipolar jets that have angular 
scale $\Gamma\theta_j\sim10$, where $\theta_j$ in the simplest model of a uniform conical jet, also referred to as a \textbf{top-hat} 
jet, represents a sharp edge. The notion of narrowly collimated jets in GRBs was first proposed by \citet{Rhoads-97} and it was 
later verified by observations of achromatic \textbf{jet-breaks} in the afterglow lightcurve that yielded $\theta_j\simeq0.05-0.4$ 
\citep[e.g.,][]{Harrison+99,Kulkarni+99,Frail+01,Berger+03}. Since $\Gamma\gtrsim10^2$ during the prompt emission phase and assuming 
that $\theta_j$ remains approximately the same, this yields $\Gamma\theta_j\sim5-40$. This geometric beaming futher implies that the 
true radiated energy of these bursts is much smaller \citep{Frail+99} with $E_\gamma = f_bE_{\gamma,\rm iso}\sim10^{48}-10^{52}\,$erg, 
where $f_b=(1-\cos\theta_j)\simeq\theta_j^2/2$ is the geometric beaming fraction with the last equality valid for $\theta_j\ll1$, 
$E_{\gamma,\rm iso}=4\pi d_L^2S_\gamma(1+z)^{-1}\sim10^{48}-10^{55}\,$erg is the isotropic-equivalent radiated energy, $S_\gamma$ [${\rm erg\,cm}^{-2}$] 
is the burst fluence, and $d_L$ is the luminosity distance. Since $f_b$ is much smaller than $4\pi$, the solid angle into which radiation 
from a spherical source is emitted, only observers whose line-of-sight (LOS) intersects the surface of the jet or passes very close to 
the jet edge can detect the GRB, which implies that the true rate of GRBs is enhanced by $\mean{f_b^{-1}}\sim500$ \citep{Frail+99} over the 
observed rate.


A top-hat jet is clearly an idealization even though it is able to explain several features of the afterglow lightcurve. Numerical simulations of 
jets breaking out of the progenitor star for the long-soft GRBs \citep{Zhang+03,Zhang+04,Morsony+07,Mizuta-Ioka-13,Gottlieb+21} and that from the 
dynamical ejecta for the short-hard GRBs \citep{Aloy+05,Lazzati+17b,Gottlieb+18,Nathanail+20,Gottlieb+21,Nathanail+21} find that these jets naturally 
develop angular structure by virtue of their interaction with the confining medium. If the true energy reservoir lies in a narrow range and the scatter 
in $E_{\gamma,\rm iso}$ is instead caused by different viewing angles, then either the jet half-opening angle 
of a top-hat jet must be different in different GRBs or the jets are not uniform and must have an underlying angular profile for both the energy per 
unit solid angle, $\epsilon(\theta)=E_{\rm iso}(\theta)/4\pi$, and the (initial) bulk LF, $\Gamma=\Gamma(\theta)$. Such jets are commonly referred to as 
\textbf{structured jets} \citep{Zhang-Meszaros-02,Kumar-Granot-03,Granot-Kumar-03,Panaitescu-Kumar-03} and can be parameterized quite generally as a 
power law with $\epsilon(\theta)\propto\Theta^{-a}$ and $\Gamma(\theta)-1\propto\Theta^{-b}$ where $\Theta = \sqrt{1+(\theta/\theta_c)^2}$ with $\theta_c$ 
being the core angle. A constant true jet energy among a sample of GRBs implies that $a=2$, 
a model referred to as a \textbf{universal structured jet} (USJ) \citep{Meszaros+98,Lipunov+01,Rossi+02,Zhang-Meszaros-02}, where it corresponds to equal 
energy per decade in $\theta$ and therefore reproduces jet breaks similar to those for a top-hat jet with $\theta_j($top-hat)\;$\sim\theta_{\rm obs}({\rm USJ})$.
This angular profile was used as an alternative model to the top-hat jet to explain the $E_{\gamma,\rm iso}\propto S_\gamma \propto t_b^{-1}$ correlation 
\citep{Frail+01} for the afterglow emission where $t_b$ is the jet-break time \citep{Rossi+02,Zhang-Meszaros-02}. Other useful parameterizations of a 
structured jet include a \textbf{Gaussian jet} with $\epsilon(\theta)\propto\Gamma(\theta)-1\propto\exp(-\theta^2/2\theta_c^2)$, which is a slightly 
smoother (around the edges) and more realistic version of the top-hat jet.

The large distances of GRBs have precluded direct confirmation and constraints of the outflow's angular structure. The main difficulty being the rather severe 
drop in fluence when they are observed outside of the almost uniform core. This changed recently with the 
afterglow observations of GRB$\,$170817A \citep{Abbott+17-GW170817-GRB170817A}, the first-ever short-hard GRB detected coincidentally with GWs 
\citep[GW$\,$170817;][]{Abbott+17-GW170817-Ligo-Detection} from the merger of two neutron stars. Helped by it's nearby distance of $D\simeq40\,$Mpc 
and an impressive broadband (from radio to X-rays) observational campaign \citep[e.g.,][]{Abbott+17-GW170817A-MMO,Hallinan+17,Troja+17b}, the afterglow 
observations lead to the first direct and significant constraint on the angular structure of the relativistic jet 
\citep[e.g.,][]{Troja+17b,DAvanzo+18,Gill-Granot-18b,Lamb-Kobayashi-18,Lazzati+18,Margutti+18,Resmi+18,Gill+19}. The afterglow from this source showed 
a peculiar shallow rise 
($F_\nu\propto t^{0.8}$) to the lightcurve peak at $t_{\rm pk}\simeq150\,$days, after which point it declined steeply ($F_\nu\propto t^{-2.2}$). 
Several useful lessons were learned. First, it was shown that a top-hat jet can only explain the afterglow lightcurve near and after the lightcurve peak 
\citep{Gill+19} and not the shallow rise for which a structured jet is needed. Second, both power-law and Gaussian structured jets can explain the afterglow 
of GW170817, where for a power-law jet the angular structure profile requires $a\sim4.5$ and $b\gtrsim1.2$ to explain all the observations \citep{Beniamini+20}.

While power-law and Gaussian structured jets remain most popular, a few other angular profiles have received some attention. Among them is the two-component 
jet model \citep{Panaitescu+98,Frail+00,Berger+03,Huang+04,Peng+05,Racusin+08} that features a narrow uniform core with initial bulk LF $\Gamma_0\gtrsim100$ 
surrounded by a wider uniform jet with $\Gamma_0\sim10-30$. Nothing really guarantees or demands the outflow to be axisymmetric and uniform, in which case 
an outflow with small variations on small ($\ll 1/\Gamma$) angular scales can be envisioned in the form of a ``patchy shell'' \citep{Kumar-Piran-00} or an 
outflow consisting for ``mini-jets'' \citep{Yamazaki+04}, with the caveat that significant variations on such causally connected angular scales are rather 
easily washed out and hard to maintain. In case such variations do indeed persist, it could have important consequences for the time-resolved 
polarization and PA. For example, patches or mini-jets can have different polarization and/or PA due to mutually incoherent B-field configurations 
which can lead to smaller net polarization and PA evolution.

\end{paracol}
\begin{figure}
    \widefigure
    \centering
    \includegraphics[width=1.0\textwidth]{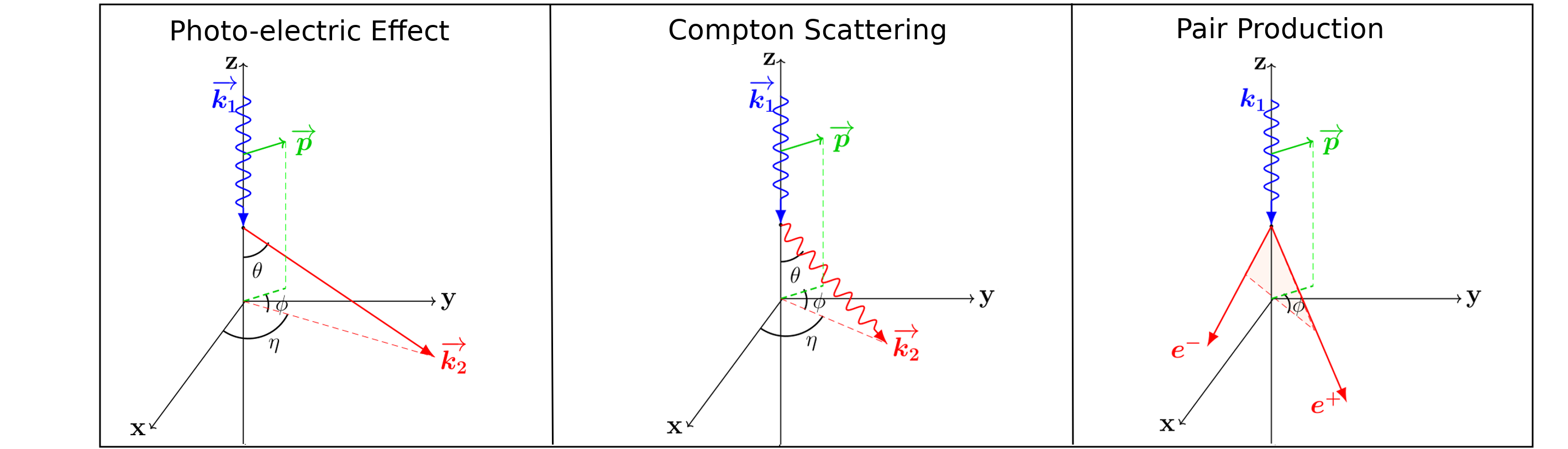}
    \caption{Illustration of the angular dependency of the interaction product on the polarization vector of the incoming photon for the three interaction mechanisms: photo-electric effect (left), Compton scattering (middle) and pair production (right). The incoming photon is shown in blue, its polarization vector in green and the secondary product(s) in red. The $\theta$ angle (as used in equations \ref{eq:phi_dsigma_dOmega} and \ref{eq:KN}) is defined as the angle between the incoming photon direction and its secondary product. The $\phi$ angle (again as used in equations \ref{eq:phi_dsigma_dOmega} and \ref{eq:KN}) is defined as the angle between the projections of the polarization vector and the momentum vector of the secondary product(s) onto the $x$-$y$ plane. The $\eta$ angle is the azimuthal angle between the $x$-axis and the projection of the momentum vector of the secondary particle onto the $x$-$y$ plane. The $\theta$ and $\eta$ angles can be directly measured in a detector, while $\phi$ is measured indirectly.}
    \label{fig:Interaction}
\end{figure}
\begin{paracol}{2}
\switchcolumn

\section{Gamma-ray Polarimetry} 
\label{sec:gamma-ray-polarimetry}
Despite the wealth of information that can be obtained from prompt GRB polarization, only a few 
measurements with modest statistical significance exist. Moreover, many of the results presented in the past were refuted by
follow-up studies. A detailed overview of many of these measurements and their respective issues is provided in \cite{McConnell+16}. 
The two most recent measurements, by POLAR \cite{Kole+20} and Astrosat CZT \cite{Chattopadhyay+19}, furthermore appear to be incompatible with one another, indicating probable issues in at least one of these results as well. 
The lack of detailed measurements, and the many issues with them, result from both the difficulty in measuring $\gamma$-ray polarization as well as 
challenging data analysis at these energies. Below we discus first the measurement principle, which causes many of the encountered issues. This is 
followed by a description of the different instruments that have been able to perform measurements to date. 

\subsection{Measurement Principles}\label{sec:measurement-principles}

The polarization of X-ray or $\gamma$-ray photons can be measured by studying the properties of the particles created during 
their interaction within the detector. For all the three possible interaction mechanisms, namely the photo-electric effect, Compton scattering 
and pair production, a dependency exists of the orientation of the outgoing products on the polarization vector of the incoming photon. This is illustrated in Fig.~\ref{fig:Interaction} for the three processes. For the photo-electric effect it is the azimuthal direction of the outgoing electron which shows a dependency on the polarization vector of the incoming photon, for Compton scattering it is the azimuthal scattering angle of the photon and for pair-production the plane defined by the electron-positron pair. 

The differential cross section for photo-absorption (via the photo-electric effect) has a dependency on $\phi$ which is defined as the 
azimuthal angle between the polarization vector of the incoming photon $\vec{p}$, as shown in Fig.~\ref{fig:Interaction}, and the projection 
of the velocity vector of the final state electron $\vec{\beta} = \vec{v}/c$ (where $\hat{\beta}=\hat{k}_2$) on to the plane normal to the 
momentum vector $\vec k_1$ of the photon,
\begin{equation}\label{eq:phi_dsigma_dOmega}
    \frac{d\sigma}{d\Omega}\propto \cos^2\phi\, \quad,\quad 
    \phi = \cos^{-1}\left(\frac{\vec{\beta}\cdot\vec{p}}{\beta p\sin\theta}\right)=\cos^{-1}\left(\frac{\vec{k}_2\cdot\vec{p}}{k_2 p\sin\theta}\right)\ ,
\end{equation}
where $d\Omega=\sin\theta d\theta d\phi$ is the unit solid angle and the polar angle $\theta$ is given by 
$\cos\theta=\hat{\beta}\cdot\hat{k}_1=\hat{k}_2\cdot\hat{k}_1$. Similarly, for the differential cross section of Compton scattering 
the dependence on $\phi$, here the angle between the polarization vector of the incoming photon $\vec{p}$ and the projection of the momentum vector 
of the outgoing photon $\vec{k}_2$ on to the plane normal to the momentum vector $\vec k_1$ of the incoming photon, where 
$\phi = \cos^{-1}\left(\frac{\vec{k}_2\cdot\vec{p}}{k_2 p\sin\theta}\right)$ as in Eq.~(\ref{eq:phi_dsigma_dOmega}), is
\begin{equation}\label{eq:KN}
    \frac{d\sigma}{d\Omega} = \frac{r_o^2}{2}\frac{E'^2}{E^2}\left(\frac{E'}{E}+\frac{E}{E'}-2\sin^2\theta \cos^2\phi\right).
\end{equation}
Here $r_0 = e^2/m_ec^2$ is the classical electron radius with $e$ being the elementary charge, $E$ is the initial photon energy, $E'$ the 
final photon energy, $\theta=\cos^{-1}(\hat{k}_2\cdot\hat{k}_1)$ the polar scattering angle.

\end{paracol}
\begin{figure}
    \widefigure
    \centering
    \includegraphics[width=0.80\textwidth]{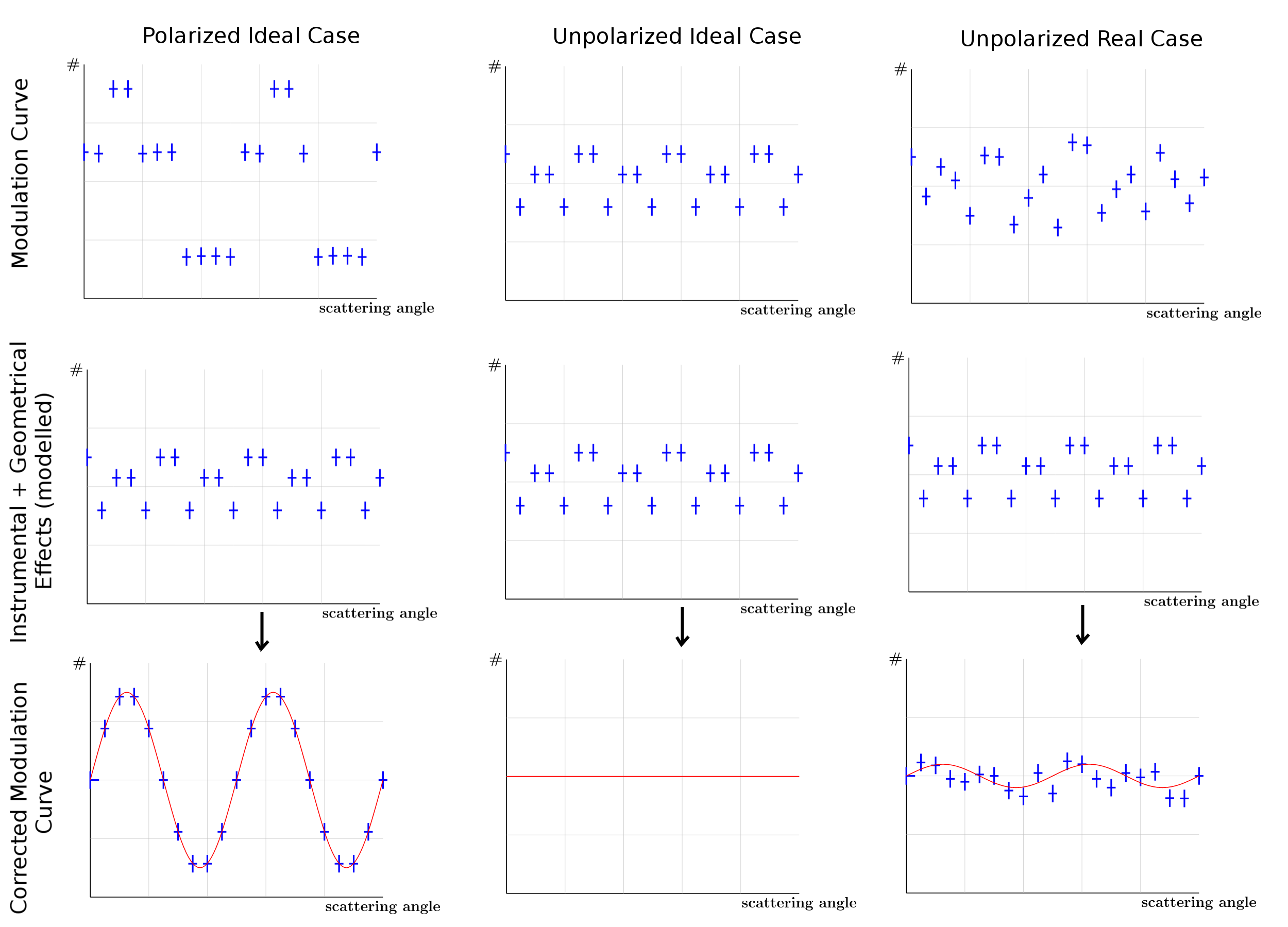}
    \caption{Illustration of recovering the polarization signal from a raw modulation curve. The left column illustrates 
    the ideal case with a high PD value, with a raw measured modulation curve (top), the perfectly simulated instrumental 
    and geometrical effects (middle) which pollute the raw modulation curve. The bottom-panel shows the modulation curve 
    after correction from the instrumental and geometrical effects which results in a perfect harmonic function. The middle 
    column illustrates the same but for an unpolarized signal resulting in a flat distribution. The right column shows the same for an unpolarized signal, however, 
    random small errors are added to the instrumental and geometrical effects, thereby simulating a non-perfect understanding 
    of the instrument. The result is a non-flat distribution which, when fitted, shows a low level of polarization.}
    \label{fig:positive_error}
\end{figure}
\begin{paracol}{2}
\switchcolumn

Finally, for pair production the differential cross section is $d\sigma/d\Omega\propto1+A(\cos{2\phi})$, where $A$ is the polarization asymmetry of the
conversion process (which has dependencies on the photon energy and properties of the target) and $\phi$ is the angle between the polarization vector of the incoming photon $\vec{p}$ and the plane defined by the momentum vectors of the electron-positron 
pair, $\vec{k}_{\pm}$.

The general concept for polarimetry in the three energy regimes 
where these cross sections dominate is therefore similar: One needs to detect the interaction itself and subsequently track the secondary particle, 
be it an electron, photon or electron-positron pair. This requirement indicates the first difficulty in polarimetry: simply absorbing the incoming 
photon flux, as is the case in, for example, standard spectrometry, is not sufficient. The requirement to track the secondary product significantly 
reduces the efficiency of the detector. 

After measuring the properties of the secondary particles, a histogram of $\phi$ can be made which shows a sinusoidal variation with a period of 
$180^\circ$ referred to as a modulation curve. The amplitude of this is proportional to the polarization degree (PD) and the phase related to the 
polarization angle (PA). As can be derived from, for example, the Compton scattering cross section, the amplitude for a $100\%$ polarized beam will 
depend on the energy of the incoming photons as well as on the polar scattering angle. Whereas the energy depends on the source, 
the polar scattering angle is indirectly influenced by the instrument design. For example, using a detector with a thin large surface perpendicular to 
the incoming flux, it is more likely to detect photons scattering with a polar angle of $90^\circ$, which have a larger sensitivity to polarization than those scattering 
forward or backward. The relative amplitude, meaning the ratio of the amplitude of the sinusoidal over its mean, is directly proportional to the PD. The relative amplitude one detects for a $100\%$ polarized beam is known as the $M_{100}$ and it depends on the source spectrum, source location in the sky, the instrument design and the analysis. Although for specific circumstances the $M_{100}$ can be measured on the ground using, for 
example, mono-energetic beams with a specific incoming angle w.r.t the detector, its dependency on the energy, incoming angle and instrument 
conditions such as its temperature, implies that the $M_{100}$ required in the analysis of real sources can only be achieved using simulations. The 
large dependency on simulations provides a source for potential systematic errors in the analysis, which can easily dominate 
the statistical error in the measurements.

Additionally, it should be noted that in practice retrieving the polarization is significantly more complicated as both instrumental and 
geometrical effects (such as the incoming angle of the photons w.r.t. the detector and the presence and orientation of materials around the detector) 
are added to the polarization induced signal in the modulation curve. In order to retrieve the polarization signal one can, for example, divide 
it by a simulated modulation curve for an unpolarized signal as illustrated in the first column of Fig.~\ref{fig:positive_error}. 
This method is often used, for example in \cite{Chattopadhyay+19}. A second option is to model these effects together with the signal and fit the uncorrected 
curve with this simulated response, as was for example done in \cite{Burgess+19}. In either case, it requires a highly detailed understanding of the 
instrument. 

In polarization analysis any imperfections in modelling the instrument will likely result in an overestimation
of the polarization. As illustrated in Fig.~\ref{fig:positive_error}, for a modulation curve resulting from an unpolarized flux, removing any 
instrumental effects from the modulation curve should result in a perfectly flat distribution. This is illustrated in the middle column of this figure. 
Any error in the model of the instrumental or geometrical effects will, however, result in a non-flat distribution, which, when fitted with a harmonic function, 
will result in some level of polarization to be detected. It is therefore in practice impossible to measure a PD of $0\%$ as it would 
require both an infinite amount of statistics, and more importantly, a perfect modelling of all the instrumental effects. On the other extreme, 
for a PD of $100\%$, imperfections in the modelling can result in a lower amplitude, but can still also increase it further resulting in measuring 
a nonphysical PD. Overall, due to the nature of the measurement, both statistical and systematic errors tend to inflate the PD rather than decrease 
it. Since it is not possible to test the modelling of the instruments when in orbit, as there are no polarization calibration sources, this issue 
exists for all measurements and can only be minimized by extensive testing of the instrument both on the ground and in-orbit.

A final figure of merit often used in polarimetry is the Minimal Detectable Polarization (MDP) \cite{Weisskopf2013}. For GRBs the MDP is best expressed as
\begin{equation}
    \mathrm{MDP} = \frac{2\sqrt{\mathrm{-ln}(1-C.L.)}}{M_{100}C_s}\sqrt{C_s+C_b}\,.
\end{equation}
Here $\mathrm{C.L.}$ is the confidence level, $C_s$ is the number of signal events and $C_b$ the number of background events. The MDP expresses 
the minimum level of polarization of the source which can be distinguished from being unpolarized for a given confidence level. It can therefore 
be seen as a sensitivity of a given polarimeter for a given observation. Whereas this is highly useful for polarimeters observing point sources, for 
GRB polarimeters there is an issue related to the $M_{100}$. For wide field of view instruments, such as polarimeters designed for GRB observations, 
the value of $M_{100}$ can start to depend on the PA of the source. For example, in POLAR, the $M_{100}$ was found to depend on the PA for GRBs 
with a large off-axis incoming angle \cite{Kole+20}. This is a result of only being able to resolve 2 dimensions of the scattering interactions 
in the detector, making it insensitive (so $M_{100}=0$) to certain values of PA when the $\gamma$-ray photons enter the detector perpendicular 
to the readout plane \cite{Kole+20}. 
As in such cases the MDP becomes dependent on PA, it loses its use as a figure of merit. However as the MDP remains highly used in the community, 
and remains the best measure of sensitvity for polarimeters we will use it here in this work as well, although with a small adaptation. In order 
to remove the PA dependence we use the mean MDP where the $M_{100}$ is averaged over all possible values of the a priori unknown PA.

\subsection{Detection Principles}

To date, the only GRB polarization measurements performed have made use of Compton scattering in the detector. The majority of these measurements were performed by making use of a segmented detector concept, for example a detector consisting of many relatively small scintillators, e.g. for GAP \cite{Yonetoku2010} and POLAR \cite{PRODUIT2018}, or a segmented semiconductor, such as INTEGRAL-SPI \cite{SPI_1999} and AstroSAT CZT \cite{CZT_1}. In either design the Compton scattering interaction can be detected in one segment of the detector while an additional interaction of the photon in a second segment can be used to reconstruct the azimuthal Compton scattering angle.  This concept is illustrated in Fig.~\ref{fig:detector}. 

\begin{figure}
    \centering
    \includegraphics[width=0.4\textwidth]{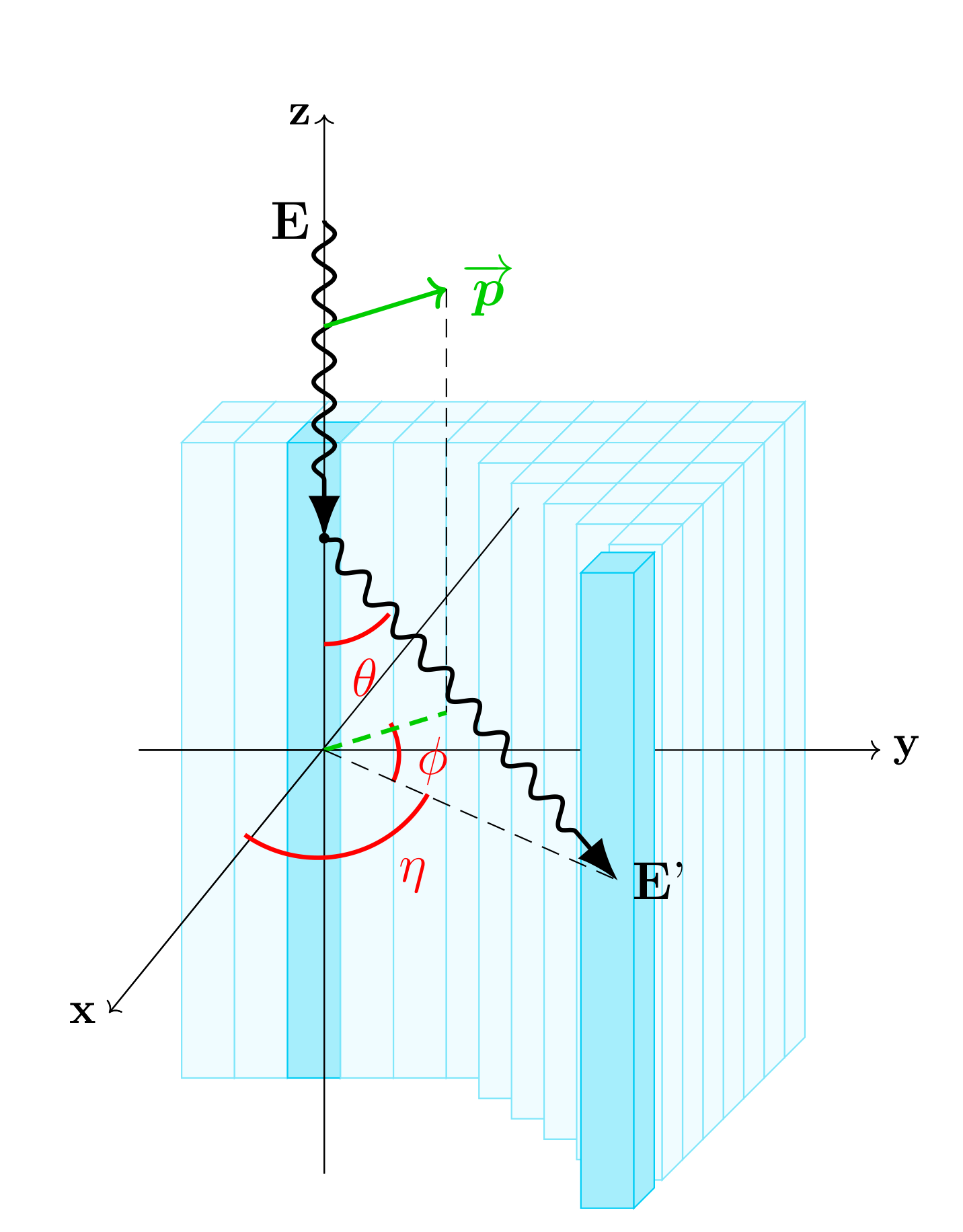}
    \caption{Illustration of the measurement principle of a polarimeter using Compton scattering. 
    The incoming $\gamma$-ray Compton scatters in one of the detector segments followed by a 
    photo-absorption (or second Compton scattering) interaction in a different segment. Using the 
    relative position of the two detector segments the Compton scattering angle can be calculated 
    from which, in turn, the polarization angle can be deduced.}
    \label{fig:detector}
\end{figure}

At energies below approximately $10\;$keV the cross section for photo-absorption dominates. 
Although no successful GRB polarization measurements have been performed using the photo-electric effect, the detection method 
has been successfully used recently to perform polarization measurements of the Crab Nebula in the 3$\,$--$\,$4.5\;keV energy range 
using the PolarLight cubesat \cite{Feng2020}. Several large scale polarimeter ideas have been developed in the past, such as 
the Low Energy Polarimeter which was part of the proposed POET mission which was dedicated to GRBs \cite{McConnell+08}. Currently 
several missions that use the same concept are currently under 
development \cite{IXPE,eXTP}. In these X-ray polarimeters the photo-absorption takes place in a thin gas detector. As the produced electron travels 
through the gas it releases secondary electrons as it ionizes the gas. These secondary electrons can be detected using finely 
segmented pixel detectors in order to track the path of the electron released in the photo-absorption interaction, allowing to 
reconstruct its emission angle.

Polarimetry in the pair-production regime is arguably the most challenging as the photon flux is low and the detection method requires 
highly precise trackers capable of separating the tracks of the electron and the positron. In spectrometry the low photon flux is often 
compensated by using large detectors with a high stopping power. For example, by combining tungsten layers with silicon detectors. Here 
the silicon serves to measure the tracks while the tungsten is used to enforce pair production in the detector. However, the use of high
Z (atomic number) materials, like tungsten, significantly increases multiple scattering of the electron and positron. Multiple 
scatterings quickly change the momentum of both products, thereby making it challenging to reconstruct their original emission direction. 
To overcome this issue detectors which use silicon both for conversion and detection,  have been proposed in the past such as PANGU \cite{PANGU}. 
Although technically possible, the large number of silicon detectors required to achieve a high sensitivity, with minimal structural material 
and a potential magnet which helps to separate the electron-positron pair, make such detectors both costly and challenging to develop for 
space. A second option is to use gas based detectors, such as in the HARPO design \cite{HARPO}. This detector, which was successfully tested 
on the ground \cite{HARPO2}, allows for precise tracking but has a low stopping power for the incoming $\gamma$-rays and therefore a low 
detection efficiency. This can be compensated with a large volume. However, as the gas volume obviously needs to be pressurized, producing 
and launching such an instrument for use in space is highly challenging. Despite these challenges several projects which follow this design are still ongoing such as the potential future space mission AdEPT \cite{Hunter+18}, which aims to use a time projection chamber to measure polarization in the  $5-200\,\mathrm{MeV}$ energy range as well as the balloon borne SMILE missions \cite{SMILE}.

Although no dedicated pair-production polarimeters are currently in-orbit, it should be noted that both the Fermi-LAT \cite{LAT} and the 
AMS-02 \cite{TING201312} instruments could, in theory, be used to perform polarization measurements in this energy range. For Fermi-LAT, 
which is a dedicated $\gamma$-ray spectrometer, consisting of silicon strip detectors combined with tungsten conversion layers, the 
aforementioned multiple scattering induced distortion is again a challenge \citep{Fermi}. The polarization capabilities of Fermi-LAT, which 
has detected many GRBs to this day, has been studied \citep{Fermi} but no results from actual data have been published to date. For AMS-02, 
which does not suffer from the use of tungsten layers and additionally contains a magnet which separates the pairs, 
measurements could be easier. However, as the instrument is designed as a charged particle detector it remains non-optimized for this purpose 
and so far no results have been published by this collaboration.

\subsection{GRB Polarimeters}\label{sec:polarimeters}

In this section we aim to provide a summary of the different instruments which have performed GRB polarization to date. 
For a detailed overview of each individual measurement (up to 2016) the reader is referred to \citep{McConnell+16}. 

As mentioned earlier, 
all polarization measurements of the prompt GRB emission have been performed by making use of Compton scattering. While in the majority of 
cases the Compton scattering takes place in the detector, there is one exception. The attempts at performing polarization measurements with 
data from the BATSE detector made use of Compton scattering from the Earth's atmosphere \cite{BATSE1,BATSE2}. The BATSE detector consisted of 
several scintillator based detectors, and by itself had no capability to directly perform polarimetry \cite{BATSE}. Instead it used several 
detectors pointing towards the Earth, each at different relative angles, to measure the relative intensity of photons scattering off different 
parts of the Earth's atmosphere. As the probability for photons to scatter off the atmosphere towards different detectors depends on their 
polarization properties, such a measurement is possible for any detector with an Earth facing sensitive surface. It does however require a highly 
detailed modelling of the Earth's atmosphere, software capable of simulating the scattering 
effects properly, and detailed understanding of the detector response as well as the location and spectra of the GRB. The large number of sources 
for systematic errors resulted in inconclusive measurements of GRB 930131 \cite{BATSE2}. Despite the initial lack of success, improvements have 
been made since then regarding Compton scattering models in software such as Geant4 \cite{MIZUNO2005}. Furthermore instruments such as 
\textit{Fermi}-GBM have measured 1000's of GRBs over the last decade, similar studies using for example this data, could prove to be successful in the future.

Systematic errors are a major issue not only for the creative polarization measurement 
solution used in BATSE, but in all GRB polarization results published thus far from different instruments. It is especially important for 
measurements performed using detectors not originally designed to perform polarimetry such as RHESSI \cite{RHESSI}, and the SPI and IBIS detectors 
on board INTEGRAL. Both RHESSI and SPI make use of a segmented detector consisting of germanium detectors and thereby allow to study 
Compton scattering events by looking for coincident events between different detectors. The IBIS instrument \cite{IBIS}, uses 2 separate sub-detectors instead, namely the ISGRI detector consisting of 16384 
CdTe detectors, and the  Pixellated Imaging CsI Telescope (PICsIT), an array of 4096 CsI scintillator detectors. Since, similar to RHESSI and SPI, IBIS 
was not originally designed to perform polarization measurements, the trigger logic in the instrument was not setup to keep coincidence events 
in the PICsIT or ISGRI alone. Rather, only coincidence events between the PICsIT and the ISGRI are kept, which although lowering the statistics 
for polarization measurements, still allows for such measurements \cite{Forot2007}.

Since all 3 instruments were not designed as polarimeters, one immediate downside of using them as such is the lack of sensitivity. A clear 
example of this is the non-optimized trigger logic of IBIS. In the 
case of RHESSI, different analyses of the same GRB \citep{Coburn-Boggs-03} resulted in vastly different results, in part due 
to the difficulty in selecting valid coincident events between different germanium detector channels, again a result of a non-optimized online event selection. The relatively imprecise time measurement 
of each event prompted a large coincidence window to be set in one of the analyses which resulted in chance coincidence events induced by 
different photons or background particles instead of the Compton scattering event \cite{Rutledge, Wigger}. If the instrument had been designed 
and tested on the ground as a polarimeter the coincidence trigger logic and time measurement would likely have been optimized and event selection 
methods tested during the calibration phase. 

The lack of on-ground calibration for polarization additionally makes verification of the detector 
response models difficult and prone to errors. For example, dead material around the detector that can affect the polarization of the incoming 
flux when it interacts with it. While such issues are important in spectrometers as well it can be argued that it is more important in a polarimeter. 
Imperfect modelling of certain detector channels for a spectral measurement can cause issues. However, if on average the channels are modelled 
correctly, having a few badly modelled channels will not greatly affect the final flux or spectral result, as over- and under-performing channels 
can cancel each other out. In a polarimeter it is the difference 
in the number of events between the detector channels which provides the final measurement and not, as in a simple spectrometer, the average of 
all the channels. For a polarimeter, however, one single over-performing detector channel would see a larger number of 
scattering events than expected, causing certain scattering angles to be favoured and thereby faking a polarization signal. 

Similarly, dead material in front of the detector channels can easily obscure certain channels more than others causing a similar effect. 
Understanding all these issues during on-ground calibration is therefore crucial to reduce systematic errors. As a result of such
difficulties, the polarization results published by the SPI collaboration clearly mention the 
possibility of significant systematic errors not taken into account in the analysis, which can affect the results \cite{McGlynn2007,Kalemci2007}.

In order to overcome such issues more recent instruments, such as GAP \cite{Yonetoku2010} and POLAR \cite{PRODUIT2018}, employ small coincidence 
windows and trigger logics optimized for polarization measurements. Most importantly, such detectors were calibrated prior to launch with polarized 
photons in different configurations, such as different photon energies and incoming angles \cite{Yonetoku2010,Kole2017,Li2018}. GAP was the first dedicated GRB polarimeter. It made use of plastic 
scintillators used to detect Compton scattering photons together with 12 CsI scintillators used to detect the photon after scattering. The instrument 
flew for several years on the IKAROS solar sail mission during which it detected a few GRBs for which polarization measurements were possible. 
The POLAR detector also uses plastic scintillators, 1600 in total, to detect the Compton scattering interaction, but uses the same scintillators 
to detect the secondary interaction. As a result the instrument is less efficient for detecting the secondary interaction and has a poorer energy 
resolution. However, it allows for a larger scalable effective area as well as a larger field of view, which in case of GAP is restricted by the 
CsI detectors that shield the plastic scintillators from a far off-axis source. The POLAR detector took data for 6 months on board the Tiangong-2 
space laboratory, which resulted in the publication of 14 GRB polarization measurements \cite{Kole+20}.

Two other detectors, which although not fully optimized for polarimetry, were calibrated on the ground for such measurements. 
They are COSI \cite{Kierans2017} and the CZTI on Astrosat \cite{CZT_1}. The balloon borne COSI detector uses two layers of germanium 
double-sided strip detectors allowing for precise measurements of the interaction locations in the instrument. During its long 
duration balloon flight in 2017 COSI saw one bright GRB for which a polarization measurement could be performed \cite{Lowell2017}. 
The CZT Imager on board the AstroSAT satellite uses, as the name suggests, a CZT semiconducter detector. As this detector is segmented 
it allows to look for Compton scattering events. The detector was calibrated with polarized beams prior to launch to study the instrument 
response to on-axis sources \cite{CZT_1}. AstroSAT CZTI has detected a large number of GRBs since its launch and has published polarization measurements of 13 of these to date while it continues to be operational.

\section{Theoretical Models of Prompt GRB Polarization}
\label{sec:theory}

The focus of this section is to present polarization predictions for the popular prompt GRB emission mechanisms, as highlighted in \S\ref{sec:key-questions}. 
Since GRBs are cosmological sources (of modest physical size in astrophysical standards), they remain spatially unresolved. Consequently, the measured 
polarization is the effective average value over the entire image of the burst on the plane of the sky. Therefore, the obtained polarization is affected 
by several effects, such as the intrinsic level of polarization at every point on the observed part of the outflow, the geometry of 
the outflow, i.e. the angular profile of the emissivity and bulk $\Gamma$, and the observer's LOS. Even though GRBs are intrinsically very luminous, 
their large distances drastically reduce the observed flux, making them photon starved. This forces observers to integrate either over the entire pulse 
or large temporal segments of a given emission episode to increase the photon count. This causes additional averaging -- time averaging over the instantaneous 
polarization from the whole source, which in many cases significantly evolves even within a single spike in the prompt GRB lightcurve.

Before presenting the model predictions for time-resolved polarization in \S\ref{sec:Pol-temp-evol}, pulse-integrated polarization is discussed first. 
In the latter any radial dependence of the flow properties is ignored for simplicity (but without affecting the accuracy of the calculation). As a result, 
pulse-integrated polarization ultimately amounts to integrating over a single pulse emitted at a fixed radius, where it's not important what that radius 
is as it doesn't enter any of the calculations.

Polarization is most conveniently expressed using the Stokes parameters $(I,~Q,~U,~V)$, where $I$ is the total intensity, $Q$ and $U$ are the 
polarized intensities that measure linear polarization while $V$ measures the level of circular polarization. In GRB prompt emission the circular 
polarization is typically expected to be negligible compared to the linear polarization ($V^2\ll Q^2+U^2$; this is usually expected to hold also 
for the reverse shock and afterglow emission) and therefore we concentrate here on the linear polarization. The local linear polarization (all 
local quantities are shown with a `bar') from a given fluid element on the emitting surface of the flow is given by \citep[e.g.,][]{Rybicki-Lightman-79}
\begin{equation}
    \bar\Pi' = \frac{\sqrt{\bar Q'^2+\bar U'^2}}{\bar I'}=\frac{\sqrt{\bar Q^2+\bar U^2}}{\bar I}=\bar\Pi\,,
\end{equation}
where
\begin{equation}
    \frac{\bar U}{\bar I} = \bar\Pi\sin 2\bar\theta_p\,, \hspace{3em} \frac{\bar Q}{\bar I} = \bar\Pi\cos 2\bar\theta_p\,, \hspace{3em} \bar\theta_p = \frac{1}{2}\arctan\fracb{\bar U}{\bar Q}\,,
\end{equation}
and $\bar\theta_p$ is the local polarization position angle (PA). When moving from the comoving frame of the jet to the observer frame, both the Stokes 
parameters and the direction of the polarization unit vector ($\hat{\bar \Pi}'=(\hat n'\times\hat B')/\vert\hat n'\times\hat B'\vert$, where $\hat n'$ and 
$\hat B'$ are the unit vectors in the comoving frame pointing along the observer's LOS and direction of the local B-field, respectively) undergo a Lorentz transformation (e.g. Eq.~(13) of \citet{Gill+20}). The degree of polarization (magnitude of the polarization vector), 
however, remains invariant (since $\bar Q'/\bar Q=\bar U'/\bar U=\bar I'/\bar I$). 
The local polarization is different from the \emph{global} one, $\Pi = \sqrt{Q^2+U^2}/I$ (all global parameters are denoted without a bar), which is 
derived from the global Stokes parameters. It is the global polarization that is ultimately measured for a spatially unresolved source. For an incoherent 
radiation field, meaning the emission from the different fluid elements is not in phase, which is also true for most astrophysical sources, the Stokes 
parameters are additive. Therefore, each global Stokes parameter is obtained by integration of the corresponding local Stokes parameter over the image of 
the GRB jet on the plane of the sky, such that 
\begin{equation}
    \left\{\begin{array}{c}
        U/I \\
        Q/I 
        \end{array}\right\} \to
        \left\{\begin{array}{c}
        U_\nu/I_\nu \\
        Q_\nu/I_\nu 
        \end{array}\right\}
        = \frac{\displaystyle\int d\Omega 
        \left\{\begin{array}{c}
        \bar{U}_\nu = \bar{I}_\nu\bar{\Pi}\sin(2\bar{\theta}_p)\\
        \bar{Q}_\nu = \bar{I}_\nu\bar{\Pi}\cos(2\bar{\theta}_p)
        \end{array}\right\}}{\displaystyle\int d\Omega\bar{I}_\nu}
        = \frac{\displaystyle\int dF_\nu 
        \left\{\begin{array}{c}
        \bar\Pi\sin(2\bar\theta_p) \\
        \bar\Pi\cos(2\bar\theta_p)\end{array}\right\}}{\displaystyle\int dF_\nu}\,,
\end{equation}
where $dF_\nu\cong I_\nu d\Omega = I_\nu dS_\perp/d_A^2$ is the flux contributed by a given fluid element, of observed solid angle $d\Omega$ 
and area $dS_\perp$ on the plane of the sky, where $d_A$ is the angular distance to the distant source. We work with the 
Stokes parameters per unit frequency for convenience, such as the specific intensity $\bar{I}_\nu=d\bar{I}/d\nu$. For simplicity, we ignore 
the radial structure of the outflow, and assume that the emission originates from an infinitely `thin-shell.' This approximation is valid if 
the time-scale over which particles cool and contribute to the observed radiation is much smaller than the dynamical time. It implies that 
the emission region is a thin cooling layer of width (in the lab-frame) $\Delta\ll R/\Gamma^2$. In this approximation, the differential flux 
density from each fluid element radiating in the direction $\hat n$, i.e. the direction of the observer, when the radiating shell is at radius 
$R$ (radial dependence included here for the general expression) can be expressed as \citep{Granot-05}
\begin{equation}
    dF_\nu(t_{\rm obs},\hat n,R) = \frac{(1+z)}{16\pi^2d_L^2}\delta_D^3L_{\nu'}'(R)d\tilde\Omega\,,
\end{equation}
where $z$ and $d_L$ are the redshift and luminosity distance of the source; $L_{\nu'}'$ is the comoving spectral luminosity of the fluid element and 
$d\tilde\Omega=d\tilde\mu d\tilde\varphi$ is its solid angle, where $\tilde\mu=\cos\tilde\theta$ with the polar angle $\tilde\theta$ measured from the 
LOS, and $\tilde\varphi$ is the azimuthal angle around the LOS. The Doppler factor of the fluid element moving with velocity $\vec\beta=\vec v/c$ is given 
by $\delta_D(R) = [\Gamma(1-\vec\beta\cdot\hat n)]^{-1} = [\Gamma(1-\beta\tilde\mu)]^{-1}$ (where the second expression holds for a radial outflow where $\hat{\beta}=\hat{r}$). In order to calculate the Stokes parameters using the differential flux density the angular structure of the outflow needs to be 
specified, as done next.

\subsection{\textbf{Polarization From Uniform Jets}}\label{sec:Pol-Uniform-Jets}

In uniform axisymmetric jets the comoving spectral luminosity, $L_{\nu'}'$ and the bulk-$\Gamma$ don't vary with polar angle $\theta$ measured from the jet axis, 
e.g. in a top-hat jet, 
\begin{equation}
    \frac{L_{\nu'}'(\theta)}{L_0'} = \frac{\Gamma(\theta)}{\Gamma_0} = 
    \begin{cases}
    1, & \theta \leq \theta_j \\
    0, & \theta > \theta_j\,.
    \end{cases}
\end{equation}
It is further assumed that $\Gamma$, $\theta_j$, $\theta_{\rm obs}$ and the spectrum (assumed here to be a power law) remain 
constant with radius during emission of the prompt GRB (while $L'_{\nu'}$ can vary with radius).
Since the emission arises in an ultrarelativistic jet ($\Gamma\gg1$), it is strongly beamed along the direction of motion primarily into a cone of angular size 
$1/\Gamma$. Consequently, most of the observed radiation arrives from angles $\tilde\theta\lesssim1/\Gamma$ around the LOS. 
If the LOS intersects the jet surface and is more than a beaming cone away from the edge of the jet, i.e. if $\theta_{\rm obs}\lesssim\theta_j-\Gamma^{-1}$ 
or equivalently if $q\equiv\theta_{\rm obs}/\theta_j\lesssim1-\xi_j^{-1/2}$ where $\xi_j\equiv(\Gamma\theta_j)^2$, then the observer remains unaware of 
the jet's edge (however, see \S\ref{sec:Pol-temp-evol}) and the 
emission can be approximated as if arising from a spherical flow. In this instance, after averaging 
over the GRB image on the plane of the sky, a finite net polarization will only be obtained if the direction of polarization is not axisymmetric around the 
LOS. Hence, it becomes necessary to break this symmetry in order to obtain any net polarization. This naturally happens if the LOS lies near 
the edge of the jet. Therefore, in such cases a special alignment between the flow direction and the observer is needed. This and other effects that 
break the symmetry and yield finite net polarization are highlighted below.

\subsubsection{\textbf{Synchrotron Emission From Different Magnetic Field Structures}}

Synchrotron emission is generally partially linearly polarized. The local polarization emerging from a given point on the outflow depends 
on the geometry of the local B-field and distribution of the emitting electrons, both in energy, $\gamma_em_ec^2$, and pitch angle, 
$\chi'=\arccos(\hat B'\cdot\hat\beta'_e)$, where $\hat\beta'_e$ is a unit vector pointing in the direction of the electron velocity. 
In the case of power-law electrons, with distribution $n_e(\gamma_e)\propto\gamma_e^{-p}$ for $\gamma_e>\gamma_{\min}$, and with isotropic 
velocity distribution so that all pitch-angles are sampled during the emission, the maximum local polarization for a locally ordered B-field 
depends on the spectrum \citep{Rybicki-Lightman-79,Granot-03}
\begin{equation}\label{eq:Pi_max}
    \Pi_{\max} = \frac{\alpha+1}{\alpha+5/3} = \frac{p_{\rm eff}+1}{p_{\rm eff}+7/3}\,.
\end{equation}
Here $\alpha(\nu)=-d\log F_\nu/d\log\nu$ is the \textit{local} spectral index and $p_{\rm eff} = 2\alpha+1$ is the effective power-law 
index of the electron distribution. Since the local value of $\alpha$ (and therefore also of $p_{\rm eff}$) smoothly varies with $\nu$, 
the maximum polarization, $\Pi_{\max}$, also varies smoothly 
with $\nu$ across the spectral breaks of the synchrotron spectrum. The \textit{asymptotic} spectral index is different for different 
power-law segments (PLSs) of the well-studied \citep{Sari+98,Granot-Sari-02} broken power-law synchrotron spectrum.
We have $\alpha=1/2$, $p_{\rm eff}=2$ and $\Pi_{\max}=9/13\approx0.692$ for $\nu_c<\nu<\nu_m$ (fast cooling); $\alpha=(p-1)/2$, $p_{\rm eff}=p$ and 
$\Pi_{\max} = (p+1)/(p+7/3)$ for $\nu_m<\nu<\nu_m$ (slow cooling); $\alpha=p/2$, $p_{\rm eff}=p+1$ and $\Pi_{\max}=(p+2)/(p+10/3)$ for 
$\nu>\max(\nu_c,\nu_m)$ (either slow or fast cooling).
For $\nu<\min(\nu_m,\nu_c)$, there's no $p_{\rm eff}$ since emission in this PLS arises from all cooling electrons that are emitting below their typical 
(optically-thin) synchrotron frequency. In this case, $\alpha=-1/3$ and $\Pi_{\max}=1/2$, the lowest local polarization obtained from synchrotron emission. 
On the other hand, shock-acceleration theory suggests that $2\lesssim p\lesssim3$ which means that the maximum local polarization in synchrotron is limited 
to $\Pi_{\max}\lesssim75\%$.

\end{paracol}
\begin{figure}
    \widefigure
    \centering
    \includegraphics[width=0.25\textwidth]{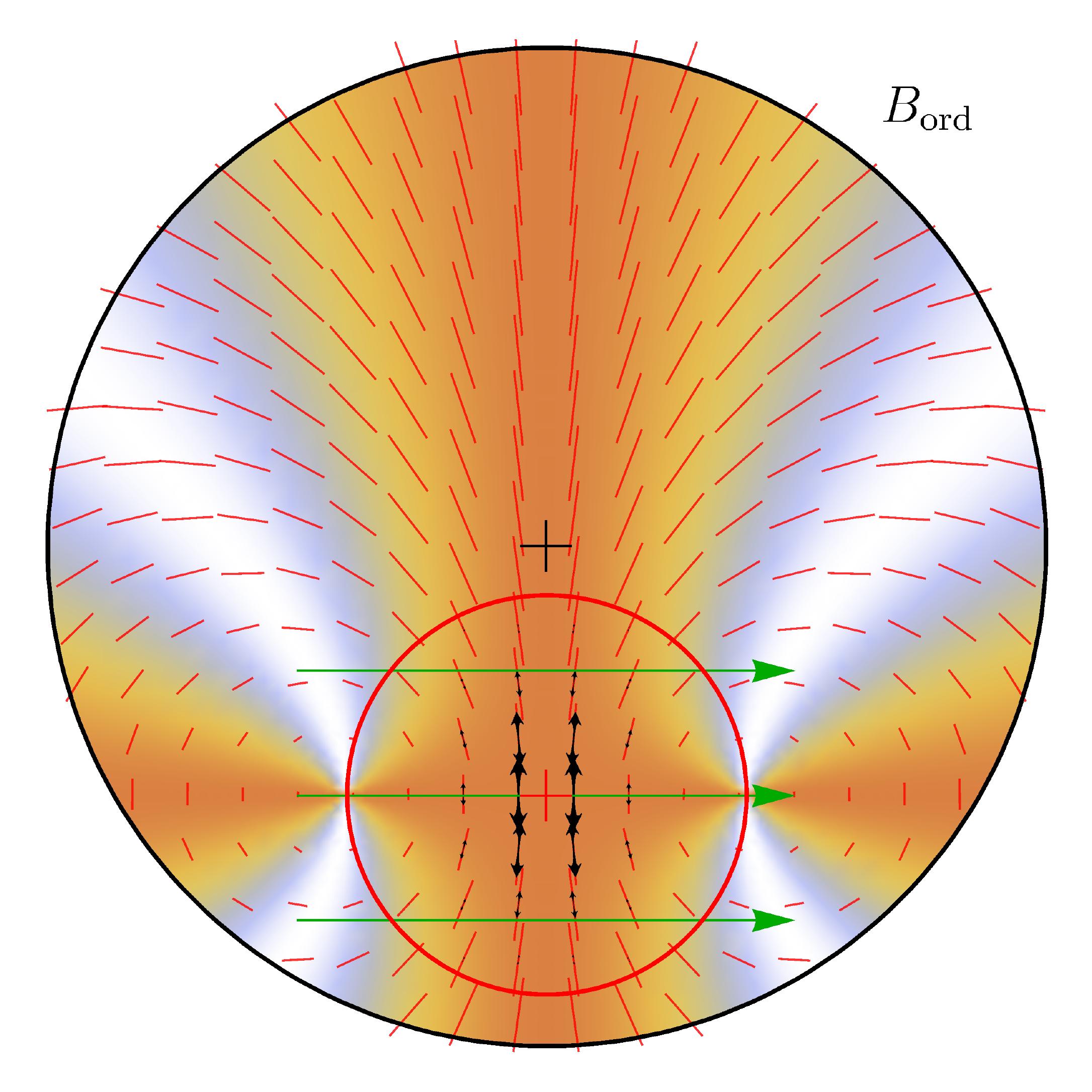}\hspace{3em}
    \includegraphics[width=0.25\textwidth]{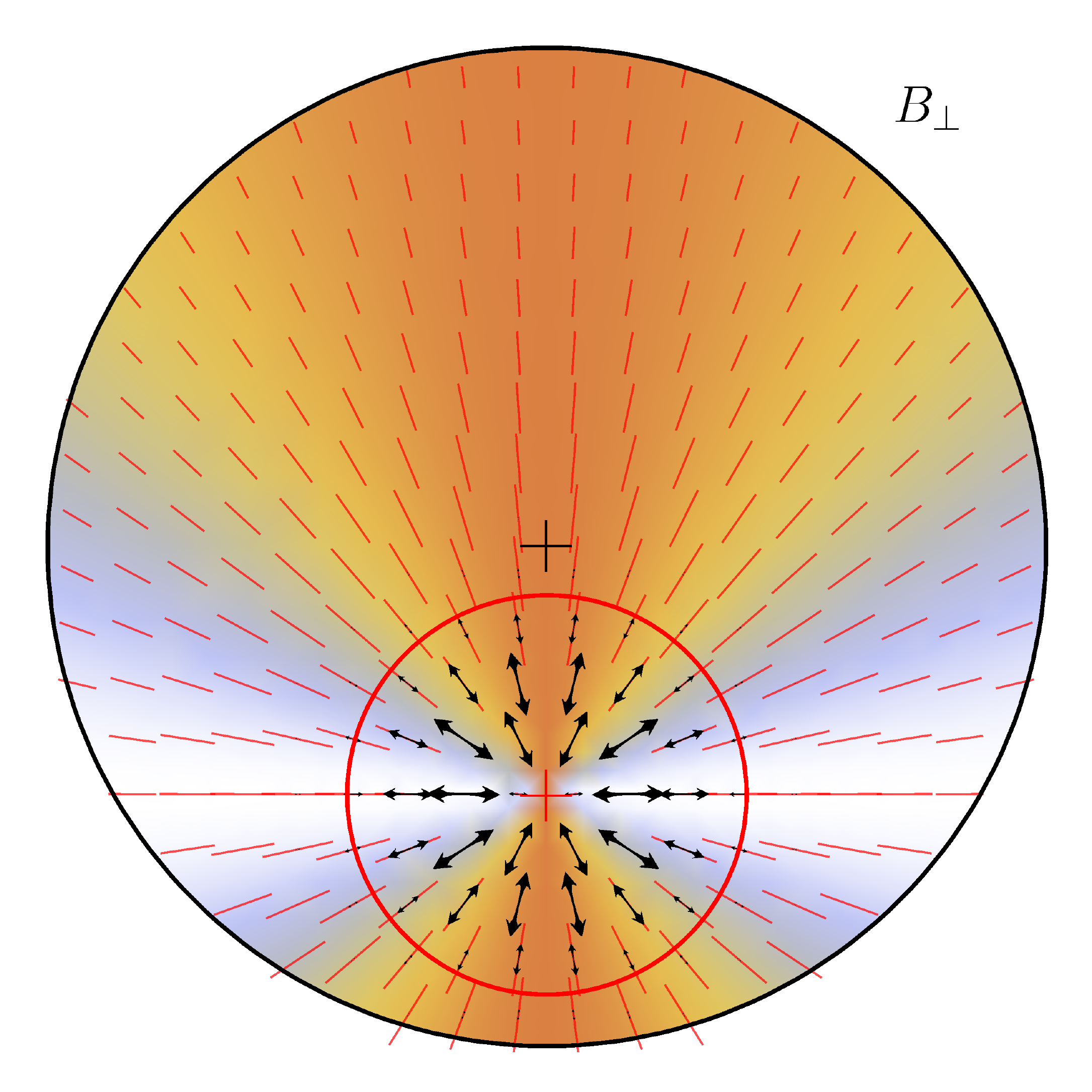}\hspace{3em}
    \includegraphics[width=0.25\textwidth]{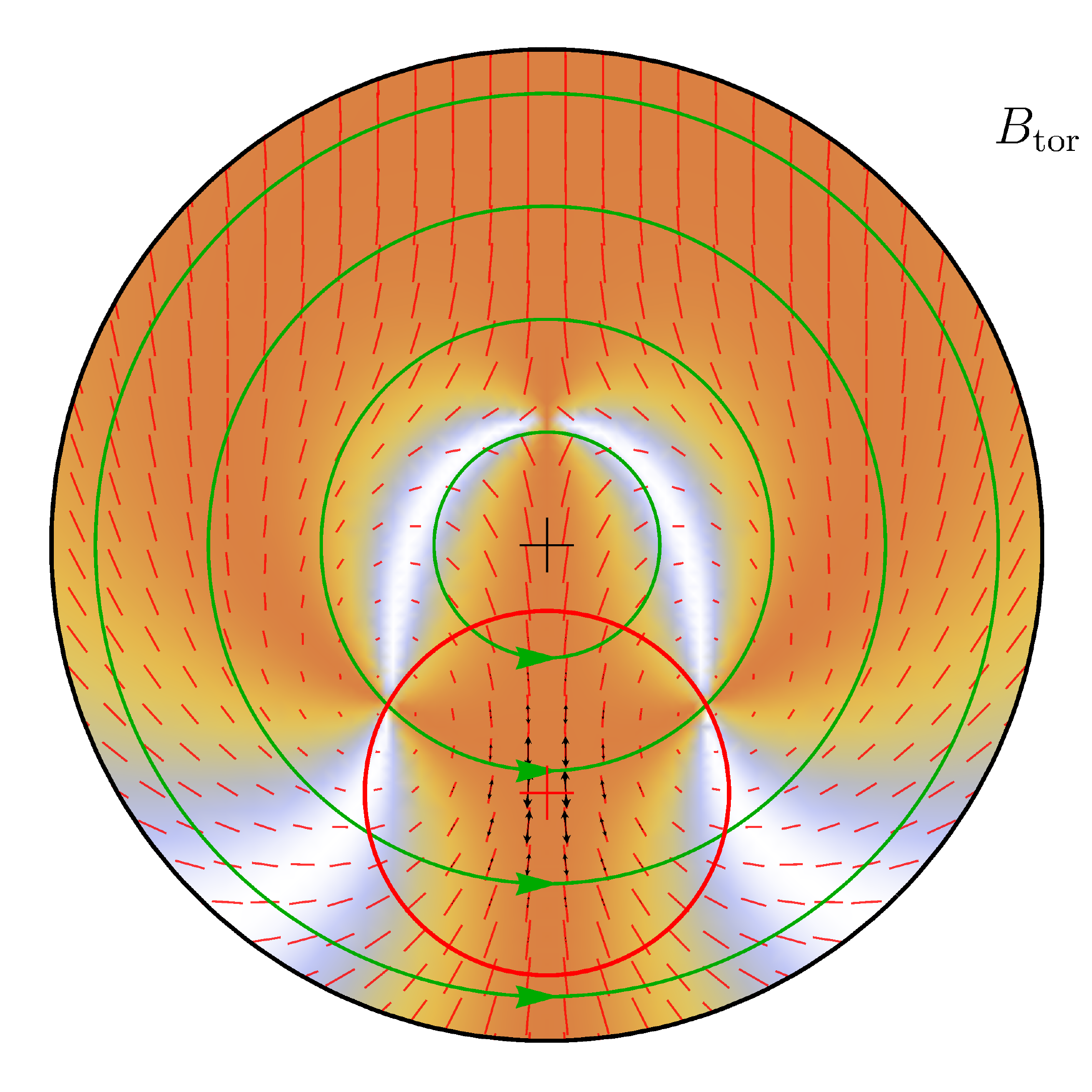}\hspace{2em}
    \includegraphics[height=0.25\textwidth]{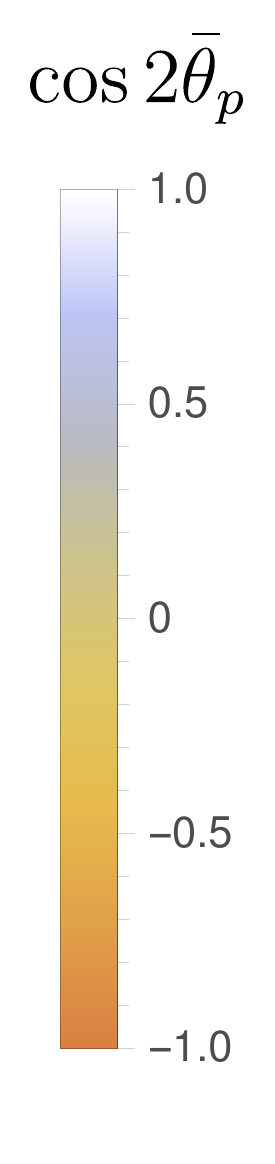}
    \caption{Polarization map for different B-field configurations shown on the surface of a top-hat jet (for $\Gamma\gg1$). 
    The jet symmetry axis marked with a black `+' symbol and the observer's LOS is marked with a red `+' symbol. The region 
    where the LOS is within the beaming cone of the local emission (i.e. from which the radiation is beamed towards us) is 
    within the red circle, outside of which the the polarized intensity (as shown by the size of the black arrows) declines 
    sharply. The red line segments show the direction and polarized intensity, now without the de-beaming suppression. Green 
    lines show the orientation of the magnetic field lines (in the cases $B_{\rm ord}$ and $B_{\rm tor}$ where it is locally 
    ordered). The color map shows $\bar{Q}\propto\cos(2\bar\theta_p)$, with $\bar\theta_p$ being the local
    polarization angle measured counter-clockwise from the horizontal axis, which corresponds to the level at which each point is polarized either along the line 
    connecting the LOS with the jet symmetry axis (orange-yellow dominated) or transverse to it (blue-white dominated).}
    \label{fig:Pol-Map}
\end{figure}
\begin{paracol}{2}
\switchcolumn

When the magnetic fields are tangled or switch direction on angular scales $\ll1/\Gamma$, e.g. in the $B_\perp$ case, the local polarization must be averaged over 
different B-field orientations. This has been calculated for an infinitely thin ultrarelativistic shell, while assuming $\alpha=1$, for a tangled B-field 
\citep{Gruzinov-99,Sari-99,Granot-Konigl-03}
\begin{equation}
    \frac{\bar\Pi_{\rm rnd}}{\Pi_{\max}} = \frac{(b-1)\sin^2\tilde\theta'}{2+(b-1)\sin^2\tilde\theta'} = 
    \begin{cases}
    \frac{-\sin^2\tilde\theta'}{1+\cos^2\tilde\theta'} & \quad(b=0,B\to B_\perp) \\
    1 & \quad(b=\infty,B\to B_\parallel)\,,
    \end{cases}
\end{equation}
where $\tilde\theta'$ is the polar angle measured from the LOS in the comoving frame (This holds for a radial flow, and more generally $\tilde\theta'\to\arccos(\hat{n}'\cdot\hat{n}'_{\rm sh})$). The level of anisotropy of the B-field is quantified by the 
parameter $b = 2\langle B_\parallel^2\rangle/\langle B_\perp^2\rangle$ which represents the ratio of the average energy densities in the two field 
orientations. The factor of 2 simply reflects the two independent directions of the $B_\perp$ component, such that $b=1$ for a field that is isotropic in three dimensions.

The polarization map over the GRB image on the plane of the sky is shown in Fig.~\ref{fig:Pol-Map} for different B-field structures (for $\Gamma\gg1$). 
Only the area contained within the beaming cone, shown by the red circle, contributes dominantly to the emission. Outside of it the intensity is strongly 
suppressed by relativistic beaming, which scales as a power of the Doppler factor. This effect is reflected by the decrease with the angle $\tilde\theta$ 
from the LOS (shown by the red `+' symbol) in the size of the black arrows, which correspond to the magnitude of the polarized intensity. When the 
jet possesses axial symmetry (and for synchrotron emission the same requirement holds also for the global magnetic field structure), then the image and 
polarization map are symmetric to reflection along the line connecting the jet symmetry axis to the LOS. Therefore, it is natural to choose a reference 
direction for measuring the local PA $\bar\theta_p$ either along this line or transverse to it (in the figure $\bar\theta_p$ as well as $\theta_p$ are 
measured from the latter, i.e. the horizontal direction). For such a choice $U=0$ i.e. the local Stokes parameter $\bar{U}\propto\sin(2\bar{\theta}_p)$ 
vanishes when integrated over the GRB jet image, and therefore, the global polarized intensity is entirely given by Stokes $Q$, i.e. the integration of 
$\bar{Q}\propto\cos(2\bar{\theta}_p)$ over the image , where the sign of $\bar{Q}$ for each fluid element depends on the local PA $\bar\theta_p$. 
The different B-field configurations produce completely different polarization maps, with distinct patterns of regions contributing predominantly 
either to polarization along the line connecting the jet symmetry axis to the LOS (orange-yellow, with local polarization $\bar\Pi<0$) or transverse 
to it (blue-white, with local polarization $\bar\Pi>0$), as shown by the color map. When averaged over the entire GRB image, these are the only two directions 
of polarization that can be obtained in an axisymmetric flow in which the magnetic field also possesses axial symmetry about the jet axis, such that it 
would represent a change of $90^\circ$ in the PA when the direction of polarization switches from one to the other. 

An example of a B-field configuration that does not possess such axial symmetry is $B_{\rm ord}$. When $B_{\rm ord}$ is not oriented along the line connecting 
the jet symmetry axis to the LOS or perpendicular to it then this breaks the symmetry of the image polarization map thereby enabling other directions of net 
global polarization to occur and the corresponding PA can vary continuously with a finite $\Pi$ \citep{Granot-Konigl-03}.


The level of net polarization after averaging over the GRB image depends on the level of symmetry of the polarization map around the LOS. In the case 
of $B_\perp$, and likewise for $B_\parallel$, the polarization map is symmetric around the LOS and therefore averaging over the GRB image would yield 
zero net polarization ($\Pi=0$) due to complete cancellation for a spherical flow (or well within a top-hat jet, $\Gamma(\theta_j-\theta_{\rm obs})\gg1$). 
This symmetry is naturally broken in $B_{\rm ord}$ and $B_{\rm tor}$ where the local B-field is ordered and provides 
a particular direction (transverse to the local B-field direction and to the propagation direction of the photon) along which the polarization vector aligns. 
Another way to break the symmetry is by having the LOS close to the edge of the jet, with 
$\theta_j-\Gamma^{-1}\lesssim\theta_{\rm obs}\lesssim\theta_j+\Gamma^{-1}\Leftrightarrow\Gamma|\theta_{\rm obs}-\theta_j|\lesssim1$ 
($1-\xi_j^{-1/2}\lesssim q\lesssim1+\xi_j^{-1/2} \Leftrightarrow\xi_j^{1/2}|q-1|\lesssim1$), so that some part 
of the beaming cone lies outside of the jet surface. The missing emission, which would otherwise contribute towards cancellation, leads to only partial 
cancellation and yields a net finite polarization, $\vert\Pi\vert>0$. The sign of net polarization is decided by whichever region, either blue-white or 
orange-yellow, makes the dominant contribution to the polarized flux. In Fig.~\ref{fig:Pol-Map}, $\Pi<0$ for both $B_{\rm ord}$ and $B_{\rm tor}$ whereas 
$\Pi\approx0$ for $B_\perp$. 

Pulse integrated polarization as a function of $q$ is shown in Fig.~\ref{fig:Pol-diff-B-fields} for different B-field configurations and 
different $\xi_j$, where the latter describes how wide or narrow the jet aperture is compared to the beaming cone. The polarization 
curves look very different for the three different field configurations, but there are some features that are worth pointing out. The 
polarization vanishes when the observer is looking down the jet axis, i.e. when $\theta_{\rm obs}=0$ ($q=0$), in all cases due to complete 
cancellation (such a cancellation would not occur for $B_{\rm ord}$, which is not shown in Fig.~\ref{fig:Pol-diff-B-fields}). For $q>0$, 
polarization grows rapidly for $B_{\rm tor}$ (for which it saturates at $\xi_j^{-1/2}\lesssim q\lesssim1-\xi_j^{-1/2}\Leftrightarrow\Gamma^{-1}\lesssim\theta_{\rm obs}\lesssim\theta_j-\Gamma^{-1}$) but slowly for both $B_\perp$ and $B_\parallel$. It reaches a local maxima when the LOS is close to the 
jet edge, i.e. as before, when $|q-1|\lesssim\xi_j^{-1/2}\Leftrightarrow\Gamma|\theta_{\rm obs}-\theta_j|\lesssim1$,
and declines sharply for $B_\perp$ and $B_{\rm tor}$ when the LOS exceeds one beaming cone outside of the jet, i.e. 
$\theta_{\rm obs}\gtrsim\theta_j+\Gamma^{-1}$ ($q>1+\xi_j^{-1/2}$). The $B_\parallel$ case yields a different behavior where $\Pi$ becomes 
maximal when the jet is viewed from outside its edge. In all cases, when $q>1+\xi_j^{-1/2}\Leftrightarrow\Gamma(\theta_{\rm obs}-\theta_j)>1$ 
the fluence drops off very sharply for a top-hat jet. So, even though a large $\Pi$ is expected for $B_\parallel$, it will be challenging to 
detect. Finally, a change in the PA by $90^\circ$ occurs when $\theta_{\rm obs}\approx\theta_j$ ($q\approx1$) for $B_\perp$ and $B_\parallel$ 
at which point $\Pi=0$.

It is clear from Fig.~\ref{fig:Pol-diff-B-fields} that only the $B_{\rm tor}$ case, an ordered field scenario, yields high levels of polarization 
when the LOS passes within the aperture of the jet. Since all distant GRBs must be viewed with $q<1$, otherwise they'll be too dim to detect, a 
measurement of $50\%\lesssim\Pi\lesssim65\%$ will strongly indicate the presence of an ordered field component. On the other hand, if the B-field 
configuration is more like $B_\perp$ or $B_\parallel$, then most GRBs will show negligible polarization.

\end{paracol}
\begin{figure}
    \widefigure
    \centering
    \includegraphics[width=0.3\textwidth]{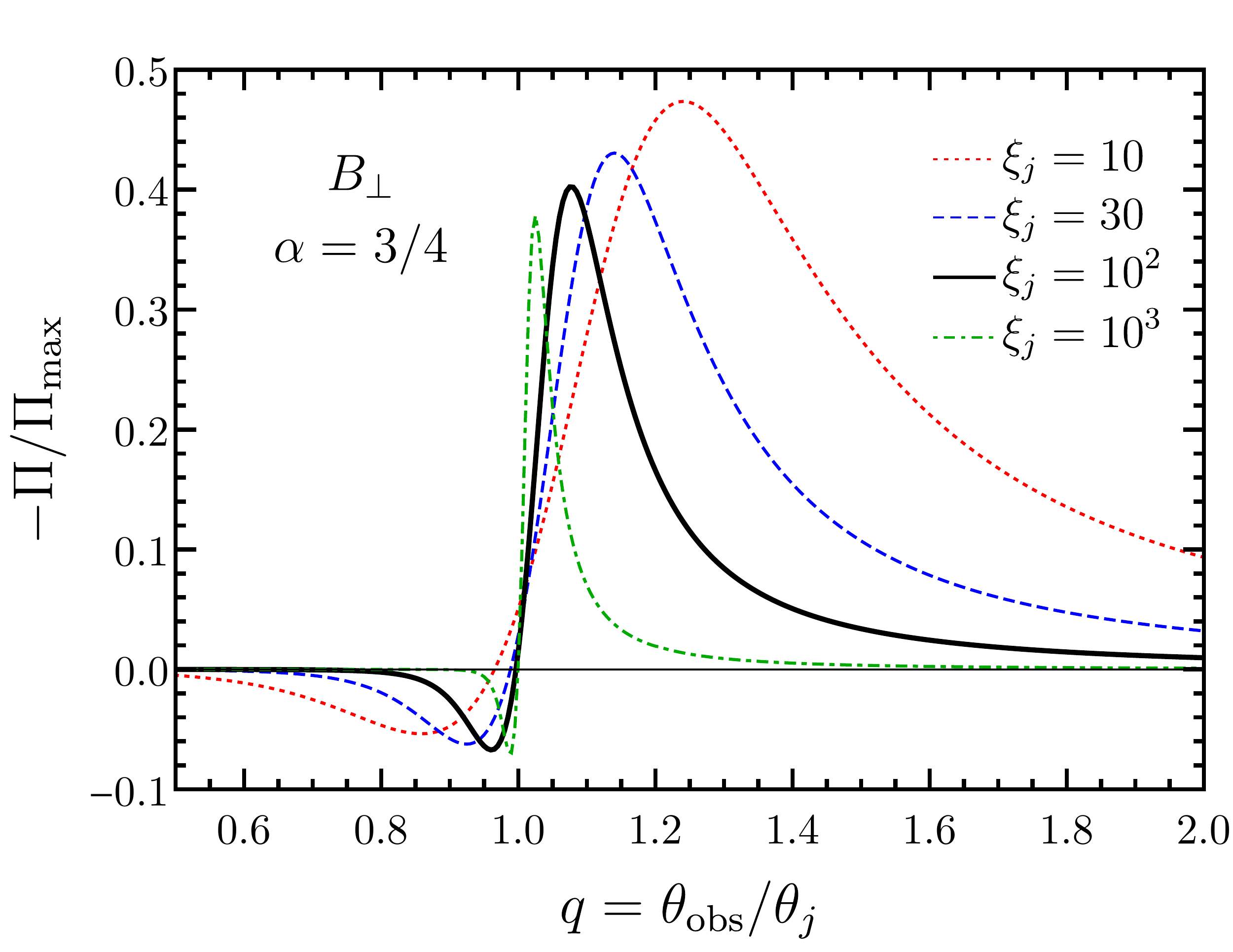}\hspace{2em}
    \includegraphics[width=0.3\textwidth]{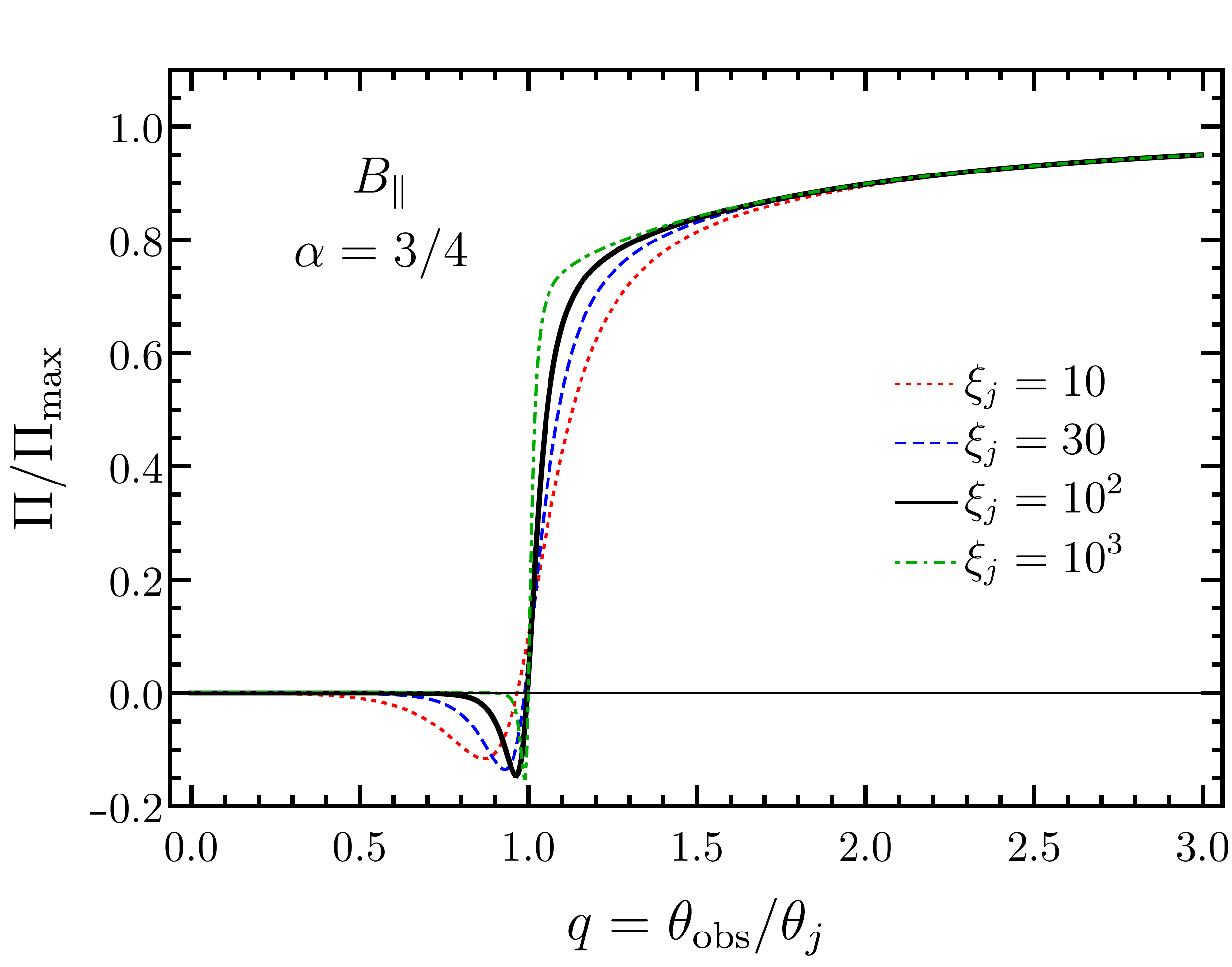}\hspace{2em}
    \includegraphics[width=0.3\textwidth]{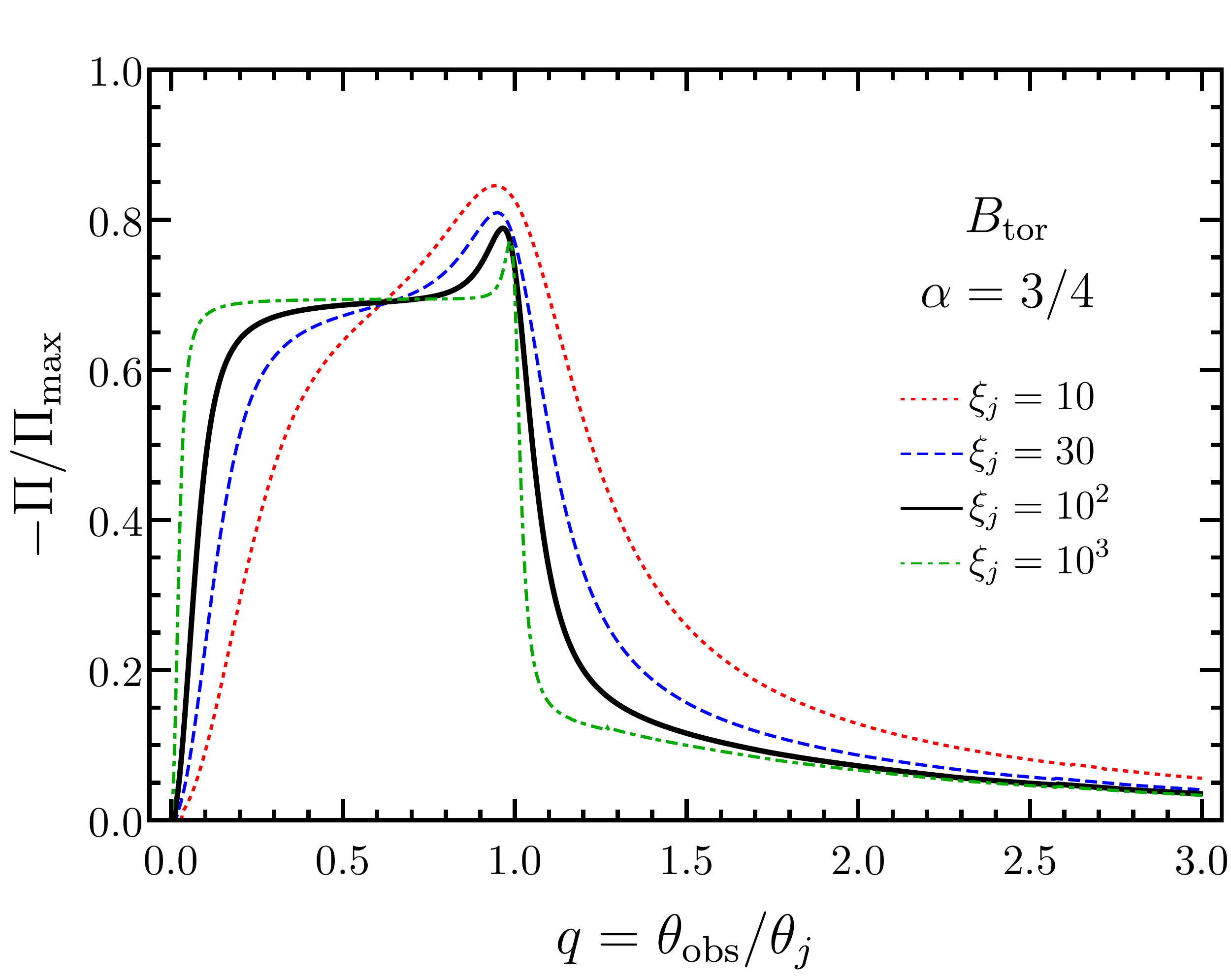}
    \caption{Pulse-integrated polarization of synchrotron emission for different B-field configurations shown for different LOSs ($q$) and size of 
    the beaming cone w.r.t to the jet aperture ($\xi_j=(\Gamma\theta_j)^2$). The spectral index is fixed to $\alpha = 3/4$, where larger $\alpha$ 
    produces larger $\Pi$. Figure adapted from \citep{Gill+20} but originally produced in \citep{Granot-03}. }
    \label{fig:Pol-diff-B-fields}
\end{figure}
\begin{paracol}{2}
\switchcolumn

\subsubsection{\textbf{Photospheric Emission From a Uniform Jet}}

A photospheric spectral component can arise and even dominate the spectral peak in scenarios where energy is dissipated below the photosphere. 
At the photosphere radiation decouples from matter and is able to stream freely towards the observer. However, in a matter dominated flow in 
which the baryon rest mass energy density, $\rho'c^2$, is much larger than that of the radiation field, $U_\gamma'$, the radiation field becomes 
highly anisotropic at the photosphere \citep{Beloborodov-11}. At the last scattering surface, this produces significant \textit{local} polarization 
at each point of the observed part of the flow. Nevertheless, upon averaging over the GRB image the net polarization is expected to be negligible in 
an axisymmetric uniform flow since there's no preferred direction for the polarization vector. To obtain finite net polarization an inhomogeneous 
outflow with gradients in bulk-$\Gamma$ (and to a lesser extent in comoving emissivity $L'_{\nu'}$) across the beaming cone are needed. This scenario 
is discussed in \S\ref{sec:Pol-Structured-Jets}.

Alternatively, if the flow is radiation dominated, i.e. $U'_\gamma\gg\rho'c^2$, as shown by \citet{Beloborodov-11} the comoving angular distribution 
of the radiation field is preserved in the ultrarelativistic limit as the flow goes from being optically thick to thin. This occurs due to the fact 
that radiation always tries to push the plasma to an equilibrium Lorentz frame in which the radiative force on the plasma vanishes. As a result, the 
radiation field accelerates the plasma to a bulk LF $\Gamma(R)\propto R$, which is a special Lorentz frame in which the (comoving) direction of freely 
streaming photons w.r.t the local radial direction remains unchanged in between successive scatterings. This means that an isotropic radiation field 
remains isotropic. Since scattering an isotropic radiation field only produces another isotropic field, the flow behaves (to leading order) as if no 
scatterings took place. Since the radiation field was necessarily isotropic when the flow was optically thick at 
smaller radii, leading to zero local polarization, it must yield the same (to leading order) when it becomes optically thin, as shown below.

The radiation field is able to accelerate the flow to $\Gamma(R)\propto R$ only if $U_\gamma'/\rho'c^2\gg1$ and while the matter maintains a small lag, 
$\Delta\Gamma=\Gamma_{\rm rad}-\Gamma_{\rm matter}$, which is initially $\ll\Gamma_{\rm matter}\equiv\Gamma\approx\Gamma_{\rm rad}$ (where these LFs are of the respective local center of momentum frames) corresponding to a relative velocity $\beta_{\gamma{\rm m}}\sim\Delta\Gamma/\Gamma\sim\rho'c^2/U'_\gamma\ll1$, but gradually increases until it eventually becomes comparable to the two near the saturation radius $R_s$ where $U'_\gamma\tau_T\sim\rho'c^2$ 
(for $\tau_T < 1$), the point beyond which matter stops accelerating and starts coasting, while the scaling $\Gamma_{\rm rad}\propto R$ remain valid as the 
radiation free streams in increasingly more radial directions. 
In a steady radiation-dominated spherical flow the comoving radiation energy density scales as $U_\gamma'\propto V'^{-4/3}\propto [R^2\Gamma(R)]^{-4/3}$, 
and the rest mass energy density of the particles scale as $\rho'\propto n_e'\propto V'^{-1}\propto [R^2\Gamma(R)]^{-1}$, where $V'$ is the comoving volume. This yields 
$U_\gamma'/\rho'c^2\propto [R^2\Gamma(R)]^{-1/3}$, which for $\Gamma(R)\propto R$ gives $U_\gamma'/\rho'c^2\propto R^{-1}$ and 
$\tau_T = n_e'\sigma_TR/\Gamma(R)\propto[R\Gamma^2(R)]^{-1}\propto R^{-3}$. This further yields $U'_\gamma\tau_T/\rho'c^2=(U'_{\gamma,\rm ph}/\rho'_{\rm ph}c^2)(R/R_{\rm ph})^{-4}$ so that $R_s\sim R_{\rm ph}(U'_{\gamma,\rm ph}/\rho'_{\rm ph}c^2)^{1/4}\sim R_{\rm ph}\beta_{\gamma{\rm m,ph}}^{-1/4}$ and $\beta_{\gamma{\rm m}}\sim\min[1,(R/R_s)^4]$. Near $R_s$ the comoving radiation anisotropy becomes significant ($\beta_{\gamma{\rm m}}\sim1$) and therefore so does the polarization of the radiation scattered at $R\sim R_s$, but this is only a fraction $\sim\tau_T(R_s)\sim (U'_{\gamma,\rm ph}/\rho'_{\rm ph}c^2)^{-3/4}\sim\beta_{\gamma{\rm m,ph}}^{3/4}\ll1$ of the photons, and therefore the overall local (i.e. from a particular fluid element) polarization is of the same order, i.e. very small.

\begin{figure}
    \centering
    \includegraphics[width=0.4\textwidth]{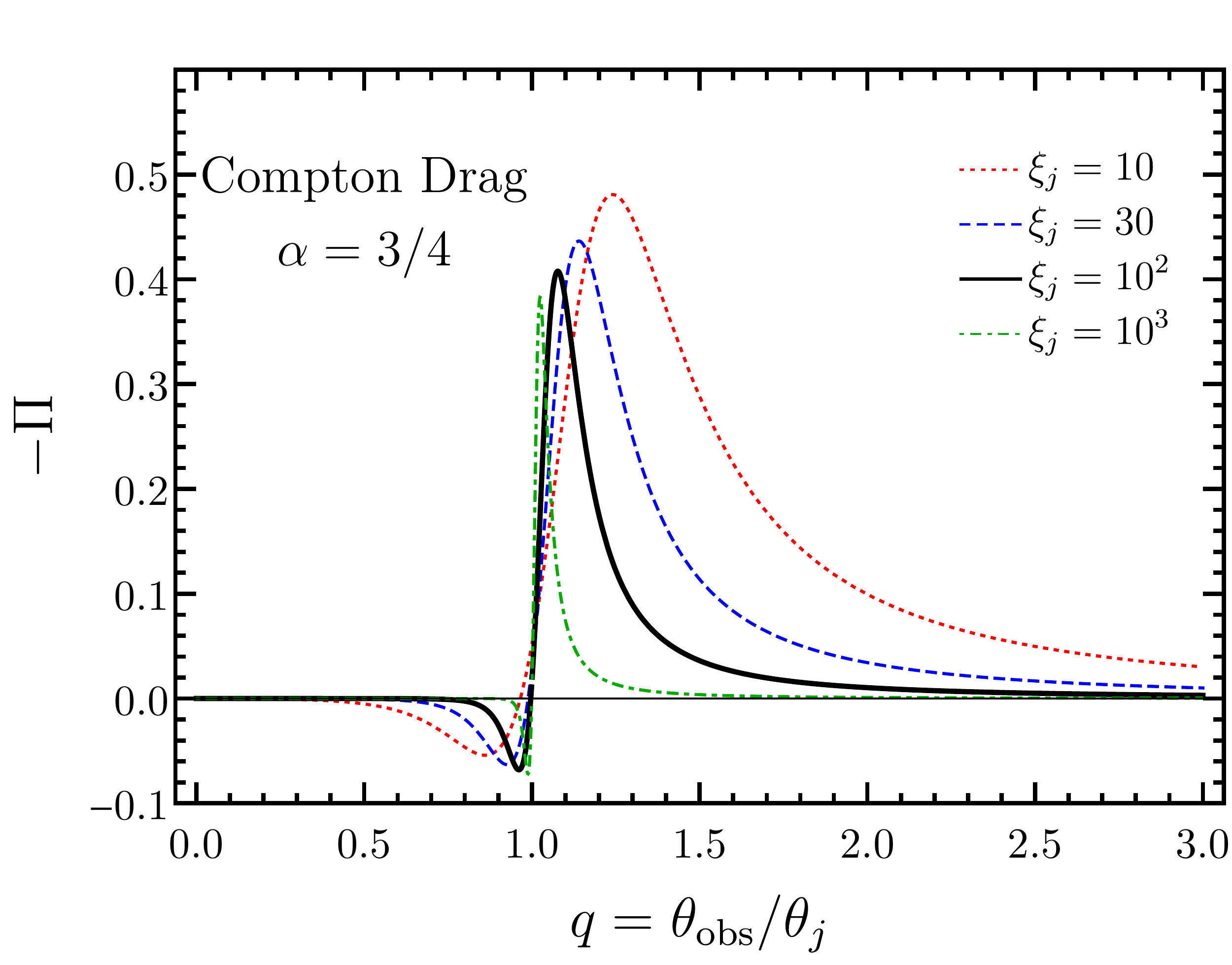}
    \caption{Pulse-integrated polarization of prompt GRB radiation generated by the Compton drag mechanism. The electrons are assumed to be cold in 
    the comoving frame. Figure adapted from \citep{Gill+20}, but also see \citep{Lazzati+04} for results for a narrower top-hat jet.}
    \label{fig:Pol-CD}
\end{figure}

\subsubsection{\textbf{Compton Drag}}

Inverse-Compton scattering of anisotropic radiation yields high levels of polarization for the scattered radiation field with $\Pi\leq100\%$. This 
is very different from Comptonization since the polarization vector of the scattered photon can now be aligned with a particular direction, which 
is transverse to the plane containing the wave vectors, $\vec k_1''$ and $\vec k_2''$, of the incoming and scattered photons, respectively, in the rest frame of 
the electron (hence the double primes). If the scattering angle is $\theta_{\rm sc}'' = \arccos(\Vec{k_1''}\cdot\Vec{k_2''})$, then Thomson scattering 
of radiation imparts local polarization
\begin{equation}\label{eq:Pi_CD}
    \bar\Pi = \frac{1-\cos^2\theta_{\rm sc}''}{1+\cos^2\theta_{\rm sc}''} \xrightarrow[{\rm electrons}]{\rm cold}
    \frac{1-\cos^2\theta'_{\rm sc}}{1+\cos^2\theta'_{\rm sc}}
    \xrightarrow[{\rm flow}]{\rm radial}
    \;\frac{1-\cos^2\tilde\theta'}{1+\cos^2\tilde\theta'}
\end{equation}
to the outgoing photon. Indeed, if $\theta_{\rm sc}''=\pi/2$ then $\bar\Pi=100\%$. Here it's assumed that the electrons are cold and therefore their 
rest frame  is the fluid frame ($\theta_{\rm sc}''=\theta_{\rm sc}'$) that is  moving with velocity $\vec v$, and if it is moving everywhere in 
the radial direction ($\hat{v}=\hat{r}$) then  $\theta_{\rm sc}'=\tilde\theta'$. In general, the local polarization depends on the angle $\theta_0'$ 
between the wave vector of the incoming photon and the velocity vector of the electron. If the electrons have a finite internal energy density, 
which means that they have a velocity distribution, then the local polarization is obtained by performing a weighted integral over all 
$\theta_0'$ \citep[see][for details]{Begelman+87}.

The expected polarization when assuming cold electrons in the comoving frame of an ultrarelativistic top-hat jet is shown in Fig.~\ref{fig:Pol-CD}. 
The polarization curves are very similar to that obtained for synchrotron emission for the $B_\perp$ field configuration, but for Compton drag the 
normalization is (nearly exactly) higher by $\Pi_{\max}^{-1}(\alpha)$ as given by Eq.~(\ref{eq:Pi_max}). Similar results were first obtained by \citet{Lazzati+04} for 
narrower jets with $\xi_j\leq25$ where they showed that when $\xi=0.04$ very high polarization with $\Pi\lesssim95\%$ can be obtained with Compton drag. 


\subsection{\textbf{Polarization From Structured Jets}}\label{sec:Pol-Structured-Jets}

\end{paracol}
\begin{figure}
    \widefigure
    \centering
    \includegraphics[width=0.42\textwidth]{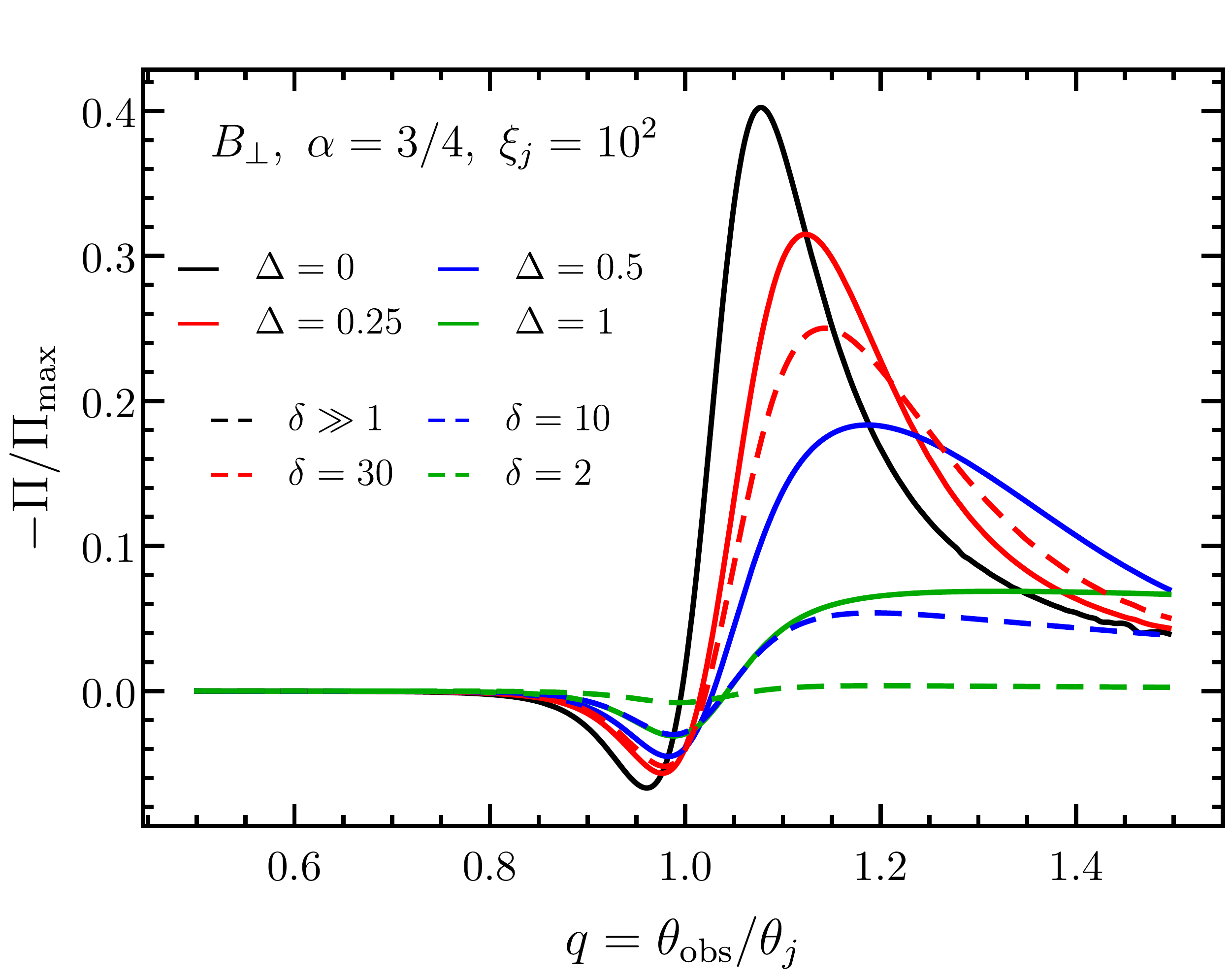}\hspace{3em}
    \includegraphics[width=0.42\textwidth]{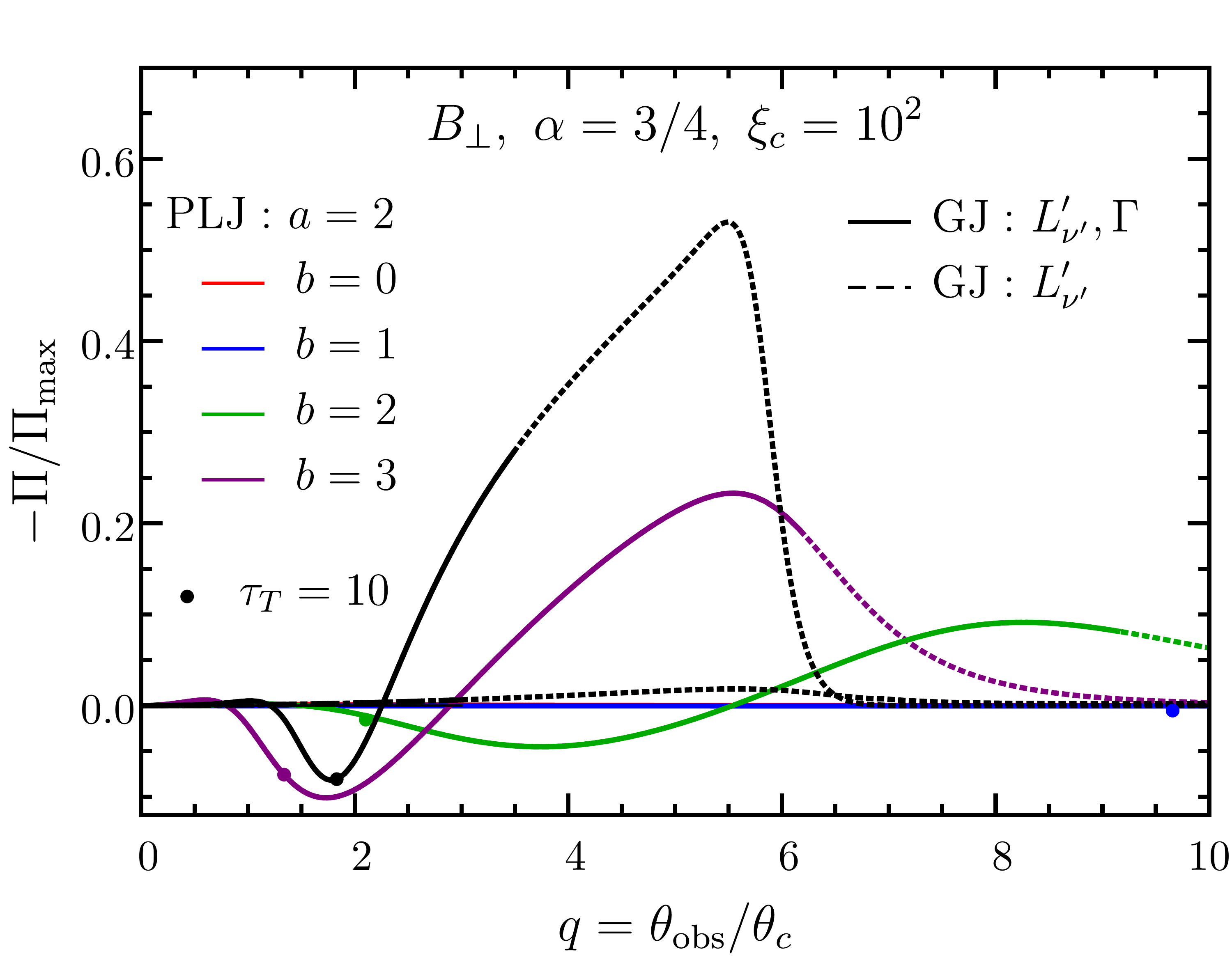}
    \includegraphics[width=0.42\textwidth]{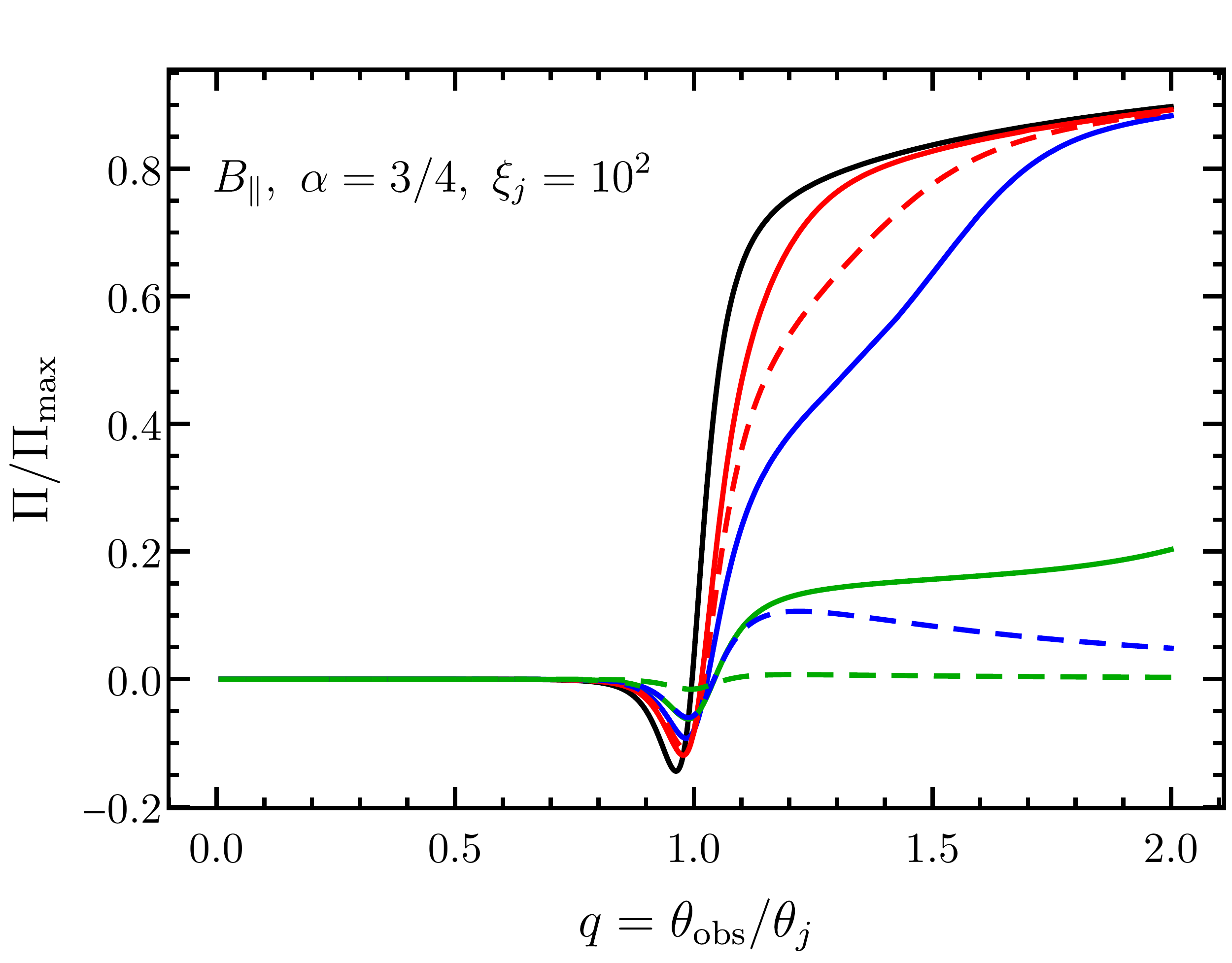}\hspace{3em}
    \includegraphics[width=0.42\textwidth]{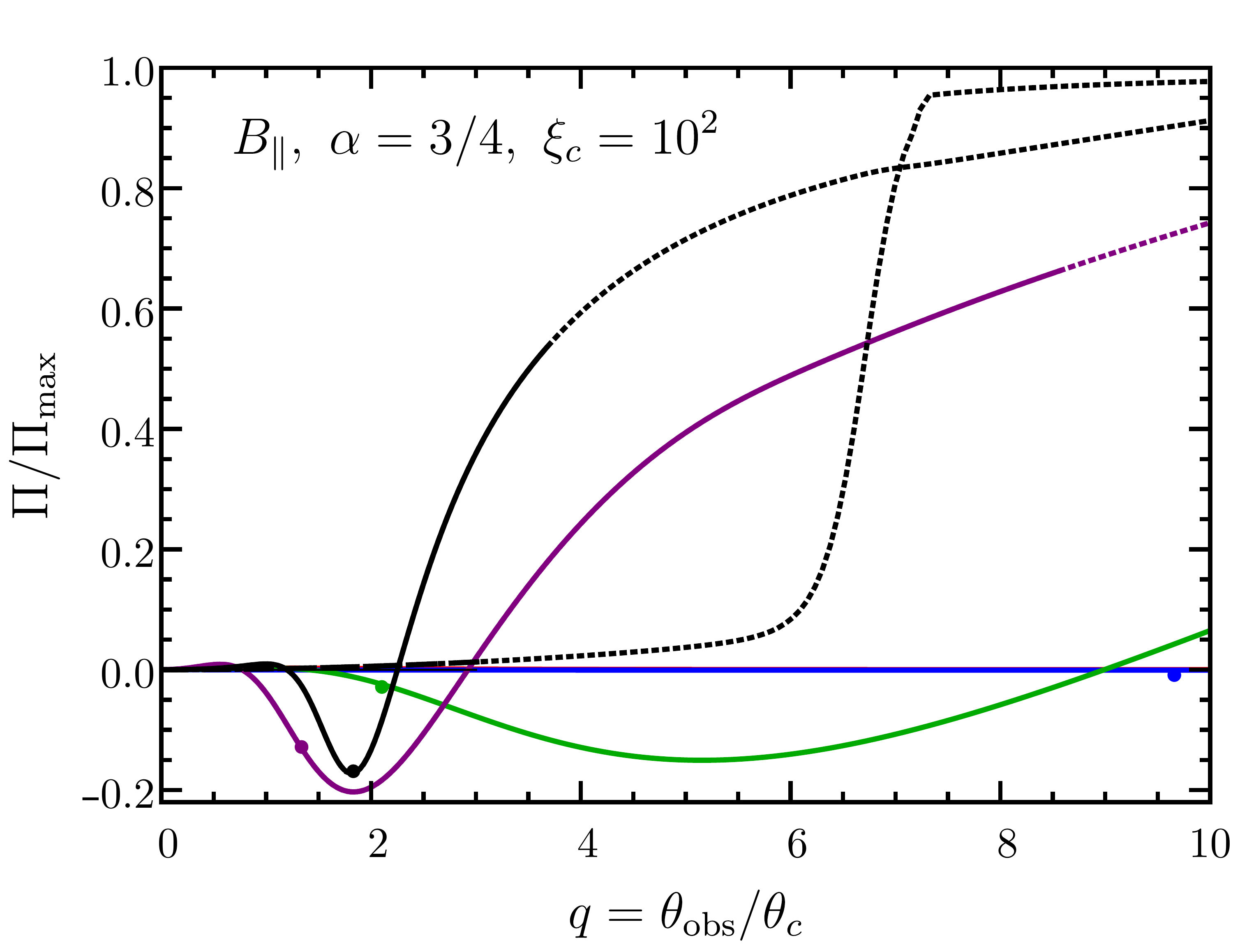}
    \includegraphics[width=0.42\textwidth]{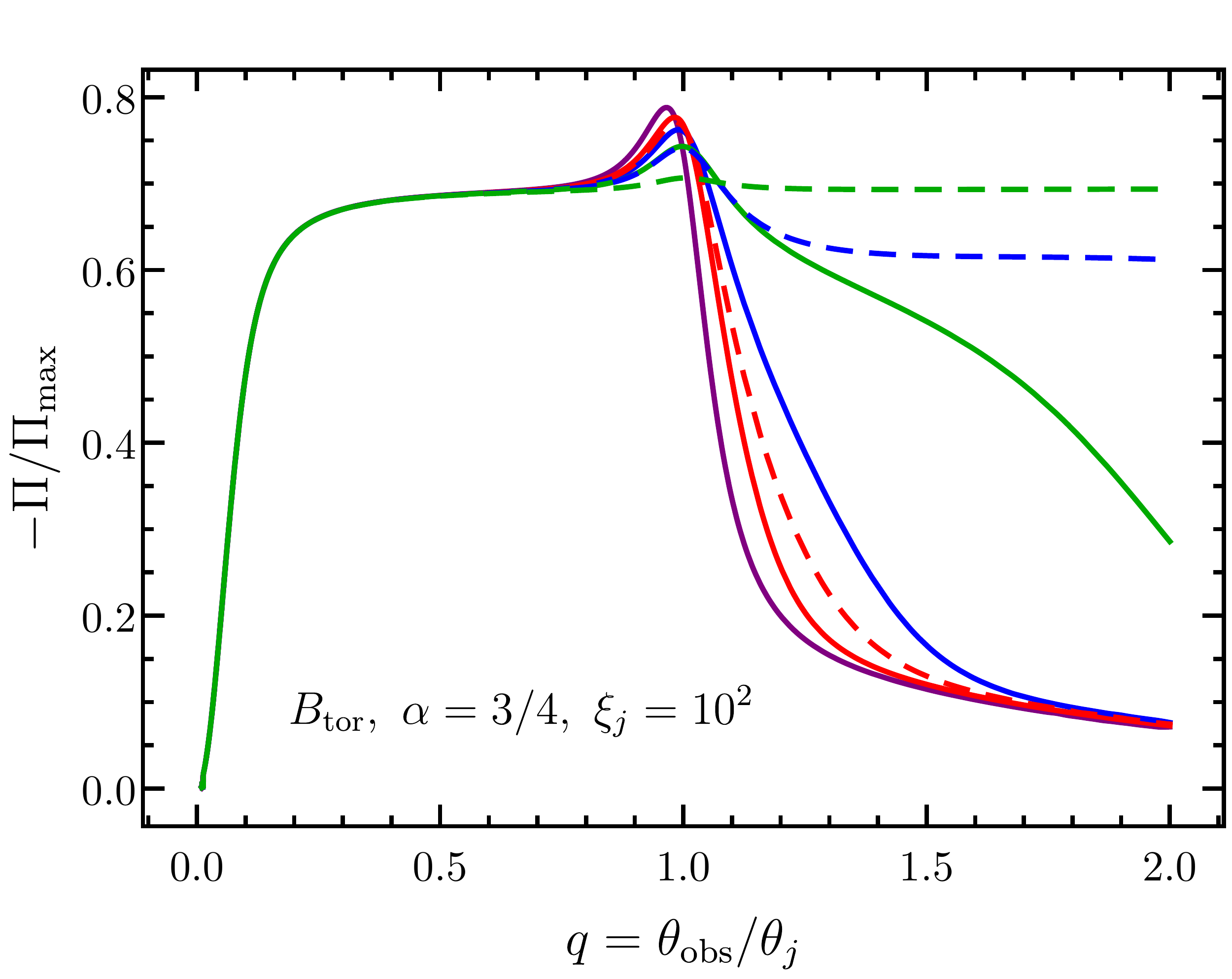}\hspace{3em}
    \includegraphics[width=0.42\textwidth]{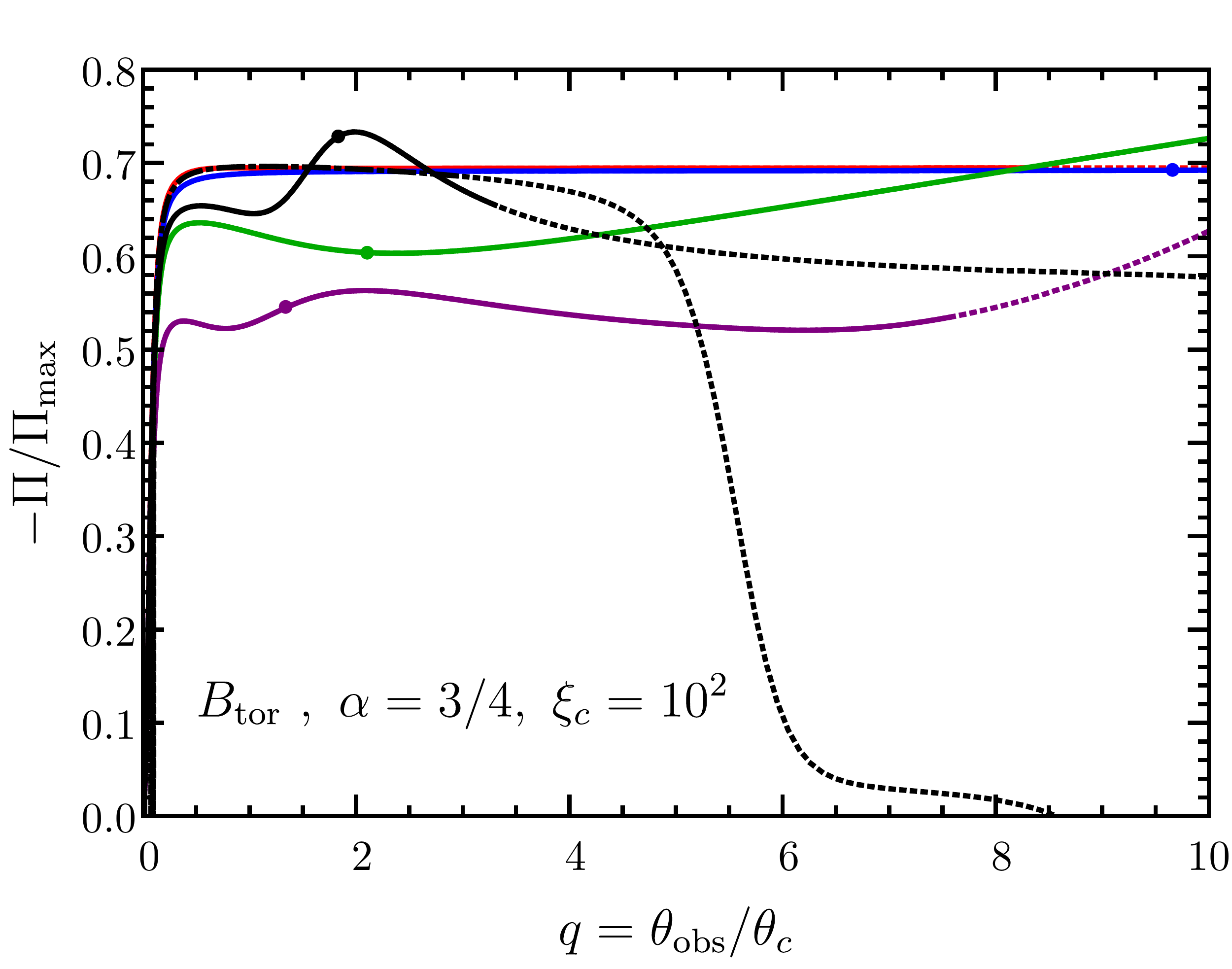}
    \caption{(\textbf{Left}) Pulse-integrated polarization for smooth jets with uniform core and exponential or power-law wings in spectral luminosity while the 
    bulk-$\Gamma$ remains uniform. The edges of the uniform jet become smoother with increasing (decreasing) $\Delta$ ($\delta$) for 
    exponential (power law) wings. (\textbf{Right}) Polarization curves for structured jets. Two cases for the Gaussian jet (GJ) are shown, where in one 
    both $L_{\nu'}'$ and $\Gamma$ vary with $\theta$ and in the other $\Gamma$ is kept uniform. For the power-law jet (PLJ), the power-law index for $L_{\nu'}'$ 
    is fixed ($a=2$), but that for the bulk-$\Gamma$ ($b$) is varied. The curve for $b=0$ is mostly overlapped by that of $b=1$. The dotted lines show the polarization curves for viewing angles at which the fluence has declined to values smaller than $1\%$ of that expected at $\theta_{\rm obs}=0$. The thick dots mark critical viewing angles beyond which the emission region becomes 
    too compact to $\gamma\gamma$-annihilation, causing the emission to be optically thick to Thomson scattering of the produced $e^\pm$-pairs. 
    Figure adapted from \citep{Gill+20} and some results for the smoothed top-hat jets were first presented in \citep{Nakar+03}
    }
    \label{fig:Pol-Smooth-Jet}
\end{figure}
\begin{paracol}{2}
\switchcolumn

The angular structure of the relativistic jet in GRBs becomes particularly important for relatively nearby events, e.g. GRB170817A ($D\simeq40\,$Mpc), 
which can be detected with the current cadre of instruments when the observer is relatively far off-axis and the emission is dim. For distant GRBs, 
as mentioned earlier, it is challenging to detect emission from significantly off-axis jets. Still, there will be some events in which the LOS is just 
outside the quasi-uniform core that may not be sharp, as found otherwise in a top-hat jet, but instead be smoother. Then, it becomes important to model 
the angular structure and compare polarization measurements with accurate theoretical models.

The first level of correction for an idealized top-hat jet model is the consideration of smooth wings of comoving spectral luminosity while the bulk-$\Gamma$ 
remains uniform \citep{Nakar+03}. Like the top-hat jet, $L'_{\nu'} = L_{\nu',0}'$ for $\xi\leq\xi_j$ ($\theta\leq\theta_j$), but outside of this 
uniform core the spectral luminosity can have either exponential or power-law wings:

\begin{equation}
    \frac{L_{\nu'}'}{L_{\nu',0}'} = 
    \begin{cases}
    \exp[(\sqrt\xi_j-\sqrt\xi)/\Delta], & \xi > \xi_j\quad(\rm{exponential~wings}) \\
    \fracb{\xi}{\xi_j}^{-\delta/2}, & \xi > \xi_j\quad(\rm{power\text{-}law~wings})\,.
    \end{cases}
\end{equation}

Here again it is assumed that $\Gamma$, $\theta_j$, $\theta_{\rm obs}$, and the spectrum do not have any radial dependence.

In a more realistic structured jet the core is no longer uniform. Instead, the spectral luminosity as well as the bulk-$\Gamma$ depend on polar angle $\theta$. 
In general, the properties of the flow can also depend on the azimuthal angle $\phi$, but here the discussion makes the simplifying and physically 
reasonable assumption of axisymmetric jets. Two different types of structured jets are considered here:

\begin{align}
    & \frac{L_{\nu'}'(\theta)}{L_{\nu',0}'} = \frac{\Gamma(\theta)-1}{\Gamma_c-1} = 
    \exp\left(-\frac{\theta^2}{2\theta_c^2}\right) &  ({\rm Gaussian~Jet}) \\
    & \frac{L_{\nu'}'(\theta)}{L_{\nu',0}'} = \Theta^{-a}, \quad \frac{\Gamma(\theta)-1}{\Gamma_c-1} = \Theta^{-b}, 
    \quad \Theta = \sqrt{1+\fracb{\theta}{\theta_c}^2} &  ({\rm Power\text{-}Law~Jet})
    \label{eq:PLJ}
\end{align}

Here $L_{\nu',0}'$ and $\Gamma_c$ are the core spectral luminosity and bulk-$\Gamma$ at $\theta=0$.

\subsubsection{\textbf{Synchrotron Emission From Structured Jets}}
The polarization curves for a smooth top-hat jet are presented in Fig.~\ref{fig:Pol-Smooth-Jet} for different B-field configurations 
as well as for different levels of smoothness of the edges. The behavior is 
similar for $\theta_{\rm obs}<\theta_j$ but significant differences between the top-hat jet case appear for $\theta_{\rm obs}>\theta_j$. Now that 
the spectral luminosity does not fall off so sharply for off-axis observers in the latter case, there is always some emission beamed along the LOS. 
For B-field configurations that show a larger degree of symmetry of the direction of polarization vectors around the LOS (e.g., $B_\perp$ and $B_\parallel$), 
the net polarization starts to decline as the edges of the jet are made smoother. This occurs due to the increase in symmetry that was broken 
sharply in the top-hat jet. A completely opposite behavior is seen in ordered B-field configurations, where the polarization increases with increasing 
smoothness. This arises since for a very sharp edge the observed flux is dominated by the core and once most of it has a similar weight (i.e. beaming 
and Doppler factor) then a lot of canceling occurs, while for a very smooth or gradual edge the flux is dominated by the region near the line of 
sight where the B-field is ordered resulting in very little averaging out of the polarization.

The right column of Fig.~\ref{fig:Pol-Smooth-Jet} shows the polarization curves for structured jets. When compared with polarization curves from top-hat 
jets or even smooth top-hat jets, these are broadly similar\footnote{Note that the $\delta=2$ smooth top-hat jet (left panel of Fig.~\ref{fig:Pol-Smooth-Jet}) is broadly similar in structure to ($a=2$, $b=0$) structured jet (right panel), where both show similar polarization behavior, and therefore a $\delta=2$ smooth top-hat jet can also be considered a structured jet.} 
but the curves are now stretched towards larger viewing angles. This means that appreciable polarization can now be measured when the LOS falls outside of the 
brighter core. In addition to that, the drop in fluence for viewing angles outside of the core is not so severe, as was found for the top-hat jet. Therefore, 
depending on the exact angular profile, off-axis observers with $q=\theta_{\rm obs}/\theta_c\lesssim$ few to several can still detect the GRB 
and measure high levels of polarization. This is demonstrated in Fig.~\ref{fig:Pol-Smooth-Jet} using a dotted line where the solid to dotted line 
transition occurs when the off-axis ($\theta_{\rm obs}>0$)
to on-axis ($\theta_{\rm obs}=0$) fluence ratio has dropped to $1\%$. Nevertheless, there are additional constraints on the detectability of such off-axis 
bursts. For example, when the bulk-$\Gamma$ is non-uniform and declines with $\theta$, the viewing angle out to which the prompt emission can be observed 
may be limited by compactness \citep[e.g.][]{Beniamini-Nakar-19,Matsumoto+19,Gill+20}. This is shown using a thick dot in the figure beyond which the Thomson 
optical depth of the $e^\pm$-pairs ($\tau_T$) produced due to $\gamma\gamma$-annihilation becomes greater than 10. As a result, the polarization is rather 
limited to $\Pi\lesssim20\%$ for $B_\perp$ and $B_\parallel$, but it can be much higher for the ordered field in $B_{\rm tor}$.

\end{paracol}
\begin{figure}
    \widefigure
    \centering
    \includegraphics[width=0.37\textwidth]{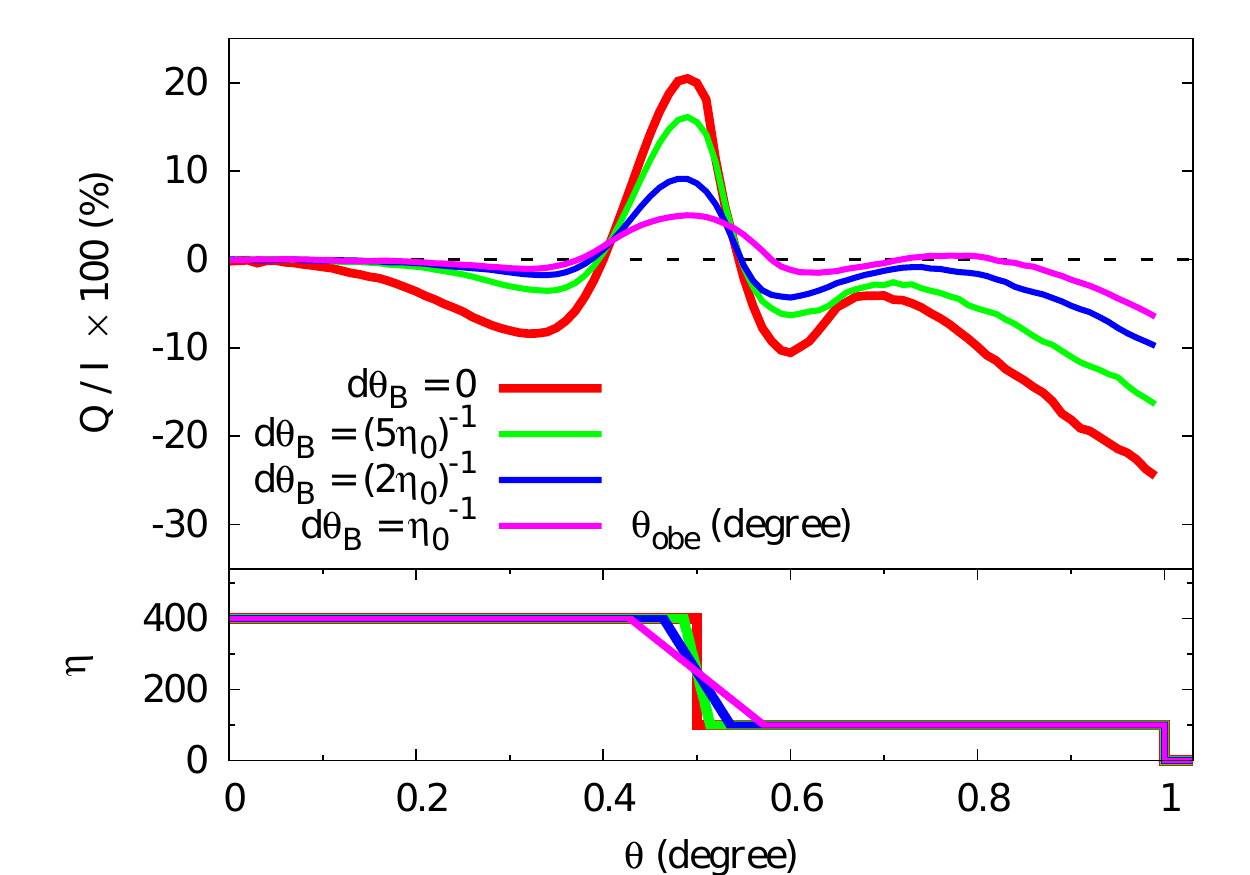}\hspace{3em}
    \includegraphics[width=0.37\textwidth]{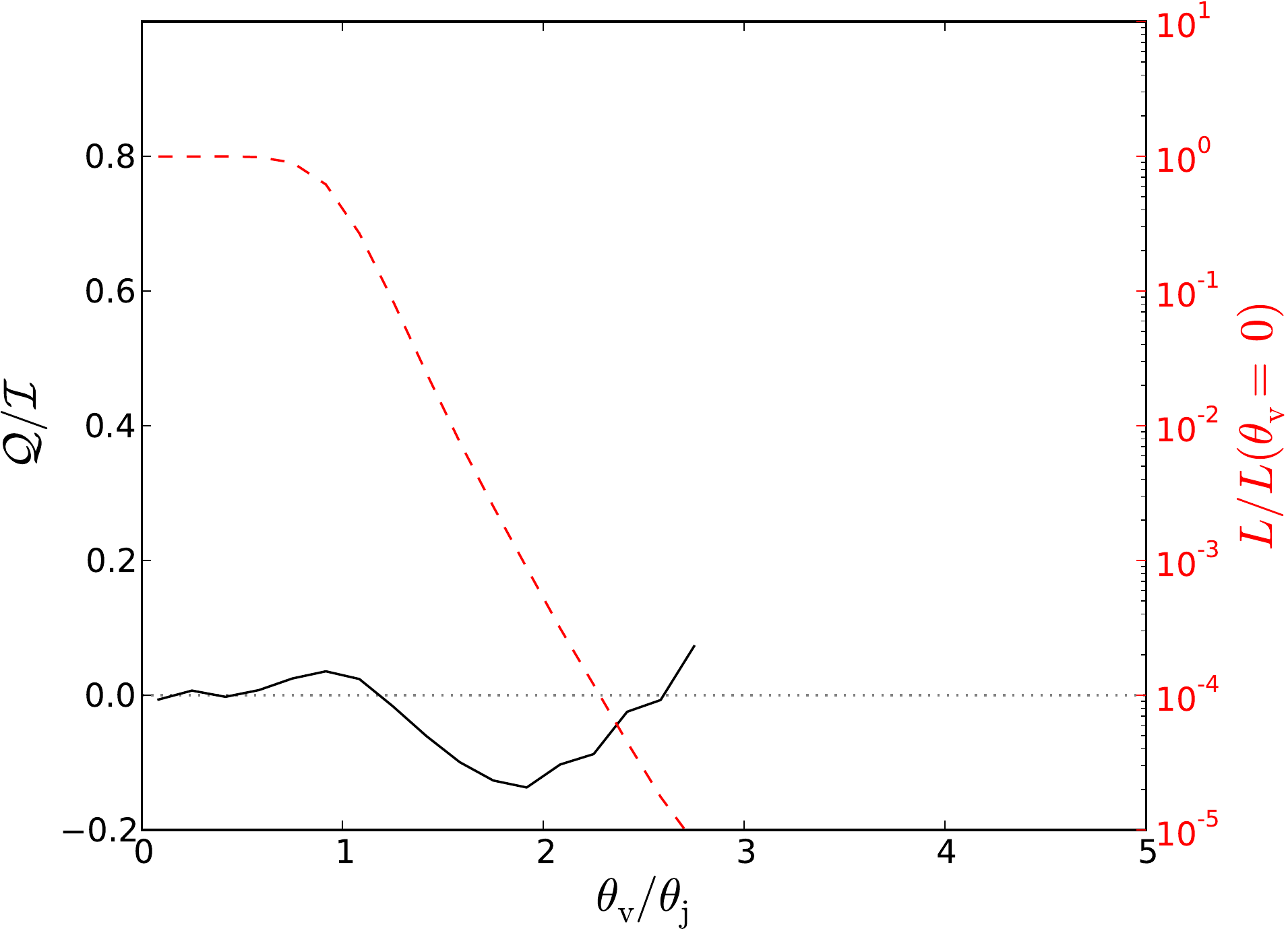}
    \includegraphics[width=0.37\textwidth]{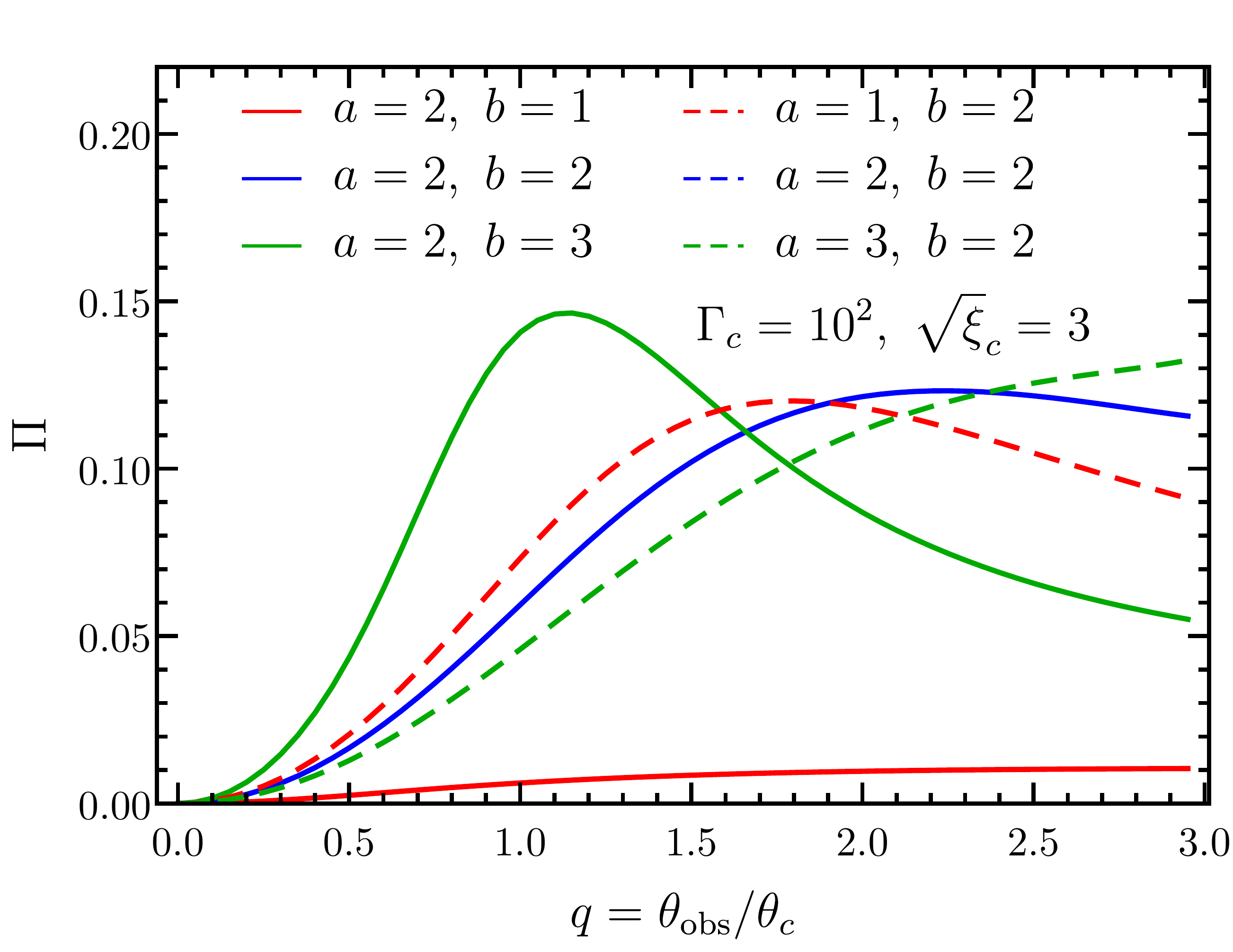}\hspace{3em}
    \includegraphics[width=0.37\textwidth]{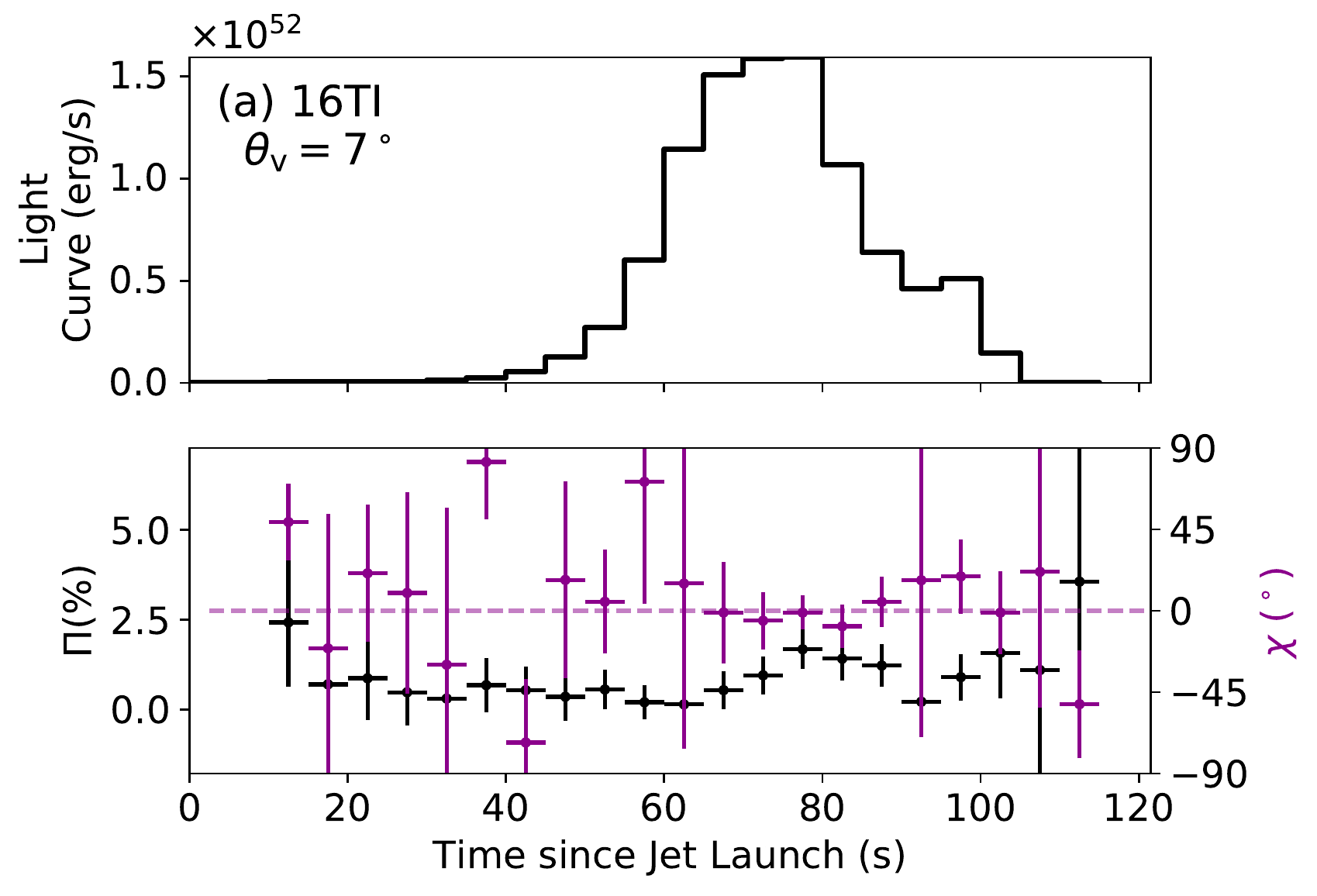}
    \caption{Polarization from non-dissipative photospheric emission model in a structured jet. 
    (\textbf{Top-left}) Polarization from the Monte Carlo (MC) simulation of \citet{Ito+14} shown for different viewing angles $\theta_{\rm obs}$ 
    and different gradients in bulk-$\Gamma$ (here $\eta$).
    (\textbf{Top-right}) MC simulation results from \citet{Lundman+14} featuring a uniform core with half-opening angle $\theta_j$ and power-law 
    shear ($\Gamma(\theta)\propto\theta^{-4}$) layer in bulk-$\Gamma$. The off-axis spectral luminosity normalized by the on-axis value 
    (viewing angle $\theta_v=0$) is shown with dashed red line.
    (\textbf{Bottom-left}) Polarization of photospheric emission from a structured jet obtained from semi-analytic radiation transfer calculation 
    of \citet{Gill+19} that features angular structure in both the comoving emissivity ($L_{\nu'}'(\theta)\propto\Theta^{-a}$, see Eq.~(\ref{eq:Pi_max})) 
    and bulk-$\Gamma$ ($\Gamma(\theta)\propto\Theta^{-b}$) with $\sqrt\xi_c=\Gamma_c\theta_c=3$ where $\theta_c$ is the core angle. The solid 
    lines fix $a=2$ and dotted lines set $b=2$ to disentangle the effect of the two profiles.
    (\textbf{Bottom-right}) Polarization derived from a MC simulation with outflow properties obtained from a 2D special relativistic hydrodynamic 
    simulation of a jet launched inside a Wolf-Rayet star (from \citet{Parsotan+20}). The top-panel shows the lightcurve and the bottom panel shows the temporal evolution of 
    $\Pi$ and position angle $\chi$. 
    }
    \label{fig:Photspheric-Pol}
\end{figure}
\begin{paracol}{2}
\switchcolumn

\subsubsection{\textbf{Photospheric Emission from Structured Jets}}
Photospheric emission yields negligible polarization in a uniform jet unless the viewing angle is less than one beaming cone away from the edge of the jet, i.e.
$|q-1|\lesssim\xi_j^{-1/2}\Leftrightarrow\Gamma|\theta_{\rm obs}-\theta_j|\lesssim1$. 
One way to obtain finite net polarization is by having a structured jet (see Fig.~\ref{fig:Photspheric-Pol}).
This was initially demonstrated in Monte Carlo (MC) simulations of photospheric emission emerging from axisymmetric relativistic outflows 
\citep{Lundman+14,Ito+14} that featured 
sheared layers outside of the uniform core with gradients in bulk-$\Gamma$ as a function of the polar angle $\theta$. It was shown that narrow jets 
with $\Gamma\theta_c\approx1$ and steep gradients in bulk-$\Gamma$ with $\Gamma(\theta)\propto\theta^{-p}$ for $\theta>\theta_c$ (some works use 
the symbol $\theta_j$ instead of $\theta_c$ to refer to the half-opening angle of the uniform core) and $p\sim4$ can yield polarization 
$\Pi\lesssim40\%$ for $q=\theta_{\rm obs}/\theta_c\gtrsim1$. A more 
realistic scenario would have $\Gamma\theta_c\approx10$ in which case $\Pi\lesssim10\%$ is expected. A similar conclusion is reached by carrying 
out a radial integration of the radiation transfer equations for the Stokes parameters in a steady flow having angular structure in the comoving 
emissivity and bulk-$\Gamma$ \citep{Gill+20}. The results of this work are shown in the bottom-left panel of Fig.~\ref{fig:Photspheric-Pol}, and 
even here it was realized that steep gradients in the bulk-$\Gamma$ profile are required to achieve significant polarization with $\Pi\lesssim15\%$.

A more realistic scenario was explored in \citet{Parsotan+20} who carried out two dimensional (2D) special relativistic hydrodynamic simulations 
of a jet launched inside a Wolf Rayet star. The flow dynamics and angular structure thus obtained from the simulation were then used with a MC 
code to obtain the polarization of photospheric emission at the last scattering surface. The results are shown in the bottom-right panel of 
Fig.~\ref{fig:Photspheric-Pol} that shows the lightcurve and temporal evolution of the polarization and PA, with the conclusion 
that $\Pi\lesssim2.5\%$ and PA remained steady within the uncertainties. In other cases, where the outflow showed more structure, a slightly 
larger time-resolved polarization of $\Pi\lesssim5\%$ and time-variable PA was obtained.

\subsection{\textbf{Temporal Evolution of Polarization}}\label{sec:Pol-temp-evol}

\end{paracol}
\begin{figure}
    \widefigure
    \centering
    \includegraphics[width=0.4\textwidth]{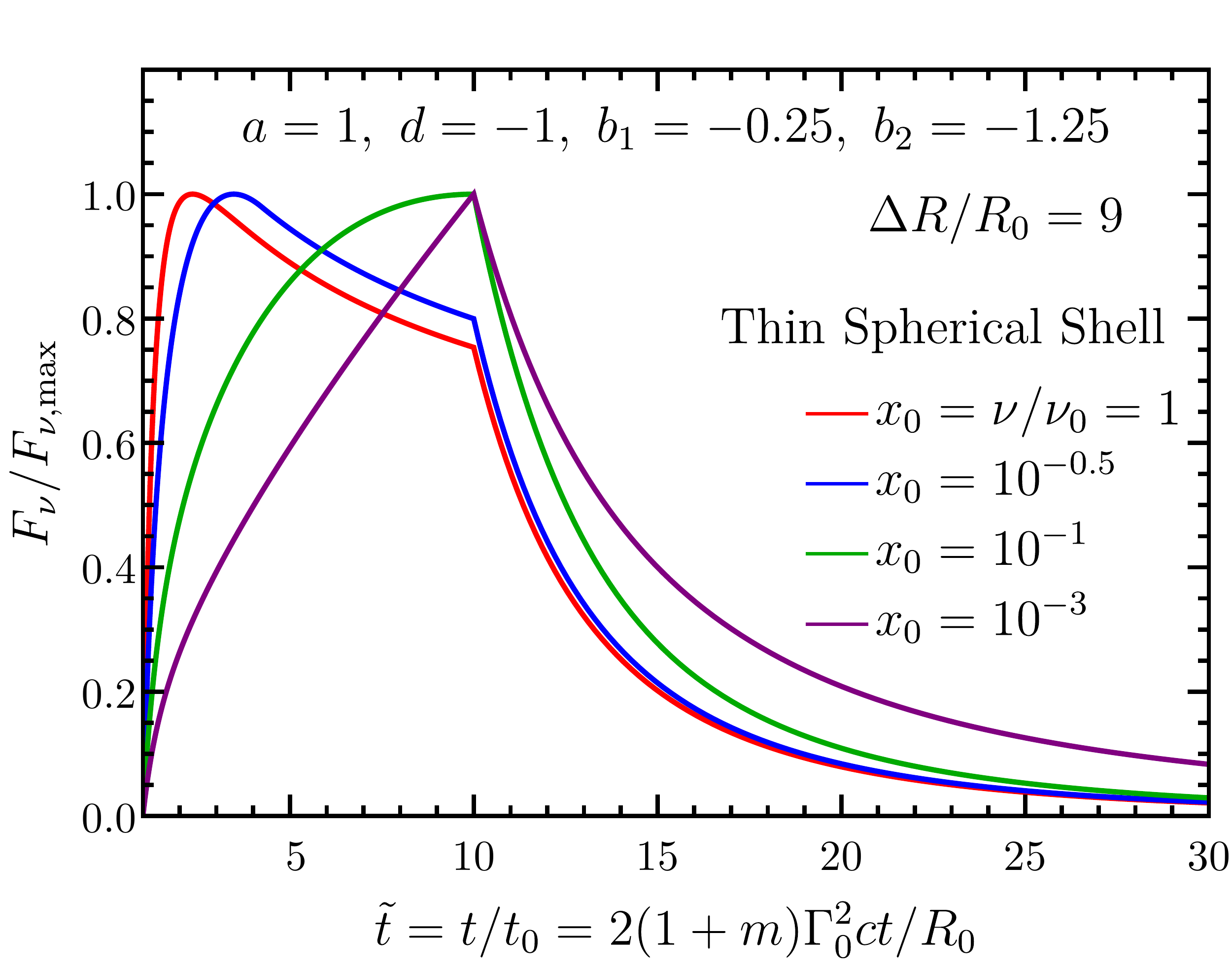}\hspace{5em}
    \includegraphics[width=0.4\textwidth]{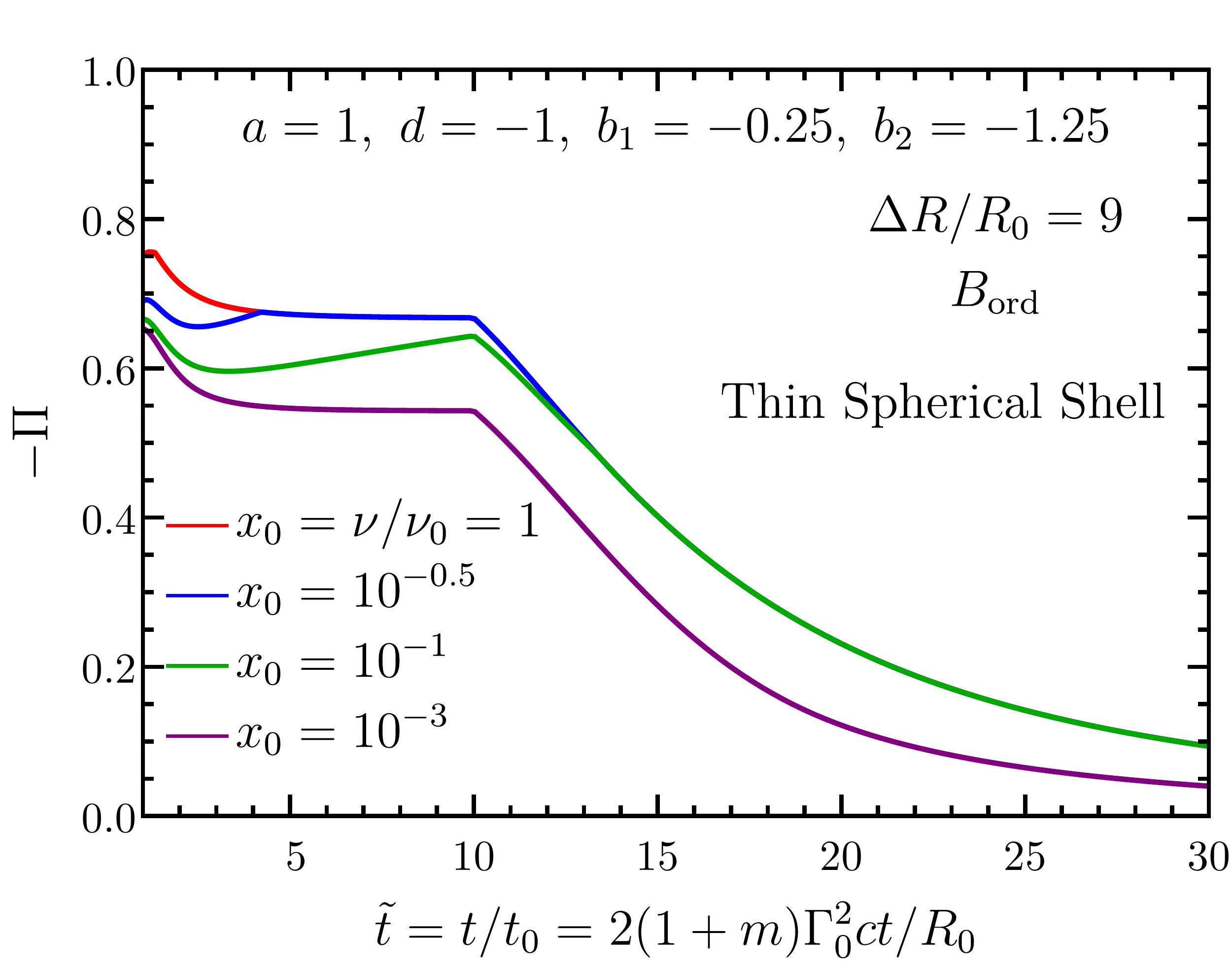}
    \caption{Pulse profile (left) and temporal evolution of polarization (right) for a coasting ($m=0$) 
    ultrarelativistic ($\Gamma_0\gg1$) thin spherical shell with an ordered field ($B_{\rm ord}$). Here energy is 
    dissipated in internal shocks in a KED flow and the emission is synchrotron, which is modeled using a Band-function 
    with asymptotic spectral indices $b_1$ and $b_2$. The shell starts to radiate at $R=R_0$ and terminates at 
    radius $R_f=R_0(1+\Delta R/R_0)$. The comoving spectral luminosity and spectral peak evolve as a power law in 
    radius with indices $a$ and $d$, respectively (see Eq.~(\ref{eq:R-evol-Lnu-nupk})). The different curves show the 
    trend at the observed frequency $\nu = x_0\nu_0$ where $\nu_0$ is the $\nu F_\nu$-peak frequency of the first photons 
    emitted along the observer's LOS from radius $R_0$, which then arrive at the apparent time $t=t_0$. The emission is 
    assumed to have a Band-function spectrum with asymptotic power-law spectral indices $b_1$ and $b_2$ below and above 
    the spectral peak energy, respectively. Figure adapted from \citep{Gill-Granot-21}.}
    \label{fig:Fnu-Pol-Sph-Shell}
\end{figure}
\begin{paracol}{2}
\switchcolumn

The earlier sections only discuss the pulse-integrated polarization, which is relevant for most GRBs that are not bright enough to be able to yield any 
time-resolved polarimetric results. However, with the upcoming more sensitive gamma-ray polarimeters in the next decade time-resolved polarimetry of prompt 
GRB emission will become possible. Therefore, in anticipation of such a development it is prudent to also construct accurate theoretical model predictions 
to compare with time-resolved polarization measurements.

When discussing time-resolved polarization it becomes important to include the radial dependence of the flow properties, which were ignored for the 
pulse-integrated discussion. We first describe a simple and very general pulse model of an accelerating, coasting, or decelerating flow (see, e.g., 
\citep{Genet-Granot-09,Willingale+10}) that is then 
used to calculate the time-resolved polarization. Consider a thin ultra-relativistic shell that starts to emit prompt GRB photons at radius $R=R_0$. The 
emission continues over a radial extent $\Delta R$ and terminates at $R_f = R_0+\Delta R$. During this time the comoving spectrum, with $\nu'L'_{\nu'}$ 
spectral peak frequency $\nu_{\rm pk}'$, and spectral luminosity evolve as a power law with radius,
\begin{equation}\label{eq:R-evol-Lnu-nupk}
    L_{\nu'}'(R,\theta) = L_0'\fracb{R}{R_0}^aS\fracb{\nu'}{\nu_{\rm pk}'}f(\theta)\quad\quad{\rm with}\quad\quad \nu_{\rm pk}' = \nu_0'\fracb{R}{R_0}^d\,,
\end{equation}
where $L_0' = L_{\nu'}'(R_0)$ and $\nu_0' = \nu_{\rm pk}'(R_0)$ are the normalizations. The factor $f(\theta)$ describes the angular profile of $L_{\nu'}'$ 
where it is normalized to unity at the jet-symmetry axis with $f(0)=1$. For a uniform spherical flow $f(\theta)=1$ and for a top-hat jet 
$f(\theta)=\mathcal{H}(\theta_j-\theta)$ with $\mathcal{H}$ being the Heaviside function and $\theta_j$ the jet half-opening angle. The comoving spectrum 
is described by the function $S(x)$, which is considered here to be the Band-function, where $x=\nu'/\nu_{\rm pk}'$. The dynamics of the thin shell are 
given by the radial profile of the bulk-$\Gamma$, such that $\Gamma^2(R)=\Gamma_0^2(R/R_0)^{-m}$ where $\Gamma_0 = \Gamma(R_0)$. The shell is coasting 
when $m=0$ and accelerating (decelerating) for $m<0$ ($m>0)$. Once the power law indices $a$ and $d$ for $L_{\nu'}'$ are provided, one has complete information 
of the temporal evolution of the pulse. These indices depend on the details of the underlying prompt GRB model, e.g., on the composition and dissipation 
mechanism. If the prompt GRB spectrum is assumed to be of synchrotron origin, then it can be shown \citep{Gill-Granot-21} that for a KED flow, where energy 
is dissipated at internal shocks ($m=0$), $a=1$ and $d=-1$. Alternatively, if the flow is PFD with a striped wind B-field structure and energy is dissipated 
due to magnetic reconnection, which also accelerates the flow with $m=-2/3$, then it's found that $a=4/3$ and $d=-2$. 

\begin{figure}
    \centering
    \includegraphics[width=0.48\textwidth]{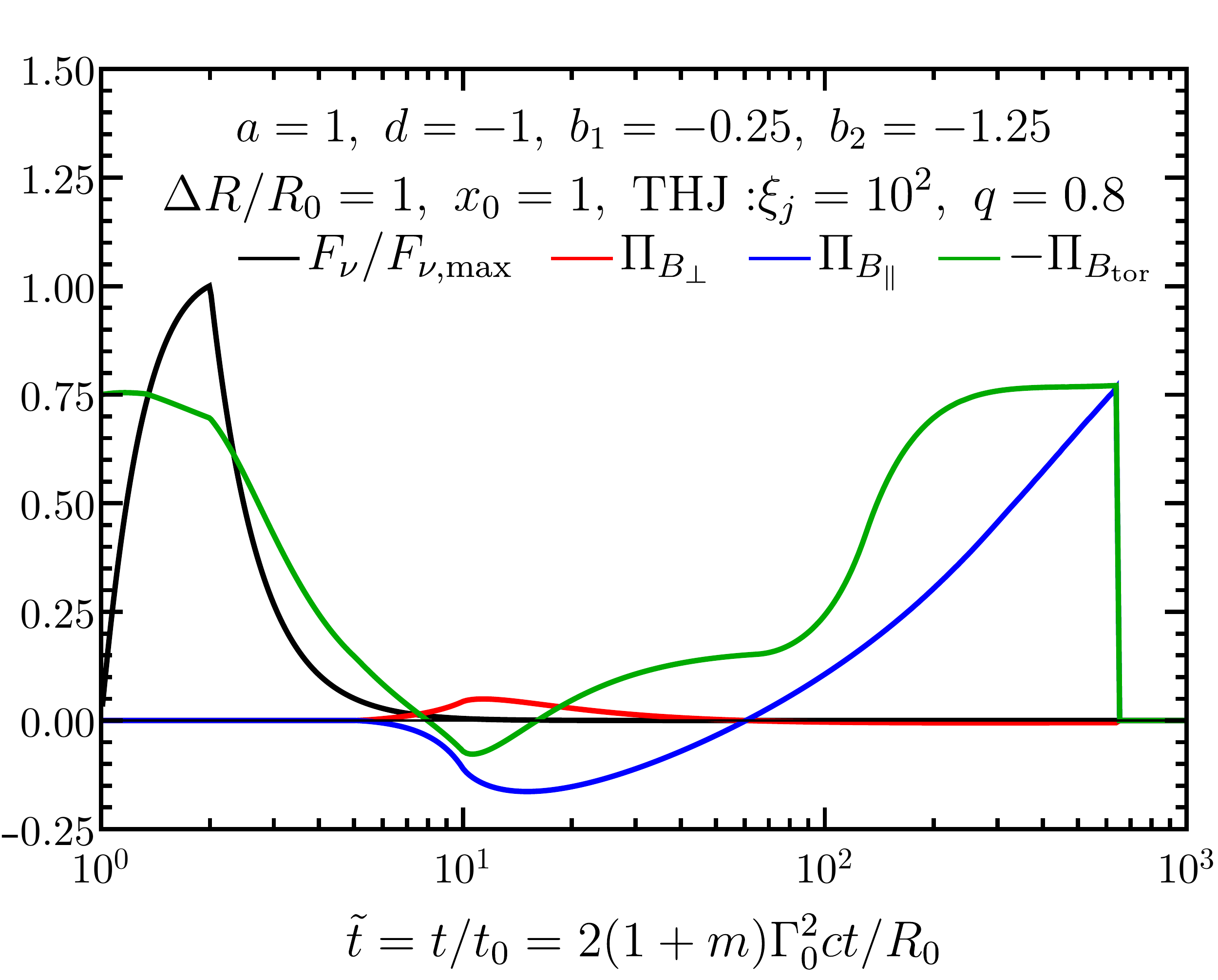}
    \caption{Pulse profile (black) and temporal evolution of synchrotron polarization in a top-hat jet 
    (THJ, with $\xi_j=(\Gamma\theta_j)^2$ and $q=\theta_{\rm obs}/\theta_j$) for different B-field configurations. 
    See caption of Fig.~\ref{fig:Fnu-Pol-Sph-Shell} for explanation of different symbols and parameters. 
    Figure adapted from \cite{Gill-Granot-21}.}
    \label{fig:Fnu-Pol-time-evol-diff-B}
\end{figure}

The pulse profile and temporal evolution of polarization for a KED flow coasting at $\Gamma_0\gg1$ is shown in Fig.~\ref{fig:Fnu-Pol-Sph-Shell} for an 
ordered B-field ($B_{\rm ord}$). The 
different curves are shown for observed frequency $\nu = x_0\nu_0$, which is a fraction $x_0$ of the peak frequency $\nu_0=2\Gamma_0\nu_0'$ of the first 
photons emitted along the LOS at radius $R_0$. The apparent arrival time of these first photons is given by $t_{0,z}\equiv t_0/(1+z) = R_0/2(1+m)\Gamma_0^2c$, 
which is the characteristic radial delay time between the shell to arrive at radius $R_0$ and the hypothetical photon that was emitted by the engine 
at the same time as the shell. For $m=0$ this is also the angular time over which radiation from within the beaming cone around the LOS arrives at the observer. 
Depending on $x_0$ the pulse profile changes and shows a peak at 
different times with the latest peak occurring at $\tilde t_f \equiv t_f/t_0 = \hat R_f^{1+m} = (R_f/R_0)^{1+m} = (1+\Delta R/R_0)^{1+m}$, the arrival 
time of last photons emitted along the LOS from radius $R_f$. At $\tilde t > \tilde t_f$ the flux density declines rapidly and the pulse becomes dominated 
by high-latitude emission that originates from outside of the beaming cone, i.e. from angles larger than $1/\Gamma_0$ from the LOS. 

The polarization curves show maximal polarization initially, corresponding to $\Pi_{\max}(\alpha)$ depending on the local value of the spectral index $\alpha$ for the 
Band-function as set by $x_0$. For $\tilde t<\tilde t_f$ the polarization first declines and then saturates which reflects the averaging of local 
polarization over the beaming cone as seen on the plane of the sky, which tends to yield a net polarization lower than $\Pi_{\max}$. For $\tilde t>\tilde t_f$, 
like the pulse profile, the polarization also declines rapidly when high-latitude emission becomes dominant. The polarization curves at different $x_0$ 
merge at $\tilde t=\tilde t_{\rm cross}(x_0)$, the crossing time of the break frequency across the observed frequency as the entire spectrum drifts 
towards softer energies over time. The merging of the polarization curves occurs due to the fact that after time $\tilde t_{\rm cross}$ all photons at 
the observed frequency $\nu$ are harder than the Band-function break frequency beyond which the Band-function features a strict power law with a given 
spectral index. Therefore, the level of polarization for all photons sampling the power law is also the same as dictated by $\Pi_{\max}(\alpha)$.

The polarization is not always maximal at the start of the pulse if the magnetic field is not ordered. This is demonstrated in 
Fig.~\ref{fig:Fnu-Pol-time-evol-diff-B} that shows the pulse profile and temporal evolution of synchrotron polarization for different B-field configurations 
in a top-hat jet. As argued earlier, in B-field configurations, e.g. $B_\perp$ and $B_\parallel$, that produce axisymmetric polarization maps around the 
LOS the net polarization vanishes. This symmetry is only broken when the observer becomes aware of the jet edge, e.g. in a top-hat jet. It is at that 
instant the magnitude of polarization begins to grow above zero. The polarization curves for the three B-field configurations also show a change in the 
PA by $\Delta\theta_p=90^\circ$ when the curves cross zero. Interestingly, this happens more than once for $B_{\rm tor}$. The reason for this can be 
understood from the polarization maps shown in Fig.~\ref{fig:Pol-Map} where the $90^\circ$ change in the PA occurs when the net polarization begins to 
be dominated by emission polarized along the line connecting the jet symmetry axis and the observer's LOS over that polarized in the transverse direction 
or vice versa. At late times, the observed emission vanishes after the arrival time of the last photons from the edge of the jet furthest from the LOS. 
Since the flux declines very rapidly at $\tilde t>\tilde t_f$, the changes in the PA are challenging to detect in practice.

\end{paracol}
\begin{figure}
    \widefigure
    \centering
    \includegraphics[width=0.48\textwidth]{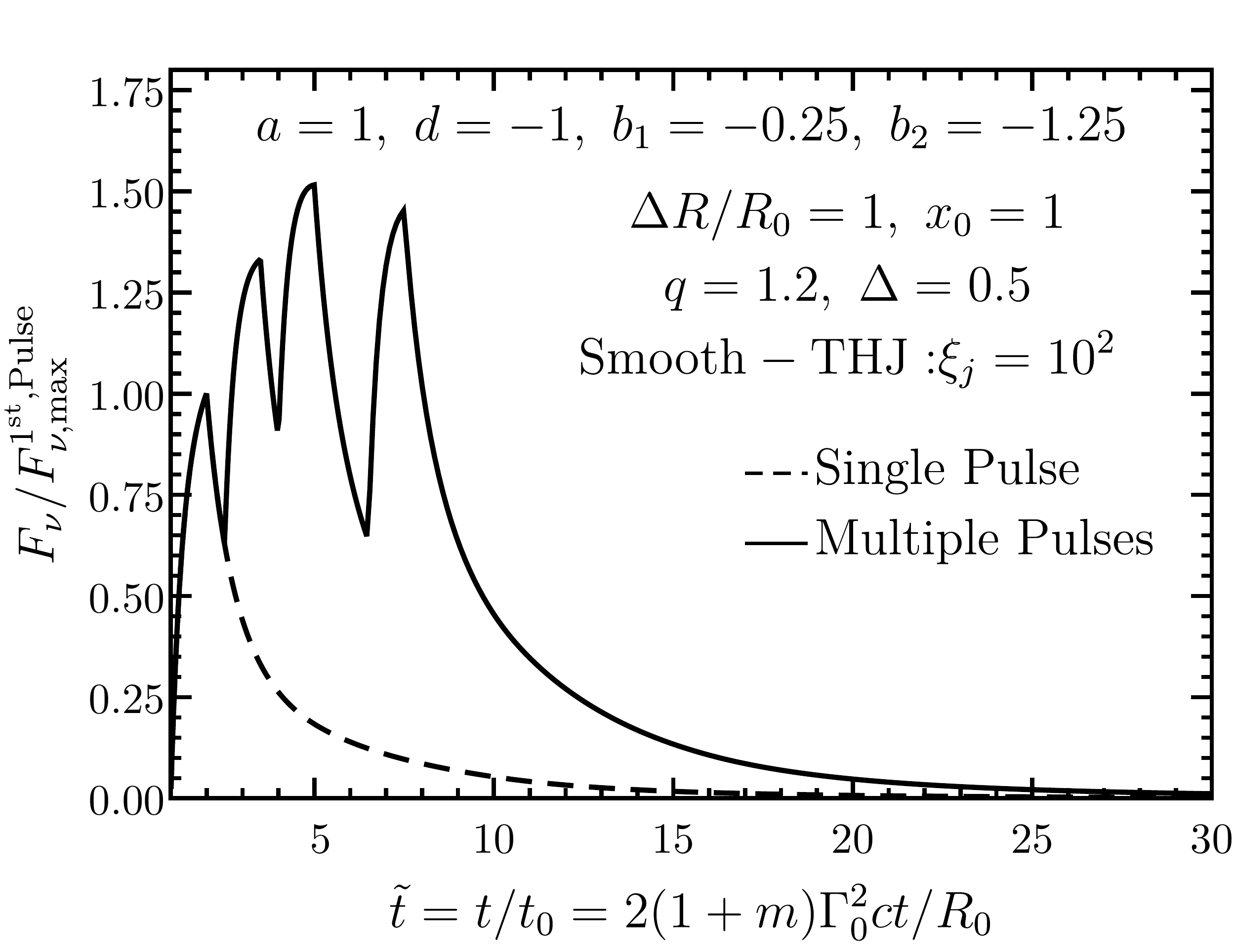}
    \includegraphics[width=0.48\textwidth]{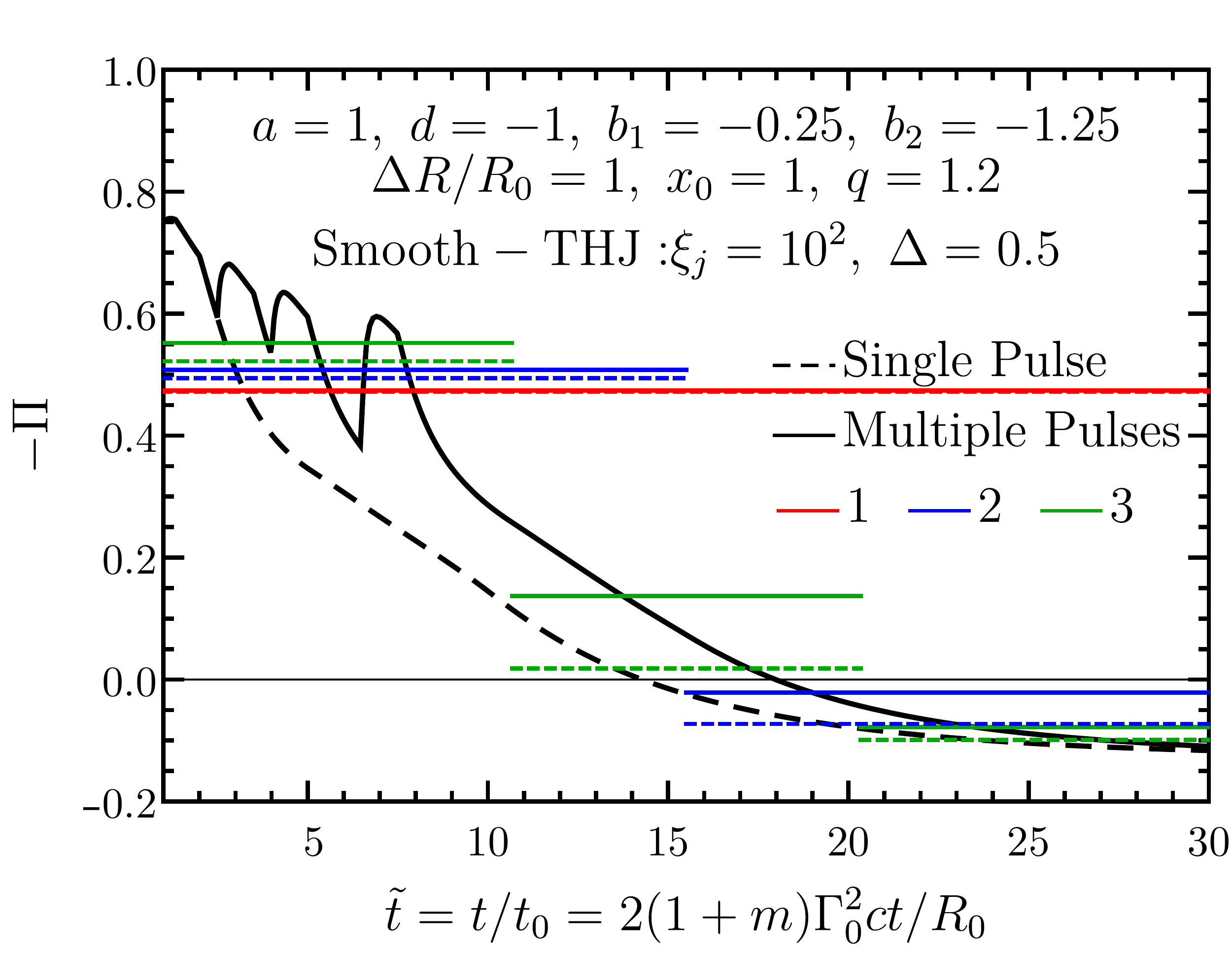}
    \caption{(\textbf{Left}) Pulse profile of multiple overlapping pulses in an emission episode, shown here for a KED smooth top-hat jet. A single pulse 
    is also shown for comparison. 
    (\textbf{Right}) Temporal evolution of the polarization for a toroidal magnetic field ($B_{\rm tor}$) shown for both the single pulse and multiple pulses. 
    Temporal segments over which polarization is obtained are calculated by dividing the pulse into one (red), two (blue), or three (green) part(s). See the 
    caption of Fig.~\ref{fig:Fnu-Pol-Sph-Shell} for explanation of different symbols. Figure adapted from \citep{Gill-Granot-21}.}
    \label{fig:Multi-Pulse}
\end{figure}
\begin{paracol}{2}
\switchcolumn

\subsection{\textbf{Polarization From Multiple Overlapping Pulses}}
\label{sec:muliple-pulses}

Since GRBs are generally photon starved, the only hope of obtaining a statistically significant polarization measurement often relies on integrating over 
broad segments of the prompt GRB lightcurve. Due to the highly variable nature of the prompt GRB emission, a given emission episode consists of 
multiple overlapping pulses. The properties of the emission region, e.g. bulk-$\Gamma$, B-field configuration, can change between different pulses 
and improper accounting of these changes in calculating the time-integrated polarization can lead to erroneous results.

In the simplest scenario multiple pulses are produced by distinct patches or mini-jets within the observed region of size $R/\Gamma$ of the outflow 
surface. These patches can be permeated by an ordered B-field the orientation of which is also mutually distinct among the different patches. 
A broadly similar B-field structure can also be obtained in both internal and external shocks due to macroscopic turbulence excited by, e.g. the Richtmyer-
Meshkov instability, that arises in the interaction of shocks and upstream density inhomogeneities \citep{Zhang+09,Inoue+11,Mizuno+11,Deng+17}. 
In the case of mini-jets the bulk-$\Gamma$ of the different jets can also be different by a factor of order unity which will affect the size of the 
individual beaming cones. Since the Stokes parameters are additive for incoherent emission the time-integrated net polarization of $N_p$ incoherent 
patches (in the visible region of angular size $1/\Gamma$ around the line of sight) is obtained from \citep{Gruzinov-Waxman-99} (where the motivation 
was afterglow emission from a shock-generated field rather than incoherent patches or mini-jets)
\begin{equation}
    \Pi = \frac{Q}{I} = \frac{\sum_{i=1}^{N_p}Q_i}{\sum_{i=1}^{N_p}I_i}\sim\frac{\Pi_{\max}}{\sqrt N_p}\,.
\end{equation}
The net polarization is significantly reduced for increasingly large number of patches due to the fact that the PA are randomly oriented and when 
added together some cancellation occurs. This essentially represents a random walk for the polarized intensity $Q$ while the total intensity adds 
up coherently. When multiple time-integrated segments of an emission episode are compared the net polarization and PA will vary between them (the 
latter is possible as this is a non-axisymmetric global configuration). Alternatively, instead of ordered B-field patches one can have a shock-produced B-field (e.g. $B_\perp$) with a patchy shell or mini-jets that give different weights to different parts of the image and thereby produce a net polarization (see, e.g. \citep{Granot-Konigl-03,Nakar-Oren-04}).

Another scenario that is worth considering is when multiple overlapping pulses are produced by episodic energization of the emission region, e.g. in 
the collision of multiple shells in the internal shock scenario where the ejection time of subsequent shells is different, such that the ejection 
time of the $i^{\rm th}$ shell in the engine frame is $t_{\rm ej,i,z} = t_{\rm ej,i}/(1+z)$. The onset time of each pulse is then given by 
$t_{\rm onset,i,z} = t_{\rm ej,i,z} + t_{0,z}$. The scenario of multiple pulses from a smooth top-hat jet is demonstrated in Fig.~\ref{fig:Multi-Pulse} 
using simplifying assumptions, where all pulses have the same $R_0$ and $\Gamma(R_0)$ (so that the radial delay time $t_{0,z}$ for emission arising 
from different pulses is the same) and radial extent $\Delta R$. In this case, the onset times of pulses is simply dictated by the different ejection 
times of the shells. The left panel shows the pulse profile and the right panel shows 
the polarization calculated for the $B_{\rm tor}$ field. Time resolved polarization obtained from multiple temporal segments, where the emission episode 
is divided into one, two, or three equal duration segments, is shown to demonstrate the different levels of polarization obtained when using the multi-pulse or 
the single pulse model. Therefore, when the emission consists of multiple overlapping pulses, it is important to compare the measurement with model 
predictions that account for multiple pulses.

\end{paracol}
\begin{figure}
    \widefigure
    \centering
    \includegraphics[width=0.35\textwidth]{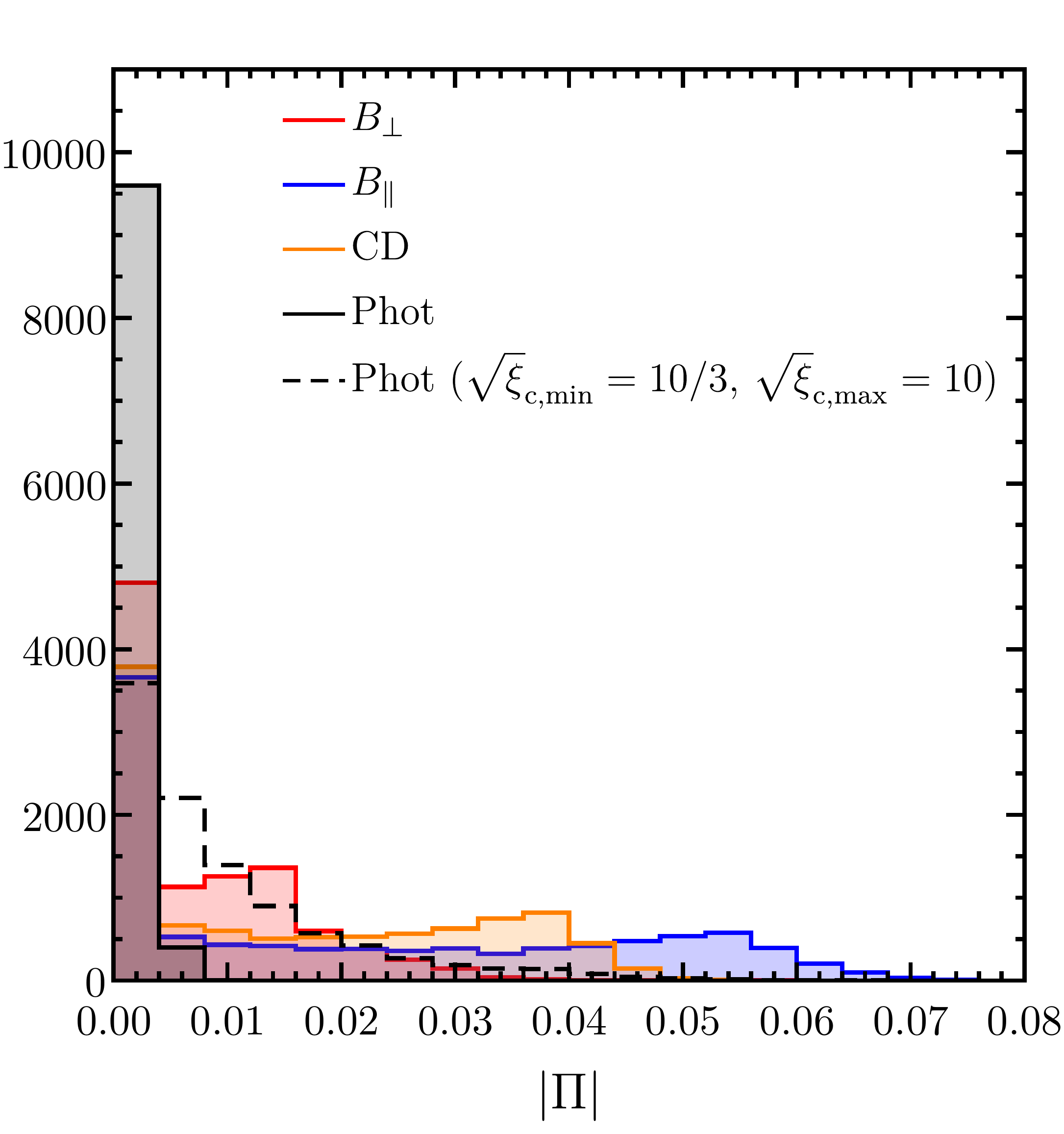}
    \includegraphics[width=0.55\textwidth]{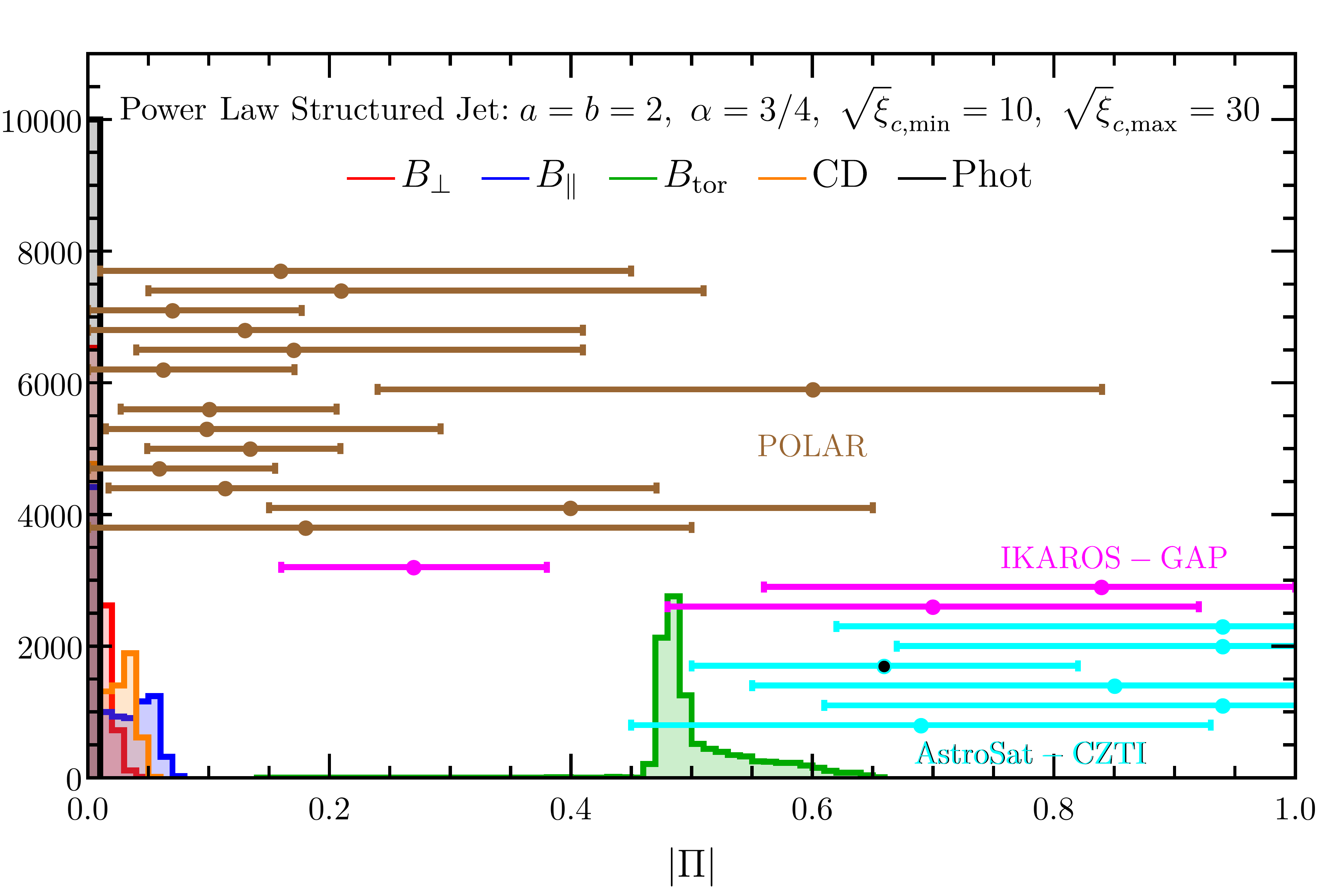}
    \caption{\textbf{\textit{Right}}: Distribution of polarization from synchrotron emission for different B-field configurations, 
    Compton drag (CD), as well as photospheric (Phot) emission in a power-law structured jet obtained from a Monte Carlo simulation 
    (with $10^4$ samples). Measured polarizations with $1\sigma$ error bars from different instruments are shown for comparison. The measurement of 
    $\Pi=66_{-27}^{+26}\%$ ($\sim5.3\sigma$) from \citep{Sharma+19} obtained using AstroSat-CZTI is shown with a black dot with cyan 
    error bars. Figure adapted from \citep{Gill+20} where more details can be found. 
    \textbf{\textit{Left}}: Zoomed-in version of the figure showing the several overlapping distributions for clarity (with 
    a bin size smaller by a factor of $0.4$).}
    \label{fig:Pol-Obs-Compare}
\end{figure}
\begin{paracol}{2}
\switchcolumn

\subsection{\textbf{Most likely polarization measurement}}
\label{sec:most-likely-pol}

As demonstrated in earlier sections, the prompt GRB polarization depends on (i) the underlying radiation mechanism, 
(ii) B-field structure (for synchrotron emission), (iii) bulk LF $\Gamma$ (top-hat jet) or $\Gamma_c$ (structured jet), 
(iv) $\theta_j$ (top-hat jet) or $\theta_c$ (structured jet), (v) viewing angle $\theta_{\rm obs}$, (vi) angular structure, 
e.g. power-law indices $a$ and $b$ for a power-law structured jet (see \S\ref{sec:Pol-Structured-Jets}). Due to variations 
in these parameters the polarization can vary between different pulses within the same GRB as well as between different GRBs. 
For an ultrarelativistic flow three basic quantities naturally arise that affect the polarization, namely (a) the normalized 
jet/core half-opening angle: $\xi_j^{1/2} = \Gamma\theta_j$ (top-hat jet) or $\xi_c^{1/2} = \Gamma_c\theta_c$ (structure jet), 
(b) the normalized viewing angle: $q=\theta_{\rm obs}/\theta_j$ (top-hat jet) or $q=\theta_{\rm obs}/\theta_c$ (structured jet), 
and (c) the normalized viewing-angle dependent fluence: $\tilde f_{\rm iso}(q,\xi_j) = E_{\gamma,\rm iso}(q,\xi_j)/E_{\gamma,\rm iso}(0,\xi_j)$ 
(top-hat jet) or $\tilde f_{\rm iso}(q,\xi_c)$ (structure jet), which is the ratio of the off-axis to on-axis isotropic-equivalent 
radiated energy or equivalently the fluence.

For different pulses emitted by the same GRB, it is natural to expect a considerable change in (iii) while the other parameters 
are likely to remain more or less fixed. In, e.g. a top-hat jet, this will change the parameter $\xi_j$ and for a given distribution 
of $\xi_j$ between several pulses the total polarization, after integrating over multiple pulses, will be different from 
that obtained for a single pulse. When adding up the Stokes parameters of different pulses an appropriate relative weight using, e.g. 
$E_{\gamma,\rm iso}$ (or more precisely the relative expected number of photons that will be detected), should be applied. 

When comparing emission from different GRBs all of the above mentioned quantities can in principle vary (or at least there is no strong 
evidence against this in the observed sample of GRBs). In this case, the fluence ratio is important in determining (i) whether for a given 
$\theta_{\rm obs} > \theta_j$ (top-hat jet) or $\theta_{\rm obs} > \theta_c$ (structured jet) the pulse will be bright enough to be observed 
by a given detector and (ii) for a given GRB out to which viewing angle it will be fluent enough for performing polarization measurements. 
For a top-hat jet the fluence is strongly suppressed due to Doppler de-beaming when $\Gamma(\theta_{\rm obs}-\theta_j)\gtrsim1$, whereas 
for a structured jet the suppression in fluence is not as severe and emission from $q\lesssim$ few to several can be detected if it is not 
suppressed due to compactness, as discussed earlier.

\end{paracol}
\begin{figure}
    \widefigure
    \centering
    \includegraphics[width=0.48\textwidth]{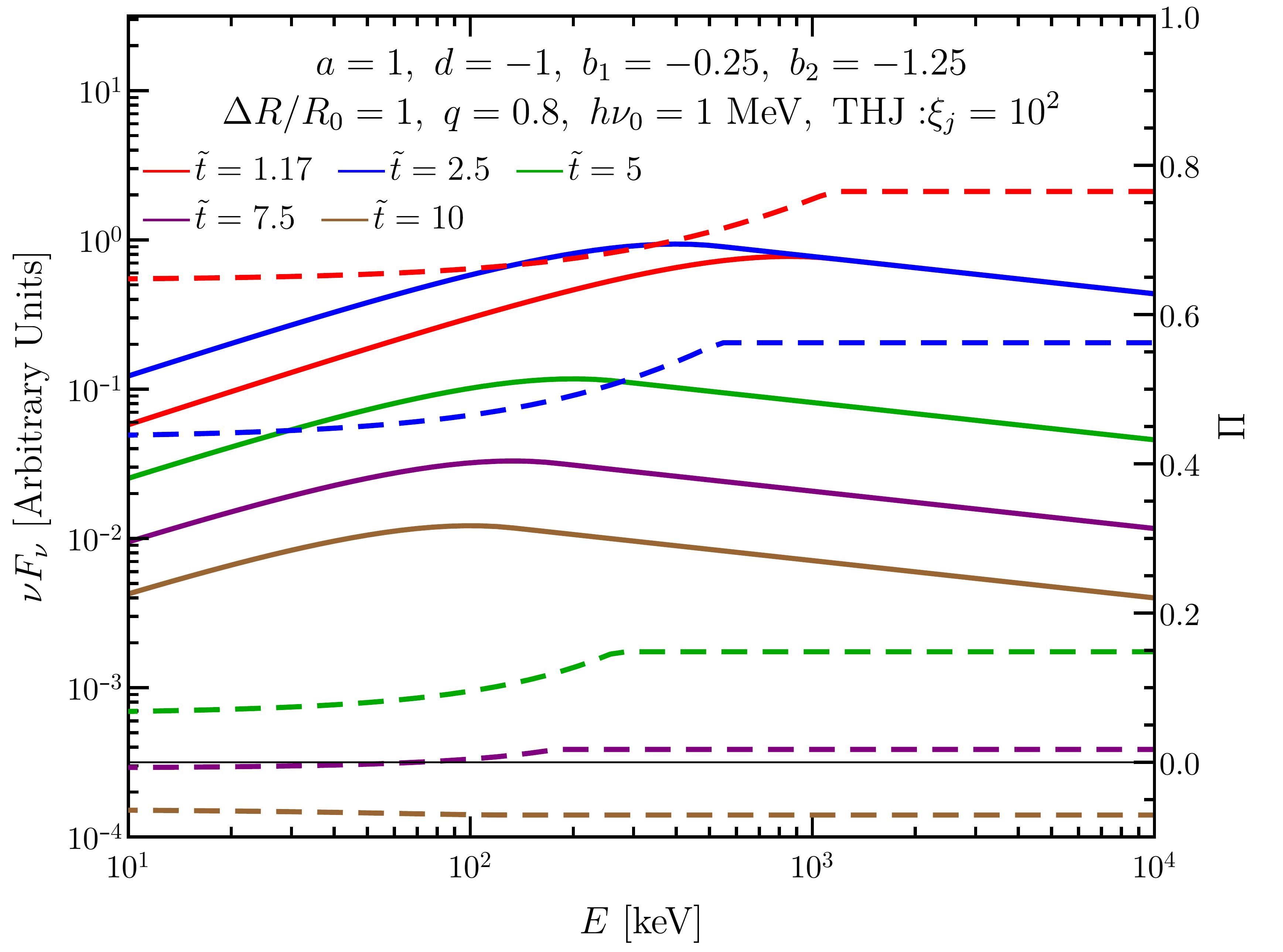}
    \includegraphics[width=0.48\textwidth,height=0.364\textwidth]{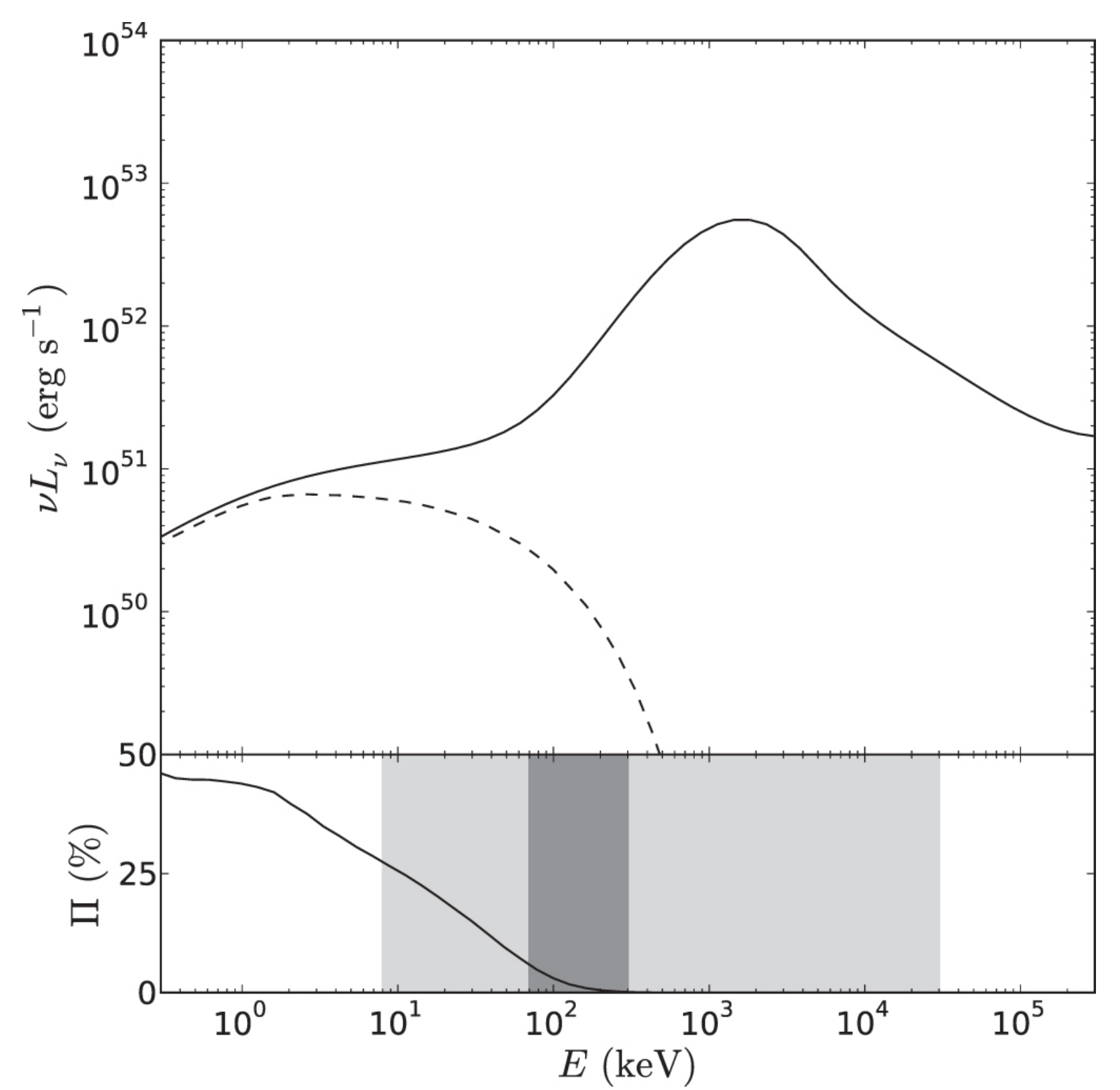}
    \caption{\textbf{Left:} Temporal evolution of the Band-like spectrum (\textit{solid lines}; left $y$-axis) and the corresponding 
    polarization (\textit{dashed lines}; right $y$-axis) from synchrotron emission with a $B_{\rm tor}$ field for a KED top-hat jet (THJ) 
    with $\xi_j=(\Gamma\theta_j)^2=10^2$ and $q=\theta_{\rm obs}/\theta_j=0.8$, and $m=0$. The different colors 
    correspond to different normalized apparent times $\tilde t=t/t_0$ where $t_0=2(1+m)\Gamma_0^2ct/R_0$ 
    is the arrival time of the initial photons emitted from radius $R_0$ along the LOS. The peak frequency 
    of the $\nu F_\nu$ spectrum at this time is given by $\nu_0$.
    \textbf{Right:} Multi-component GRB spectrum and its energy-resolved polarization. While the photospheric 
    component dominates both spectral peak and at higher energies, the low-energy spectrum is produced by 
    synchrotron emission. As a result, the polarization grows towards lower energies as the fraction of synchrotron 
    photon grows. The light and dark shaded regions correspond to the energy ranges of \emph{Fermi} GBM 
    (NaI + BGO detectors, ($8-30\,$MeV) and GAP ($70-300\,$keV), respectively. Figure from \citep{Lundman+18}.
    }
    \label{fig:Specto-Pol}
\end{figure}
\begin{paracol}{2}
\switchcolumn

A distribution of polarization for a given radiation mechanism, while accounting for variations in the aforementioned quantities between different pulses 
from the same source and different GRBs, and its comparison with actual measurements can be used to answer some of the key questions of GRB physics. Such 
a distribution obtained from a Monte Carlo simulation (see \citep{Gill+20} for more details) is shown in Fig.~\ref{fig:Pol-Obs-Compare} for a power-law jet 
and for different radiation mechanisms as well as different B-field configurations. As expected, the $B_{\rm tor}$ field being ordered yields the highest 
polarization with $45\%\lesssim\Pi\lesssim60\%$. 
Therefore, if GRB jets feature a large-scale toroidal field then most GRBs that are emitting synchrotron radiation will show $\Pi\sim50\%$. For the 
other two B-field configurations, $B_\perp$ and $B_\parallel$, the expected polarization is small with $\Pi\lesssim10\%$ and one is most likely to find 
GRBs with negligible polarization. The same conclusion can be drawn for the Compton drag and photospheric radiation mechanisms. The polarization 
in the photospheric emission model can be $\Pi\lesssim15\%$ when the flow features a much steeper bulk-$\Gamma$ angular profile with 
$\sqrt{\xi_c}=\Gamma_c\theta_c\sim\,$few (see Fig. 14 of \citep{Gill+20} for more details). When comparing with 
observations, some of which have at least $3\sigma$ detection significance, no firm conclusions can be drawn at this point. Measurements made by 
IKAROS-GAP and AstroSat-CZTI find highly polarized GRBs with $\Pi\gtrsim50\%$, although with large $1\sigma$ error bars. On the other hand, the POLAR data 
appears to indicate that GRBs are more likely to have significantly smaller polarization with most of their sample consistent with unpolarized sources. 
The apparent discord between the results of these works not only highlights the challenges involved in obtaining a statistically significant polarization 
measurement but also calls for the need to build more sensitive detectors.

\subsection{\textbf{Energy Dependence of Polarization}}
\label{sec:energy-dependence}
Polarization is energy dependent. This can be easily seen in emission mechanisms where the local polarization depends on the spectral index, e.g. in optically-thin 
synchrotron radiation (see Eq.~\ref{eq:Pi_max}).  The energy dependent spectro-polarimetric evolution in this case is shown in the left panel of 
Fig.~\ref{fig:Specto-Pol}; temporal evolution of polarization at a given energy and the pulse profile for the same case was shown earlier in 
Fig.~\ref{fig:Fnu-Pol-time-evol-diff-B}. The polarization is sensitive to the local spectral index which, for a Band-like spectrum, changes near the 
spectral peak and asymptotes far away from it.

Energy dependent polarization is possible also in emission mechanisms where the local polarization is independent of energy, such as Compton drag in the 
Thomson regime (where the energy-independent Eq.~(\ref{eq:Pi_CD}) holds). A featureless power-law spectrum will have no energy dependence, but the 
energy-independent polarization would still depend on the spectral power-law index, $\Pi=\Pi(\alpha)$. This occurs since different $\alpha$-values give 
different weights to different parts of the image between which the Doppler factor varies such that the same observed frequency corresponds to different 
comoving frequencies. For a non-featureless spectrum the same effect can cause energy dependence in the polarization, e.g. for a Band spectrum the relative 
weights of different parts of the image (and therefore also the polarization) will depend on the initial location of the observed frequency relative to 
the peak frequency along the LOS (i.e. on $x_0=\nu/\nu_0$).

Alternatively, if multiple spectral components from different radiation mechanisms having different levels of polarization contribute to the observed spectrum, 
the polarization of the total spectrum will change with energy. This is expected in some photospheric emission models \citep{Lundman+18} that posit that the spectral 
peak is dominated by the quasi-thermal photospheric component while the low and/or high energy wings may come from synchrotron emission (see, e.g., 
Fig.~\ref{fig:spectrum-evolution} and discussion in \S\ref{sec:hybrid-spectrum}). The right panel 
of Fig.~\ref{fig:Specto-Pol} presents such a case, where the polarization grows with decreasing energy owing to the dominance of flux by the synchrotron 
component. Near the spectral peak, the polarization vanishes. In this way, energy-resolved polarization measurements can be invaluable in understanding 
the GRB radiation mechanism.

\section{Observations}
\end{paracol}
\begin{table}
\begin{center}
  \begin{tabular}{ | l | c | c | c | c |}
    \hline
    \textbf{GRB} & \textbf{Instr./Sat.} & \textbf{Pol. ($\%$)} & \textbf{Energy (keV) } & \textbf{Remark} \\ \hline
    171010A \cite{Chand+19} & AstroSAT/CZT & $<42$ & 100-300 & Significant systematics in mod. curve \\ \hline
    170320A \cite{Kole+20} & POLAR & $18\substack{+32 \\ -18}$ & 50-500 & N.A. \\ \hline
    170305A \cite{Kole+20} & POLAR & $40\substack{+25 \\ -25}$ & 50-500 & N.A. \\ \hline
    170210A \cite{Kole+20} & POLAR & $11.4\substack{+35.7 \\ - 9.7}$ & 50-500 & N.A. \\ \hline
    170207A \cite{Kole+20} & POLAR & $5.9\substack{+9.6 \\ - 5.9}$ & 50-500 & N.A. \\ \hline
    170206A \cite{Kole+20} & POLAR & $13.5\substack{+7.4 \\ - 8.6}$ & 50-500 & N.A. \\ \hline
    170127C \cite{Kole+20} & POLAR & $9.9\substack{+19.3 \\ - 8.4}$ & 50-500 & N.A. \\ \hline
    170114A \cite{Kole+20} & POLAR & $10.1\substack{+10.5 \\ - 7.4}$ & 50-500 & PA evolution \\ \hline
    170101B \cite{Kole+20} & POLAR & $60\substack{+24 \\ - 36}$ & 50-500 & N.A. \\ \hline
    170101A \cite{Kole+20} & POLAR & $6.3\substack{+10.8 \\ - 6.3}$& 50-500 & Hint of PA evolution  \\ \hline
    161229A \cite{Kole+20} & POLAR & $17\substack{+24 \\ - 13}$& 50-500 & N.A. \\ \hline
    161218B \cite{Kole+20}  & POLAR & $13\substack{+28 \\ - 13}$& 50-500 & N.A. \\ \hline
    161218A \cite{Kole+20} & POLAR & $7.0\substack{+10.7 \\ - 7.0}$ & 50-500 & N.A. \\ \hline
    161217C \cite{Kole+20} & POLAR & $21\substack{+30 \\ - 16}$& 50-500 & N.A. \\ \hline
    161203A \cite{Kole+20} & POLAR & $16\substack{+29 \\ - 15}$& 50-500 & N.A. \\ \hline
    160910A \cite{Chattopadhyay+19} & AstroSAT/CZTI & $94\pm32$ & 100-300 & N.A. \\ \hline
    160821A \cite{Sharma+19} & AstroSAT/CZTI & $21\substack{+24 \\ -19}$ & 100-300 &time interval T0+115 to T0+155 s$^\dagger$ \\ \hline
    160821A \cite{Chattopadhyay+19} & AstroSAT/CZTI & $54\pm21$ & 100-300 & time interval T0+130 to T0+149 s \\ \hline
    160802A \cite{Chattopadhyay+19} & AstroSAT/CZTI & $85\pm 33$& 100-300  & N.A. \\ \hline
    160703A \cite{Chattopadhyay+19} & AstroSAT/CZTI & $<55$ & 100-300 & Best fitted PD $>80\%$ in contour \\ \hline
    160623A \cite{Chattopadhyay+19} & AstroSAT/CZTI & $<46$& 100-300  & N.A. \\ \hline
    160607A \cite{Chattopadhyay+19} & AstroSAT/CZTI & $<77$ & 100-300 & Best fitted PD $>60\%$ in contour \\ \hline
    160530A \cite{Lowell2017} & COSI & $<46$ & 100-1000 & N.A. \\ \hline
    160509A \cite{Chattopadhyay+19} & AstroSAT/CZTI & $<92$& 100-300  & Best fitted PD $>90\%$ in contour \\ \hline
    160325A \cite{Chattopadhyay+19} & AstroSAT/CZTI & $59\pm28$ & 100-300 & N.A. \\ \hline
    160131A \cite{Chattopadhyay+19} & AstroSAT/CZTI & $94\pm33$ & 100-300 & N.A. \\ \hline
    160106A \cite{Chattopadhyay+19} & AstroSAT/CZTI & $69\pm24$ & 100-300 & N.A. \\ \hline
    151006A \cite{Chattopadhyay+19} & AstroSAT/CZTI & $<84$ & 100-300 & Best fitted PD $>80\%$ in contour \\ \hline
    140206A \cite{Gotz+14} & IBIS/INTEGRAL & $\ge48$ & 200-400 & not calibrated on ground  \\ \hline
    110721A \cite{Yonetoku+12} & GAP/IKAROS & $84^{+16}_{-28}$& 70 – 300 & N.A.  \\ \hline
    110301A \cite{Yonetoku+12} & GAP/IKAROS & $70\pm22$& 70 – 300 & N.A.  \\ \hline
    100826A \cite{Yonetoku+11} & GAP/IKAROS & $27\pm11$& 70 – 300 & Pol. Angle evolution \\ \hline 
    061112  \cite{McGlynn2009} & SPI/INTEGRAL & $<60$ & 100-1000 &   not calibrated on ground \\ \hline
    061112  \cite{Gotz+13} & IBIS/INTEGRAL & $>60$ & 250-800 &   not calibrated on ground \\ \hline
    041219A  \cite{Gotz+2009} & IBIS/INTEGRAL & $\le4$ and $43\pm25$& 200-800 & Separated first and second peak  \\ \hline
    041219A \cite{Kalemci2007} & SPI/INTEGRAL & $99\pm33$ &100-350& potential  systematic error  \\ \hline 
    041219A \cite{McGlynn2007a} & SPI/INTEGRAL & $60\pm{35}$ &100-350& potential systematic error  \\ \hline 
    021206 \cite{Coburn-Boggs-03} & RHESSI & $80\pm20$& 150-2000 & potential systematic errors  \\ \hline
    021206 \cite{Rutledge} & RHESSI & $<100$& 150-2000 & too low signal to background  \\ \hline
    021206 \cite{Wigger} & RHESSI & $41^{+57}_{-44}$ & 150-2000 & potential systematic error  \\ \hline
    960924 \cite{BATSE2} & BATSE/CGRO & $\ge50$& 20-1000 & potential systematic errors \\ \hline
    930131 \cite{BATSE2}& BATSE/CGRO & $\ge35$ & 20-1000 & potential systematic errors \\ \hline

  \end{tabular}
\end{center}
\caption{The list of all GRBs for which a measurement has been published to date.\\ $^\dagger$For GRB 160821A several analyses were published by members of the AstroSAT collaboration. For this GRB a time resolved analysis found high levels of polarization with varying PA as well.
}
\label{tab:Measurements}
\end{table}
\begin{paracol}{2}
\switchcolumn
\noindent

\subsection{Time-Integrated Polarization Measurements}
\label{sec:time-integrated-measurements}

To date the $\gamma$-ray polarization of a total of 31 GRBs has been published. For several GRBs different analyses have been 
published, either by different groups using the same data or, in one case, using data from two different instruments.  The time and energy integrated polarization 
parameters from these measurements are shown in Table~\ref{tab:Measurements}, together with the energy range in which they were performed. It is important to note, that the energy ranges mentioned here are those stated in the respective publications but their definitions differ between experiments. The energy ranges stated by SPI for example come from an event selection based on the deposited energy, whereas for POLAR, which cannot perform measurements of the incoming photon energy directly, the stated range is based on the energy dependent effective area to polarization.

As can be seen from  Table~\ref{tab:Measurements}, especially for the earliest measurements, at the bottom of the table, the results indicate typically 
high levels of polarization, although, as explained earlier, this can in some cases be attributed to an error in the analysis. Additionally, publications 
of GRB polarization measurements have focused on those measurements for which a non-zero PD was found. At least several GRB measurements exist, such 
as some detected by GAP, for which the PD was found to be compatible with $0\%$, however, these were not published but only presented at conferences.\footnote{https://ttt.astro.su.se/groups/head/cost14/talks/Yonetoku.pdf} This causes an additional bias towards higher PD values found in the list. 

In recent years data from GAP, POLAR, and Astrosat CZTI, have significantly increased the number of measurements, however, the measured PD 
shows a large range between the different instruments. POLAR finds results which are mostly compatible with a low or unpolarized flux, whereas 
Astrosat CZTI reports high levels of polarization in \citep{Chattopadhyay+19}, with best fitting PD for 10 out of the 11 GRBs exceeding $50\%$. 
Although in numerical form only an upper limit is provided for some of these GRBs by Astrosat CZTI (which are the numbers reported in 
Table~\ref{tab:Measurements}) the contour plots for these GRBs in figure 13 of \citep{Chattopadhyay+19} indicate 
that high levels of PD are favoured for all. In most cases the best fitting PD is close to the upper limit. The only exception is 160623A where a best fitting 
PD of approximately $30\%$ is found. It should be noted though that for GRB 160821A two separate analyses provided different results for the main emission period. 
The first from \cite{Chattopadhyay+19} indicates a rather high level of polarization, whereas \cite{Sharma+19} finds a time integrated PD compatible with a lowly 
or unpolarized flux. The analysis method used for both analyses was different, while additionally the selected time intervals differed (a period with low 
fluence was added in \cite{Sharma+19}). Although the interval selection is not discussed in detail, in \cite{Chattopadhyay+21} it is mentioned that 
the intervals used in \cite{Chattopadhyay+19} are optimized to maximize the significance of the PD detection, giving a possible explanation. The same 
analysis as applied in \cite{Chattopadhyay+19} was applied in \cite{Chand+18} for GRB 171010A where an upper limit of $42\%$ is reported. 

The overall impression given by the Astrosat CZTI results is that GRBs are rather highly polarized. From the POLAR results this is not the case as 
no significant PD was detected and all results are compatible with an unpolarized flux within the $99\%$ confidence interval. The POLAR results favors low polarization degrees, with PD values exceeding $50\%$ excluded by 5 of the brightest GRBs with a $99\%$ confidence level. The results from GAP show both GRBs with a high level of polarization, 
as well as those with a low level, while COSI, the last of the 4 detectors which was well calibrated on ground, additionally excludes high values of PD.

Despite the significant increase in available measurements no clear conclusion on the PD of GRBs has emerged. 
It therefore appears that simply continuing to push for more measurements with the current generation of instruments might not be the best way forward. Rather, detailed studies scrutinizing
the different results found by different instruments are an easier and more promising way forward. One way to achieve this, which will 
be discussed later on, is the use of more standardized analyses methods as well as by making the polarization data public for an independent analysis by 
different groups.

\subsection{Time Resolved measurements}
\label{sec:time-resolved-measurements}

Time resolved analysis was performed on a range of different GRBs by different collaborations. POLAR only finds hints of an evolving PA for 2 GRBs in their 
catalog, GRB 170114A and 170101A \cite{Kole+20}. Out of the 14 GRBs studied by POLAR these are the only two with a single Fast Rising Exponential Decay (FRED) 
like structure. Such GRBs are of interest as they are typically considered to originate from a single emission region with no contamination from multiple 
overlapping pulses that complicate the analysis. For any other GRB in the POLAR catalog, all being multi-pulsed, no signs of an evolving PA were found.
It is possible that the PA varies between different overlapping pulses and integrating over different temporal segments of the emission episode results 
in an approximately fixed PA and also a lower PD due to cancellations. Alternatively, it is equally possible that the PA does not vary between pulses for many GRBs 
and the PD is intrinsically low.
 
The data of POLAR does not allow to determine the nature of the PA evolution for the 2 GRBs for which hints of it were found. The data is compatible with both random variations as well as a single $90^\circ$ change \cite{Burgess-19}.  Finer time binning or higher statistics within the time bins are required to fully resolve this. 

Both the IBIS, GAP, and AstroSAT CZTI collaborations have reported an evolution of the PA over more complex GRBs consisting either of multiple separated pulses 
(100826A, 160325A) or overlapping pulses (041219A, 160821A) \cite{Yonetoku+11,Sharma+20,Gotz+2009,Sharma+19}. For both 100826A and 160821A the evolution is 
reported to be compatible with PA changes of $90^\circ$. For 160325A, for which a high PD was found in time integrated analysis presented in \cite{Chattopadhyay+19}, the time resolved analysis finds that the first emission episode shows no or low polarization whereas the second episode shows a PD above $43\%$ with a $1.5\,\sigma$ 
confidence level. For 041219A the evolution in PA during the first emission period could explain the low PD observed with time integrated analysis.

Neither GAP or AstroSAT has reported any studies of PA evolution for GRBs with FRED-like pulses. Therefore, similar 
to the time-integrated polarization, currently existing results do not allow to draw any strong conclusions for PA evolution. 
This is due to the limited number of measurements, lower precision as well as 
the disagreement between results found by different groups.

\subsection{Energy Resolved measurements}
\label{sec:energy-resolved-measurements}

To date no energy resolved polarization measurements for GRBs have been performed. This is mainly a result of the low statistical significance found for 
the existing measurements. Dividing the data into energy bins would further reduce the available statistics and therefore not allow for constraining 
measurements to be performed. A secondary issue with such measurements is the difficulty in the analysis for many of the polarimeters. Unlike in spectrometers 
a significant number of the detected photons in a polarimeter are not fully contained in the detector. After a first Compton scattering interaction a 
second scattering interaction can follow after which the photon escapes. As a result, there is a large uncertainty on the incoming photon energy. The analysis 
is therefore not as simple as dividing the available polarization events based on the energy they deposited in the detector.  Instead, one needs to take into 
account the energy dispersion and use, for example, forward folding methods using a energy dependent polarization response. Although possible, such methods 
have not yet been applied to date. It should be noted that for certain instruments, such as COSI, Compton kinematics can be applied to ensure only fully 
contained photons are selected in the analysis. This significantly reduces the issue of energy dispersion, however, for a proper handling of the data an energy-dependent polarization response is still required for energy-dependent polarization measurements.

\section{Other Polarization Measurements}
\label{sec:other-pol}

So far we have concentrated on the polarization of the prompt GRB emission. While this is indeed the main focus of this review, here we briefly outline some of the main features and prospects of polarization measurements from other phases of GRB emission. Such polarization measurements can be very complimentary to prompt GRB polarimetry and provide vital additional constraints on the jet angular structure, our viewing angle and the magnetic field structure within the GRB outflow or in the shocked external medium. Some of the polarization measurements from these other GRB emission components are performed in the optical, NIR, sub-mm or radio bands, and are therefore technically less challenging and more reliable.
We outline the different relevant emission phases in approximate order of increasing time from the GRB onset.\\

\noindent\textbf{X-ray flares}\\

X-ray flares -- flaring and re-brightening behavior in the X-ray emission from GRBs -- were discovered by the Neil Gehrels 
\textit{Swift} space observatory \citep{Gehrels+04} and are detected in about a third of \textit{Swift} GRBs 
\citep{Burrows+05,Falcone+07,Chincarini+10,Margutti+11a,Margutti+11b,Yi+16}. They typically display a characteristic shape with a sharp 
rise in flux followed by a smoother decay, eventually fading back to the pre-flare flux level, and also show a different spectrum (typically 
harder) compared to the underlying emission. X-ray flares typically occur at times $10^2\,{\rm s}\lesssim t\lesssim10^5\,$s after the GRB 
onset. Their temporal and spectral properties appear to be a smooth continuation of the prompt GRB emission spikes 
\citep{Krimm+07,Margutti+11a,Margutti+11b}. While during the prompt GRB emission the typical width or spectrum of the different spikes 
typically does not show a clear systematic evolution, the X-ray flares gradually become wider (with FWHM $\Delta t$ satisfying 
$\Delta t/t\sim0.1-0.3$), less luminous ($\mean{L}\propto t^{-2.7\pm0.1}$), and softer with time $t$. Their overall properties strongly suggest 
that X-ray flares have a common origin with the prompt GRB emission and likely share similar dissipation and/or emission mechanisms.

Therefore, studying the polarization properties of X-ray flares may provide new insights both for their origin, as well as on the  emission 
and/or dissipation mechanisms that are common with the prompt emission. There are some theoretical predictions for their polarization properties 
\citep[e.g.][]{Fan+05,Geng+18} but there is still much room for more detailed and realistic predictions that could be tested against future 
observations. Their observed similarities to prompt GRB pulses suggests that many of the models for prompt GRB polarization may be generalized 
to apply also for X-ray flares. The fact that X-ray flares last up to hours or sometimes even days after the GRB onset allows pointed observations 
by sensitive instruments, while their softer spectrum makes them prime targets for future pointed X-ray polarimeters such as eXTP with a 
Polarimetry Focusing Array at 2$-$10\,keV energies \citep[e.g.][]{Zhang+19a,in't_Zand+19}.\\

\end{paracol}
\begin{figure}
    \widefigure
    \centering
    \raisebox{6pt}{\includegraphics[width=0.511\textwidth]{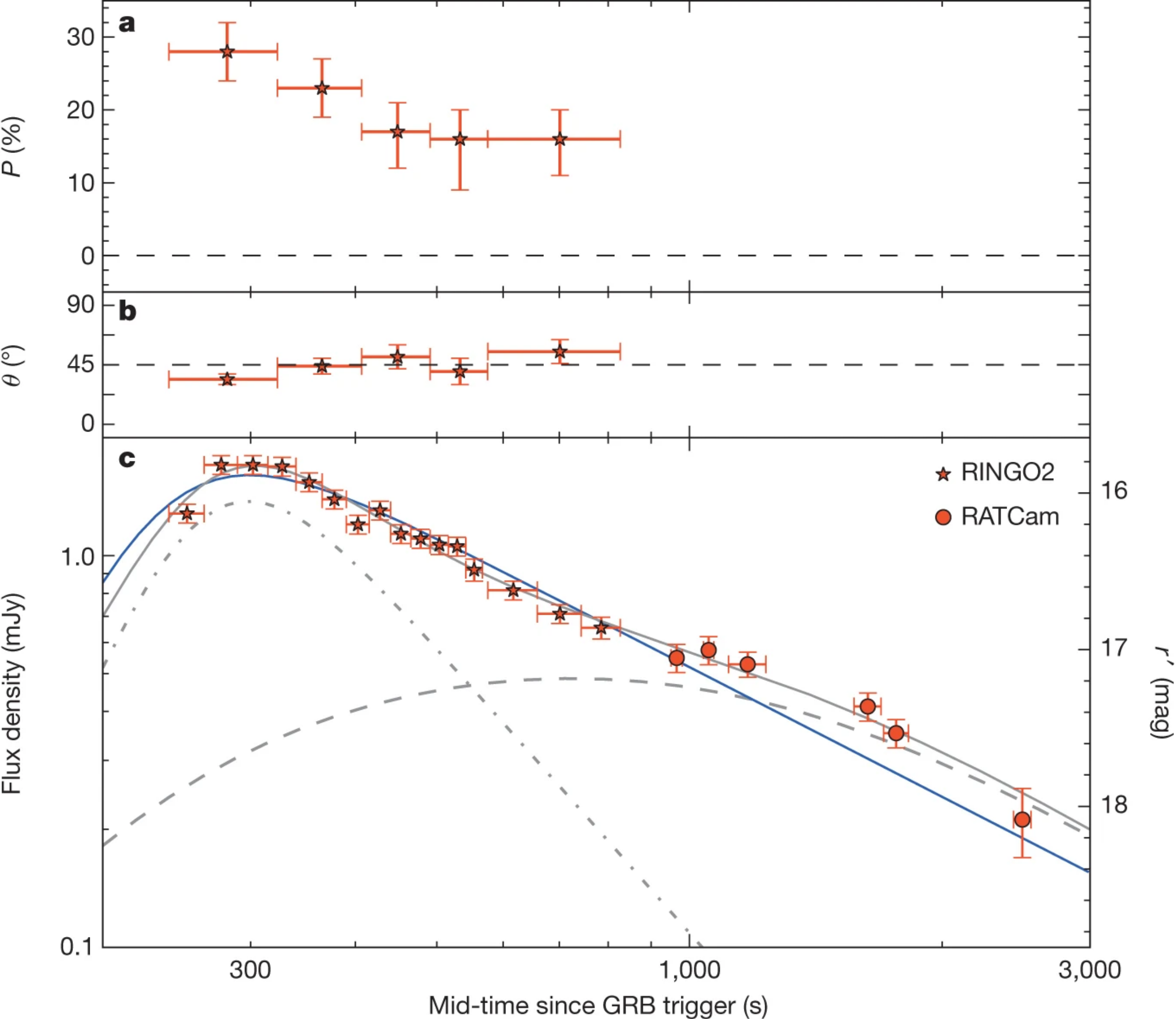}}
    \hfill
    \includegraphics[width=0.35\textwidth]{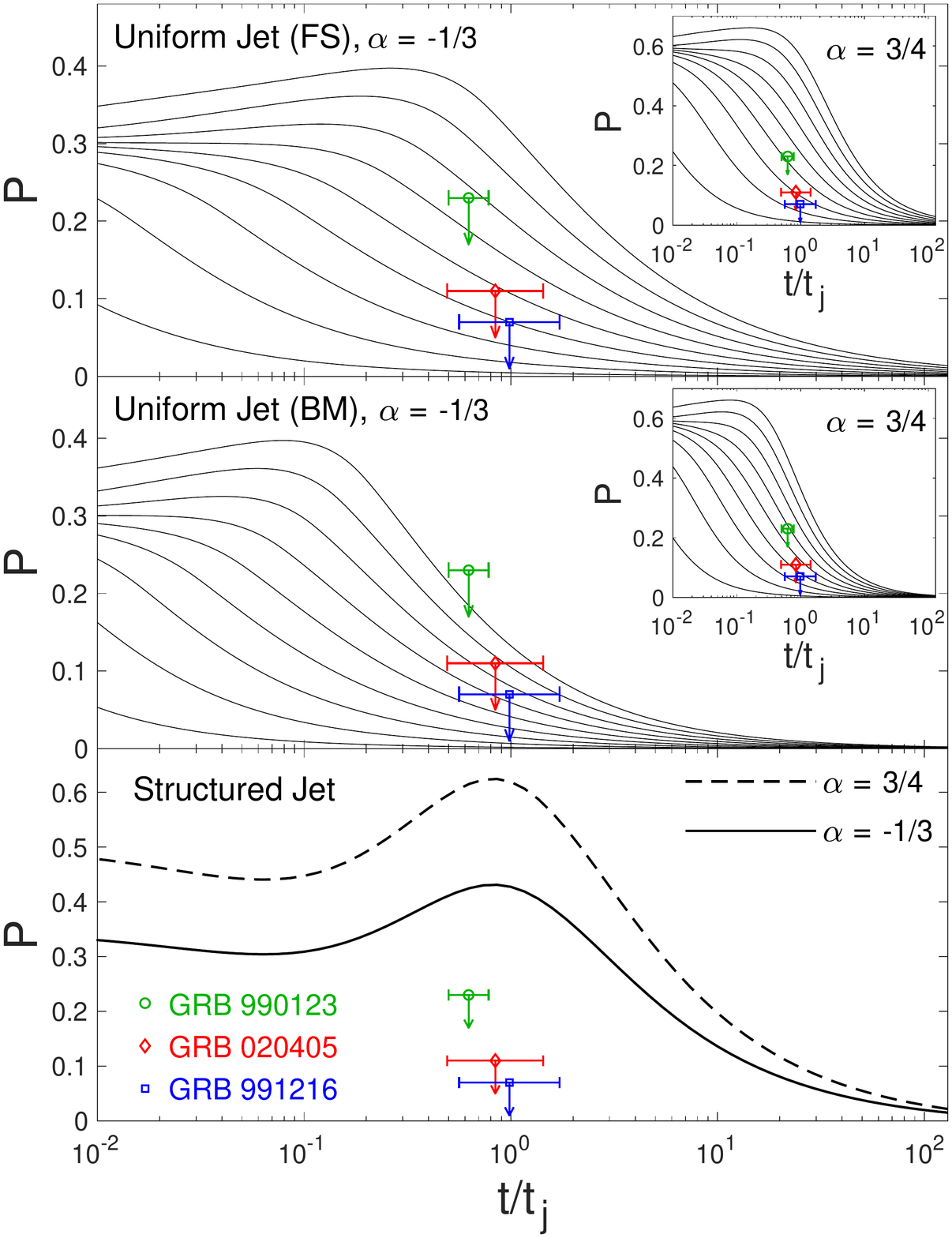}
    \caption{\textbf{\textit{Left}}:
    Evolution of optical polarization [degree P (\textbf{a}), and position angle $\theta$ (\textbf{b}; degrees east of north)] and 
    brightness [(\textbf{c}) in red (555--690\,nm) light] using RINGO2 and RATCam, in GRB 120308A \citep[from][]{Mundell+13}.   
    \textbf{\textit{Right}}: 3$\,\sigma$ upper limits on the linear polarization of the radio flare emission from three different 
    GRBs overlaid on the theoretical polarization light curves for a toroidal magnetic field in the GRB ejecta \citep[from][]{Granot-Taylor-05}. 
    The top two panels are for a uniform (top-hat) jet where the different lines, from top to bottom, are for $\theta_{\rm obs}/\theta_j= 0.9,0.8,...,0.1$, 
 while $\alpha=-d\log F_\nu/d\log\nu$ is the spectral index (in the observed radio band) and $\Pi_{\rm max}=(\alpha+1)/(\alpha+5/3)$. In the top panel 
 the Lorentz factor of the ejecta is assumed to remain equal to that of the freshly shocked fluid just behind the forward shock (``FS''), while in the 
 middle panel it is assumed to follow the \citet{BM76} self-similar solution. The bottom panel is for a ‘‘structured’’ jet, in which the energy per solid 
 angle drops as $\theta^{-2}$ outside some small core angle.
}
\label{fig:Pol-opical_flash-radio_flare}
\end{figure}
\begin{paracol}{2}
\switchcolumn

\noindent\textbf{Reverse shock emission}\\

As the GRB outflow sweeps up enough external medium, it is decelerated by a reverse shock\footnote{If the GRB ejecta are still highly magnetized at the deceleration radius $R_{\rm dec}$, $\sigma(R_{\rm dec})\gtrsim1$, this may suppress the reverse shock, making it weak or even completely nonexistent.} while a strong relativistic forward shock propagates into the external medium powering the long-lived afterglow emission. Most of the outflow's energy is transferred to the shocked external medium when the reverse shock finishes crossing the ejecta shell at the deceleration radius, $R_{\rm dec}$, corresponding to the deceleration (apparent) time, $t_{\rm dec}$, which therefore signals the peak or onset of the afterglow emission \citep[e.g.][]{Sari-Piran-95,Sari-Piran-99,Kobayashi-Sari-00,Kobayashi-Zhang-03,Nakar-Piran-04,Granot-12a}. For the "thick shell" case where the reverse shock is at least mildly relativistic this time is comparable to the prompt GRB duration, $t_{\rm dec}\sim t_{\rm GRB}$, while for the "thin shell" case (where the reverse shock gradually transitions from Newtonian to mildly relativistic) $t_{\rm dec}>t_{\rm GRB}$. 
For frequencies that are above the cooling break frequency $\nu_c$ of the reverse shock emission at $t_{\rm dec}$, which may include the optical for a 
sufficiently large $n_{\rm ext}(R_{\rm dec})$ (e.g. as expected for the stellar wind of a massive star progenitor in long GRBs), once the reverse shock 
finishes crossing the ejecta shell the emission from the LOS sharply drops and the flux decays rapidly ($\sim t^{-3}$) corresponding to 
high-latitude emission. Otherwise, for frequencies in the range $\max(\nu_a,\nu_m)<\nu<\nu_c$, where $\nu_a$ is the break frequency corresponding to 
synchrotron self-absorption, a slightly less steep flux decay of about $t^{-2}$ is expected, as the emission is dominated by the material along the line 
of sight where the shocked electrons cool adiabatically. Therefore, the optical 
emission typically peaks on a timescale of tens of seconds and then sharply drops -- the so called \textit{\textbf{optical flash}}
\citep[e.g.][]{Akerlof+99,Sari-Piran-99a,Fox+03,Racusin+08,Uehara+12,Vestrand+14}. The radio, however, is typically below the self-absorption frequency $\nu_a$ at 
$t_{\rm dec}$ (while $\nu_m<\nu_a$), and its flux keeps rising until $\nu_a$ sweeps past the radio band, roughly after a day or so -- the so called 
\textit{\textbf{radio flare}} \citep[e.g.][]{Kulkarni+99,Frail+00,Berger+03,Laskar+13,Perley+14,Laskar+19}.

In terms of the polarization properties of the reverse shock emission, it is important to keep in mind the following points:
\begin{enumerate}
\item The reverse shock emission comes from the shocked ejecta, and therefore provides important information about the magnetic field structure 
within the GRB outflow.
\item In contrast with the prompt GRB emission where the dominant emission mechanism is uncertain, in the reverse shock radio to optical emission is almost certainly synchrotron radiation (given its large emission radius and broadband SED). 
\item Measuring polarization in the optical or radio is generally more reliable than in gamma-ray or X-ray energies, mainly because it is technically less challenging (despite the rapid response robotic telescopes needed for the optical flash). 
\item As the ejecta decelerates by sweeping up the external medium, the lower bulk Lorentz factor $\Gamma$ implies a larger visible region of angle $\sim1/\Gamma$ around our LOS, in which the structure of the jet and of the magnetic field in the ejecta can be probed.
\end{enumerate}
The {\bf optical flash} emission typically peaks on a timescale of $\sim10-100\;$s and the ejecta Lorentz factor $\Gamma$ is only somewhat lower than 
during the prompt GRB emission\footnote{The ejecta are decelerated by the reverse shock, typically reducing $\Gamma$ down to $\sim\frac{1}{2}\Gamma_\infty$, where $\Gamma_\infty$ is its value during the coasting phase (it can be lower than this for a highly relativistic reverse shock). However, the prompt GRB emission in photospheric models can arise from $\Gamma<\Gamma_\infty$, at which point the outflow is still accelerating and hasn't yet reached $\Gamma_\infty$.} with $\Gamma\sim10^2-10^{2.5}$. Therefore, it is expected to probe a comparable (i.e. only somewhat larger) region of angle $\sim1/\Gamma\sim10^{-2.5}-10^{-2}\,$rad 
around our line of sight. Nonetheless, optical polarization measurements are more reliable than in gamma rays, and the optical flash is almost 
certainly synchrotron, which enables a cleaner and more robust inference of the ejecta magnetic field structure within this region.

From the observational perspective, since the optical flash usually has significant temporal overlap with the early optical afterglow emission 
from the shocked external medium, this requires a detailed modeling of both the total flux and the polarized flux as a function of time from these 
two distinct emission regions in order to properly disentangle between them and derive stronger and more robust constraints on the underlying 
properties of the GRB ejecta and its magnetic field structure. Most (but not all, e.g. \citep{Troja+17a}) of the early optical polarimetric 
observations relevant for the optical flash were done by the RINGO polarimeters on the Liverpool telescope \citep{Steele+06,Mundell+07,Steele+09,Steele+10,Arnold+12,Mundell+13,Kopac+15,Steele+17,Jordana-Mitjans+20}. Combining photometric and polarimetric 
observations \citep{Kobayashi-17} they conclude that their data clearly indicates that all epochs in which significant (linear) polarization was 
measured were dominated by emission from the reverse shock (while the optical afterglow emission from the forward external shock was sub-dominant). 
Here are a few examples. In GRBs 101112A and 110205A \citep{Steele+17} a polarization of $\Pi=6^{+3}_{-2}\,\%$ and $13^{+13}_{-9}\,\%$, respectively, were 
measured at the optical peak time of $T_{\rm dec}\sim299\;$s and $\sim1027\;$s, respectively, which appear to be dominated by the reverse shock 
because of the sharp rise to the peak (as $\sim t^{4.2}$ and $\sim t^{4.6}$, respectively). In both GRBs $T_{\rm dec}\gg T_{\rm GRB}$, 
indicating a thin shell (with $T_{\rm GRB}\approx T_{90}\sim9.2\;$s and $249\;$s, respectively). One of the best examples so far is GRB 120308A 
\citep{Mundell+13}, in which $\Pi=28\%\pm4\%$ was detected at $240\,{\rm s}<t<323\,{\rm s}$, which gradually decreased down to $\Pi=16^{+4}_{-5}\%$ 
at $575\,{\rm s}<t<827\,{\rm s}$, as the emission gradually transitioned from reverse-shock to forward-sock dominated (see \textit{left} panel of 
Fig.~\ref{fig:Pol-opical_flash-radio_flare}).

The {\bf radio flare} emission \citep[e.g.][]{Kulkarni+99,Frail+00,Berger+03,Laskar+13,Perley+14,Laskar+19} typically peaks on a timescale of a day 
or so ($\sim10^5\;$s) with\footnote{By this time the shocked GRB ejecta shell settles in the back of the \citet{BM76} self-similar solution, and its 
$\Gamma$ is smaller by a factor of up to $\sim1.5-1.8$ compared to the material just behind the forward shock that dominates the afterglow emission 
at the same observed time \citep{Kobayashi-Sari-00,Granot-Taylor-05}.} $\Gamma\sim5-10$. This corresponds to a visible region of angle 
$\sim0.1-0.2\,$rad around our line of sight, which is significantly larger than during the optical flash. Moreover, it often includes the entire jet 
(for a simple top-hat jet model) as suggested by the fact that the radio flare peak time is often comparable to the jet break time in the afterglow 
lightcurve. \citet{Granot-Taylor-05} have used VLA data of radio flares from three GRBs (990123, 991216, 020405) to constrain its polarization, finding 
only upper limits for both linear and circular polarization. Their best limits are for GRB 991216, for which they find $3\sigma$ upper limits on the 
linear and circular polarization of 7\% and 9\%, respectively. These limits provide interesting constraints on GRB models and in particular are hard to 
reconcile with a predominantly ordered toroidal magnetic field in the GRB outflow together with a ‘‘structured’’ jet, where the energy per solid angle 
drops as the inverse square of the angle from the jet axis (see \textit{right} panel of Fig.~\ref{fig:Pol-opical_flash-radio_flare}). The polarization 
of the radio flare may be affected by the location of the observed frequency $\nu$ relative to the synchrotron self-absorption break frequency $\nu_a$ (polarization is suppressed when $\nu < \nu_a$, during the rising phase of the radio flare) or by Faraday depolarization on the way from the source to us (both are discussed in \citep{Granot-Taylor-05}) and may also be subject to plasma propagation effects within the source (as discussed below, at the end of this section).

Comparing the polarization of the optical flash and radio flare for the same GRB would enable us to study the magnetic field in the GRB ejecta over a wide 
range of angular scales, probing magnetic structures with a coherence length over this angular range, $10^{-2.5}\lesssim\theta_B\lesssim10^{-1}$. 
Measuring the reverse shock emission polarization at intermediate times and frequencies, such as at sub-mm with ALMA \citep[e.g.][]{Laskar+13,Laskar+19}, 
would provide a better coverage of this wide range. A particularly interesting example is GRB 190114C, which was also detected at TeV energies 
\citep{MAGIC_GRB190114C}. ALMA measured its sub-mm ($97.5\,$GHz) total intensity and linear polarization at 2.2\,$-$\,5.2$\,\,$hr after the burst, when 
the emission was dominated by the reverse shock \citep{Laskar+19}, detecting\footnote{This was the first detection and measurement of the temporal evolution 
of polarized radio/millimeter emission in a GRB.} linear polarization at $\approx5\sigma$ confidence, decreasing from $\Pi=0.87\%\pm0.13\%$ to 
$\Pi=0.60\%\pm0.19\%$ while the position angle evolved from $10^\circ\pm5^\circ$ to $-44^\circ\pm12^\circ$. Using the measured linear polarization 
\citet{Laskar+19} constrained the coherence scale of tangled magnetic fields in the ejecta to an angular size of $\theta_B\approx10^{-3}\;$rad, while the 
rotation of the polarization angle rules out the presence of large-scale, ordered axisymmetric magnetic fields, and in particular a large-scale toroidal 
field, in the jet. \\

\noindent\textbf{Afterglow emission} \\

Linear polarization at the level of a few percent has been detected in the optical or NIR afterglow of 
about a dozen GRBs \citep{Covino+99,Wijers+99,Rol+00,Bjornsson+02,Masetti+03,Covino+03,Barth+03,Rol+03,Lazzati+03,Gorosabel+04,Maiorano+06,Covino-Gotz-16}.
Higher levels of polarization ($10\%\lesssim\Pi\lesssim30\%$) have been measured mostly in the very early afterglow, likely being dominated by reverse 
shock emission, as discussed above (see, however, \citep{Bersier+03}). The linear polarization of the afterglow emission was considered as a confirmation 
that it arises primarily from synchrotron radiation, as was already suggested by its spectral energy distribution. 

A variety of \textbf{models} have been suggested for GRB afterglow polarization: emission from different patches of uniform but mutually uncorrelated 
magnetic field, either with microlensing \citep{Loeb-Perna-98} or without it \citep{Gruzinov-Waxman-99}, or emission from a random magnetic field within 
the plane of the afterglow shock together with scintillation in the radio \citep{Medvedev-Loeb-99} or with a jet viewed not along its symmetry axis 
\citep{Ghisellini-Lazzati-99,Sari-99,Rossi+04}, possibly with the addition of an ordered component that pre-exists in the external medium and is 
compressed by the afterglow shock and/or a tangled magnetic field that is not purely in the plane of the shock and may even be predominantly in the direction 
of the shock normal \citep{Granot-Konigl-03,Teboul-Shaviv-21} or due to clumps in the external medium or a similarly inhomogeneous outflow \citep{Granot-Konigl-03,Nakar-Oren-04}. 

\end{paracol}
\begin{figure}
    \widefigure
    \centering
    \raisebox{-160pt}{
    \includegraphics[width=0.4\textwidth]{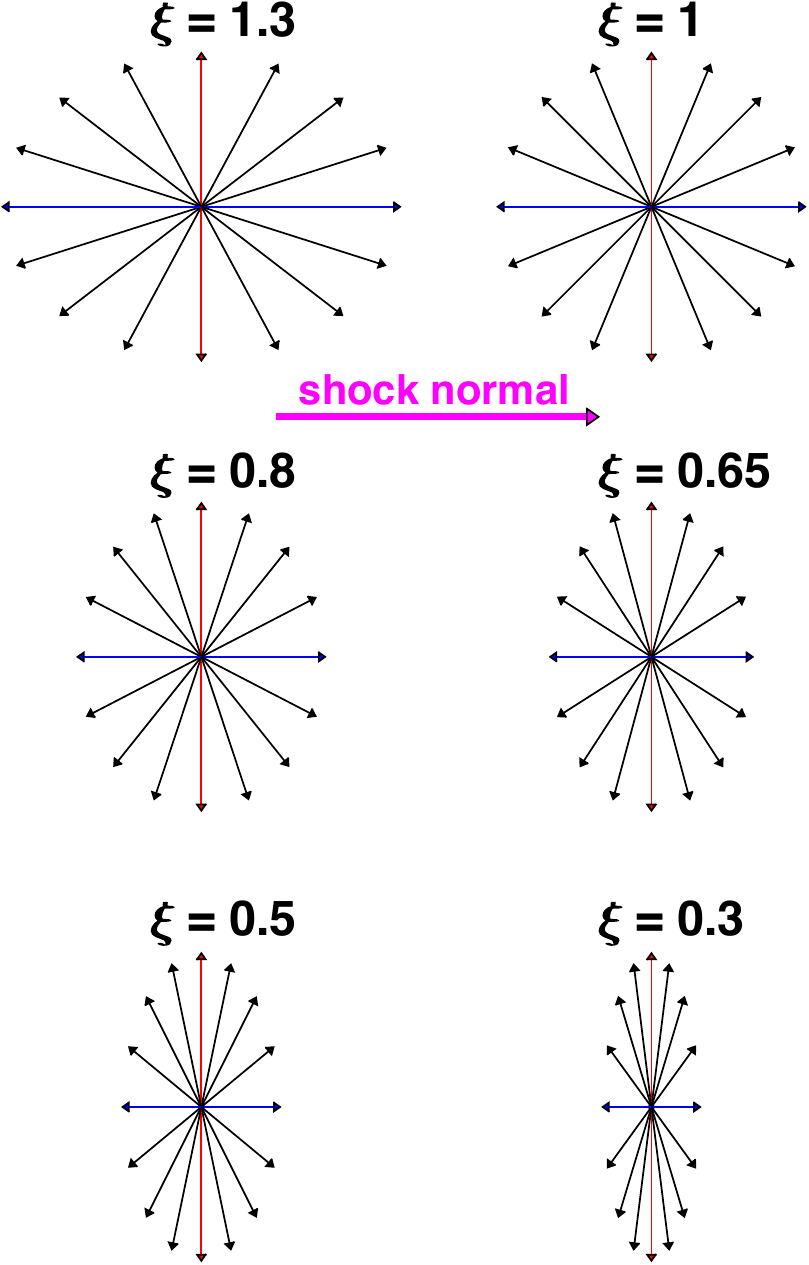}
    }
    \hspace{1cm}
    \raisebox{0pt}{
    \begin{minipage}[r]{0.52\textwidth}
\hspace{0.39cm}\includegraphics[width=0.782\textwidth]{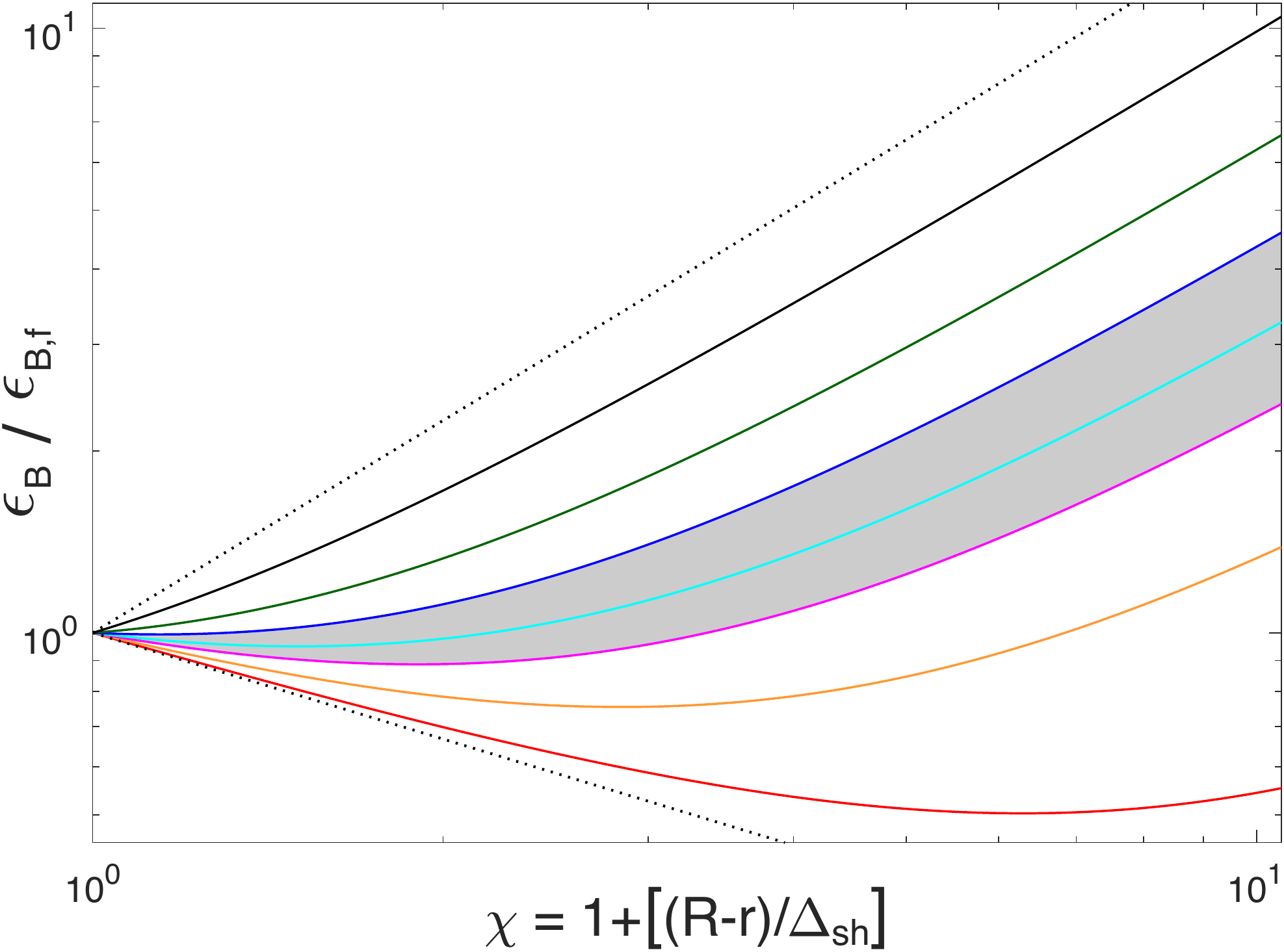}
\\ \vspace{-0.07cm}\\
\includegraphics[width=0.84\textwidth]{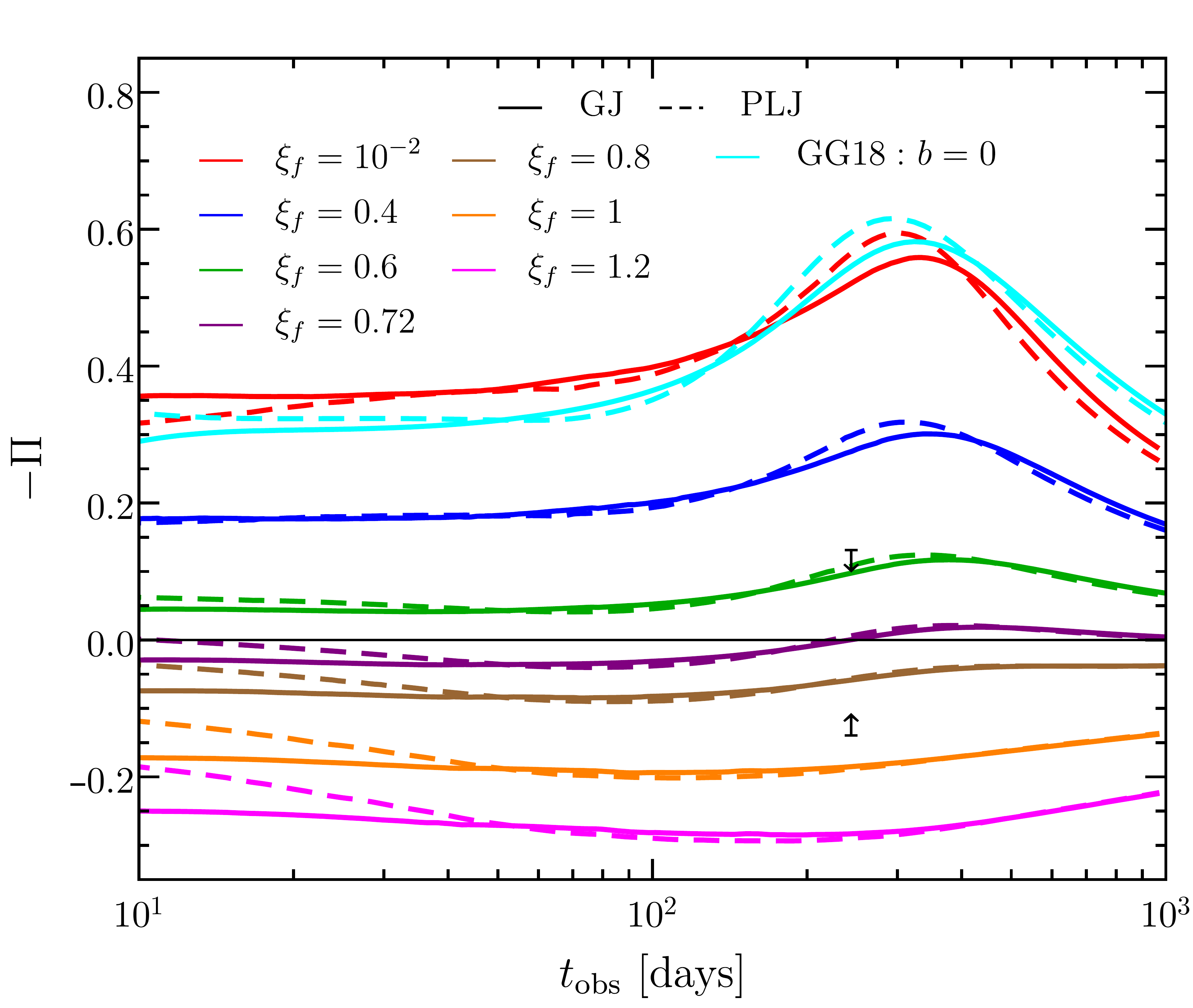}
\end{minipage}}
\caption{Constraining the magnetic field structure in collisionless relativistic shocks from a radio afterglow linear polarization upper 
limit in GRB\,170817\,/\,GW\,170817 \citep{Gill-Granot-20}. \textbf{\textit{Left}}: Schematic of post-shock magnetic field geometry for 
different values of the local anisotropy parameter $\xi\equiv B_\parallel/B_\perp=\xi_f\chi^{(7-2k)/(8-2k)}$, whose initial value just behind 
the shock is $\xi_f$, for an external density profile $\rho_{\rm ext}\propto R^{-k}$, where $\chi=1+2(4-k)\Gamma_{\rm sh}^2(1-r/R)$ is the 
\citet{BM76} self-similar variable, $r$ is the radial coordinate while $R$ and $\Gamma_{\rm sh}$ are the local radius and Lorentz factor of 
the afterglow shock front. 
\textbf{\textit{Top Right}}: The corresponding evolution of the magnetic field equipartition parameter, $\epsilon_B$, with the distance 
behind the shock (as parameterized through $\chi$) for $\xi_f= 0, 0.2, 0.4, 0.57, 0.7, 0.89, 1.2, 2, \infty$ (from bottom to top). The two extreme values of $\xi_f = 0, \infty$ are shown as dotted (straight) lines. The light-grey shaded region corresponds to the allowed range found in \citep{Gill-Granot-20}, 0$.57 \lesssim\xi_f\lesssim0.89$.
\textbf{\textit{Bottom Right}}: 
The linear polarization evolution, $\Pi(t)$, obtained from a volume integration of the flow, shown for different values of $\xi_f$. The two 
arrows mark the polarization upper limit, $|\Pi|<12\%$. Comparison is made between two jet structures -- a Gaussian jet (GJ) and a power-law 
jet (PLJ). The result from \citep{Gill-Granot-18b}, which assumed an infinitely thin shell geometry as well as locally isotropic synchrotron 
spectral emissivity, is also shown (labeled GG18) for the magnetic field anisotropy parameter $b=0$. 
}
\label{fig:GG20}
\end{figure}
\begin{paracol}{2}
\switchcolumn

The most popular models for GRB afterglow polarization feature an axis-symmetric jet viewed not along its symmetry axis along with a tangled 
shock-produced magnetic field that is symmetric about the local shock normal  \citep{Ghisellini-Lazzati-99,Gruzinov-99,Sari-99,Granot+02,Granot-Konigl-03,Rossi+04,Gill-Granot-18a,Gill-Granot-20,Shimoda-Toma-21,Birenbaum-Bromberg-21}. 
In such models the only preferred direction on the plane of the 
sky is that connecting the jet symmetry axis and our LOS, and therefore the net polarization of the unresolved image must lie either 
along this direction or transverse to it. Indeed, the tell-tale signature of such models for a uniform top-hat jet is a $90^\circ$ change in 
the polarization PA $\theta_p$ as $\Pi$ vanishes and reappears rotated by $90^\circ$, around the time of the jet break in the afterglow 
lightcurve \citep{Ghisellini-Lazzati-99,Sari-99}. On the other hand, for a structured jet viewed from outside of its narrow core a constant 
$\theta_p$ is expected. Overall, in such models the linear polarization and its temporal evolution depend on: \textbf{(i)} the 
\textbf{jet's angular structure}, \textbf{(ii)} the local \textbf{structure of the shock-generated magnetic field} about the shock normal, and 
\textbf{(iii)} our viewing angle $\theta_{\rm obs}$ from the jet symmetry axis. Therefore, afterglow linear polarization observations can teach us 
both about the jet's angular structure and about the shock-produced magnetic field structure. However, there is a significant \textbf{degeneracy} 
between the two, which usually requires making large assumptions about one of them in order to significantly constrain the other.

The exceptional case of the short GRB 170817A, which was associated with the first gravitational wave detection of the binary neutron star merger, 
GW 170817, has allowed us to break this degeneracy. This event was observed from a large off-axis viewing angle and its low-luminosity prompt 
gamma-ray emission and  subsequent long-lived afterglow emission could be observed thanks to its relatively small distance ($D\approx40\;$Mpc). 
The combination of an extremely well monitored afterglow from radio to X-rays \citep[e.g.,][]{Abbott+17-GW170817A-MMO,Hallinan+17,Troja+17b}, and 
the super-luminal motion of its radio flux centroid ($\mean{\beta_{\rm app}}=\mean{v_{\rm app}}/c=4.1\pm0.5$ between 75 and 230 days after the burst; 
\citep{Mooley+18}) has allowed a good determination of our viewing angle and of the jet's angular structure 
\citep[e.g.,][]{DAvanzo+18,Gill-Granot-18b,Granot+18a,Lamb-Kobayashi-18,Lazzati+18,Margutti+18,Gill+19,Resmi+18,Troja+18,Beniamini+20,Nathanail+20,Nathanail+21}. 
This has enabled making robust predictions for the linear polarization that depend on the shock-produced magnetic field structure \citep{Gill-Granot-18b}. 
Shortly thereafter a linear polarization upper limit, $|\Pi|<12\%$ (99\% confidence), was set in the radio ($2.8\;$GHz) at $t=244\;$days \citep{Corsi+18}. 
Assuming emission from a two-dimensional surface identified with the afterglow shock front, this has led to a constraint of $0.7\lesssim b\lesssim1.5$ 
on the magnetic field anisotropy parameter, $b\equiv2\mean{B^2_\parallel}/\mean{B^2_\perp}$ \citep{Gill-Granot-18b,Corsi+18}, which was introduced by 
\citep{Granot-Konigl-03}, where $B_\parallel$ and $B_\perp$ are the magnetic field compenents parallel and perpendicular to the shock normal direction 
$\hat{n}_{\rm sh}$, respectively, and $b=1$ corresponds to an isotropic field in 3D (for which the local and global polarizations vanish). A more detailed 
analysis \citep{Gill-Granot-20} accounted for the emission from the whole 3D volume behind the afterglow shock, with the global angular jet structure 
implied by the GRB 170817A/GW 170817 observation and a local radial hydrodynamic profile set by the \citet{BM76} self-similar solution. The magnetic field 
was modeled  as an isotropic field in 3D that is stretched along $\hat{n}_{\rm sh}$ by a factor\footnote{Defining for convenience a local coordinate 
system where $\hat{n}_{\rm sh}=\hat{z}$, in the above definition of $b$ we have $\mean{B^2_\parallel}=\mean{B^2_z}$ and 
$\mean{B^2_\perp}=\mean{B^2_x+B^2_y}=2\mean{B^2_x}$ due to the $B$-field's symmetry about $\hat{n}_{\rm sh}$. Here in the definition of $\xi$, however, 
$B_\perp$ represents either $B_x$ or $B_y$, but not $(B^2_x+B^2_y)^{1/2}$ (while $B_\parallel=B_z$).} $\xi\equiv B_\parallel/B_\perp$, whose initial value 
$\xi_f=B_{\parallel,f}/B_{\perp,f}$ describes the field that survives downstream on plasma scales $\ll R/\Gamma_{\rm sh}$, and it is evolved downstream 
according to the \citep{BM76} solution assuming flux freezing (i.e. no further magnetic dissipation or amplification far downstream of the shock front). 
\citet{Gill-Granot-20} find that the shock-produced magnetic field has a finite, but initially sub-dominant, parallel component: $0.57\lesssim\xi_f\lesssim0.89$
(see Fig.~\ref{fig:GG20}).

\textbf{Circular polarization} at the level of $\Pi_{\rm circ}=0.61\%\pm0.13\%$ has been reported in the optical afterglow of GRB 121024A \citep{Wiersema+14} 
at $t=0.15\;$days after the burst, when the linear polarization was $\Pi_{\rm lin}\approx4\%$ implying a relatively high circular to linear polarization ratio 
of $\Pi_{\rm circ}/\Pi_{\rm lin}\approx0.15$. \citet{Nava+16} performed a detailed analysis of the expected $\Pi_{\rm circ}$ and $\Pi_{\rm lin}$ in GRB 
afterglows, finding that while ad-hoc configurations may allow large local $\Pi_{\rm circ}$ values, after transformations to the observer frame and integration 
over the whole visible region are performed, $\Pi_{\rm circ}/\Pi_{\rm lin}$ remains vanishingly small in any realistic optically thin synchrotron afterglow 
emission model and thus concluding that the origin of the observed $\Pi_{\rm circ}$ in GRB 121024A cannot be intrinsic.

\textbf{Plasma propagation effects} due to the presence of cooler thermal electrons, that are not shock accelerated and represent a fraction $1-\xi_e$ of the total 
number, may be important if a significant ordered magnetic field component is present in the emitting region \citep{Matsumiya-Ioka-03,Sagiv+04,Toma+08}. Such effects 
are most prominent in the early afterglow and around the self-absorption frequency, and may 
therefore potentially affect the reverse shock emission (the ``optical flash'' or ``radio flare''), as well as the forward shock emission in the radio 
up to a day or so \citep{Matsumiya-Ioka-03,Sagiv+04,Toma+08}. These effects may include Faraday conversion of the linear polarization of the emitted 
radiation to circular polarization or Faraday depolarization of the emitted linear polarization. For typical GRB afterglow microphysical parameters, the 
latter effect may strongly suppress the linear polarization in the radio but preserve that in the optical. Therefore, simultaneous observations yielding 
statistically significant measurements of polarization in both optical and radio can be extremely useful to confirm the population of thermal electrons as well 
as the existence of an ordered B-field. In some GRBs, this effect may manifest in the sub-mm band where comparison between ALMA and VLA measurements can 
constrain the value of $\xi_e$ \citep{Toma+08}. In fact, \citet{Urata+19} argued that the unusually low afterglow polarization ($\Pi=0.27\%\pm0.04\%)$ of 
GRB 171205A in the sub-mm band, as compared to the typical late-time optical polarization, may have been the result of Faraday depolarization. Since the 
true afterglow shock kinetic energy is given by $E'=E/\xi_e$ \citep{Eichler-Waxman-05}, where $E$ would be the true energy for $\xi_e=1$, a constraint on 
$\xi_e$ would lead to better constraints on the burst energetics.


\section{Outlook for 2030} 
\label{sec:outlook}
The handful of successful $\gamma$-ray polarimeters has shown over the previous decade that although challenging, GRB polarization measurements are possible. 
With this new success a range of new instruments with not only a higher sensitivity but also a wider energy range are foreseen to be launched over 
the coming decade. As mentioned earlier, however, simply increasing the number of measurements does not improve our understanding if different instruments provide 
incompatible results. Below we will first discuss the promising advances in detector development for the coming decade. This will be followed by a discussion 
on the need for improvements and standardization of the analysis.

\subsection{Future Instruments}
\label{sec:future-instruments}

\subsubsection{POLAR-2 and LEAP}
In the Compton energy range of $\sim\,$10$\,$--$\,$1000$\;$keV four instruments are proposed. 
Both the LEAP \cite{MCConnell+20} and SPHiNX \cite{Pearce+18} instruments have been proposed for launches in the coming decade, 
while the POLAR-2 project has already been accepted for launch in 2024 \cite{Kole+20b}. Additionally, the Daksha mission, a larger scale full sky monitor follow-up mission based on the Astrosat CZTI is proposed to be launched in the coming decade as well \cite{Chattopadhyay+21}. Out of these four, the POLAR-2 and LEAP projects 
aim to make the next step in this field by producing instruments with an effective area an order of magnitude larger than the 
POLAR instrument. The SPHiNX project instead has an effective area similar to that of POLAR and will therefore have to make gains 
over currently existing measurements by aiming for a longer mission life time. For Daksha the effective area is planned to be an order of magnitude larger than that of Astrosat CZTI. As the experiment will consist of 2 satellites, each observing half the sky, this increase in effective area is evenly distributed over the full sky. The design allows for a significant increase in the number of GRBs for which polarization measurements are possible, while also increasing the precision of each such measurement, although not by one full order of magnitude.

The POLAR-2 instrument is similar in design to POLAR with, apart from several minor design improvements, a focus on an improvement 
in 3 parts. The first is the size, which is 4 times larger than POLAR, resulting in a total geometrical area of approximately $2500\,$cm$^2$. 
Secondly, the scintillator readout technology is improved to decrease the low-energy threshold of the instrument 
from $50\,$keV to $20\,$keV, giving a total energy range of 20$\,$--$\,$800$\;$keV for polarization measurements. Finally, POLAR-2 will be 
equipped with spectrometers making it independent of other instruments for spectral and location parameters of GRBs, which reduces the 
systematic error on many GRB measurements. The instrument is approved for launch in early 2024 towards the Chinese Space Station (CSS).

The LEAP instrument is similar to POLAR-2 both in size and in the detection mechanism that uses plastic scintillators. Contrary to POLAR-2, 
the LEAP instrument will also use high Z scintillators, which increase the absorption cross section. Therefore, the instrument will have a 
larger sensitivity to polarization and better spectral response, but a reduction in its effective area and field of view. Whereas the total effective area for LEAP, that useful for spectrometry, is $\sim3500\,\mathrm{cm}^2$ at $250\,\mathrm{keV}$, for polarization it is around $\sim1000\,\mathrm{cm}^2$ \cite{LEAP}. For POLAR-2 the effective area of the polarimeter usable for spectrometry is $\sim2000\,\mathrm{cm}^2$, and therefore significantly smaller than LEAP. For polarization it is however $\sim1400\,\mathrm{cm}^2$ and therefore larger than LEAP. The reduction in effective area of the polarimeter for spectrometry in POLAR-2 is compensated by separate spectrometers which will increase this by at least $50\%$.

The two instruments therefore have different strengths. With a proposed launch in 2025 for LEAP
towards the International Space Station (ISS), the combination of both of these instruments in orbit would allow for detailed polarization measurements 
of the majority of GRBs with fluences (as measured in the 10$\,$--$\,$1000$\;$keV energy range) above $10^{-6}\,{\rm erg\,cm}^{-2}$.

\subsubsection{Low Energy Polarimeters}
Apart from adding significant sensitivity in the energy range of $\sim10\,$--$\,1000\;$keV, missions are also 
proposed to perform the first GRB polarization measurements at keV energies.

As previously mentioned the first polarization measurements at these energies, albeit of point sources, were recently performed by the small 
scale PolarLite mission \cite{Feng2020}. The IXPE mission \cite{IXPE}, which uses a similar measurement technology as PolarLite, is planned 
to be launched in 2021. However, as it is optimized for point sources it has a narrow FoV. This in combination with a long slewing time makes 
it unlikely to measure any GRBs. The larger scale eXTP mission, however, will still be optimized for point sources, but is designed to also 
observe targets of opportunity such as GRBs using a shorter slewing time. As such eXTP will be capable of measuring the polarization of the 
afterglow of GRBs in the 2$\,$--$\,$10$\;$keV energy range as well as any potential X-ray flares occurring in the afterglow \cite{eXTP}. eXTP 
is a joint Chinese European mission and is currently foreseen to be launched in 2028.

The above mentioned instruments are optimized for point sources and therefore have a small field of view to optimize the signal to noise.  
In order to measure the polarization of the prompt emission from GRBs, which appears at random positions in the sky, a relatively 
large field of view is required. A mission under consideration with this capability at keV energies is the Low energy Polarimetry Detector (LPD) under development at the GuangXi University \footnote{Private communication with Prof. Hongbang Liu}. The instrument is foreseen to have a sensitivity to polarization in the energy range of 2$\,$--$\,$30$\;$keV and maximum effective area of $\sim300\;\mathrm{cm^2}$ around 10$\;$keV by using a similar technology as that used by Polarlight with an optimization of the gas for higher energies. The instrument is under consideration to be placed along side POLAR-2 on the CSS allowing to perform combined measurements of the prompt emission from 2$\;$keV to 800$\;$keV.

\subsubsection{High Energy Polarimeters}
In the MeV energy range one possible mission to be launched in the coming decade is AMEGO \cite{AMEGO}. The AMEGO mission makes use of many 
layers of silicon placed on top of a calorimeter. This makes it ideal to perform polarization measurements using Compton scattering in the 
$\sim100\,$keV to $5\,$MeV energy range. AMEGO will yield polarization measurements for the brightest $1\%$ of GRBs that it will observe.

A second instrument under development is a satellite version of the COSI balloon mission \cite{Tomsick:2019wvo}. This instrument will make use of 
germanium strip detectors capable of measuring the three-dimensional interaction position of incoming photons. The energy range is similar 
to that of AMEGO ($200\,$keV to $5\,$MeV). Thanks to its large field of view it will observe around $\sim40$ GRBs per year with a fluence exceeding 
$4\times10^{-6}\;{\rm erg\,cm}^{-2}$ for which it can perform measurements with an MDP of around $50\%$. 

A highly promising instrument concept for polarimetry at MeV energies is the Advanced Particle-astrophysics Telescope \cite{APT}. The instrument is designed to maximize the effective area for photons in the MeV to TeV energies without using passive materials for photon conversion. The detector aims to use high Z scintillator crystals for the conversion in combination with scintillating fibres. This allows for a large scale detector with precise measurements of both electron positron pairs and Compton scattered photons. The current mission concept would be an order of magnitude more sensitive as a gamma-ray detector than Fermi-LAT and be capable of performing polarization measurements at MeV energies for GRBs as weak as 170817A for which an MDP of $\sim 40\%$ was simulated. The project is in its early stages and currently a path finder mission is planned for a balloon flight.

Apart from these two instruments the earlier mentioned HARPO detector \cite{HARPO} will be capable of performing polarization measurements in the MeV energy range using pair production in a gas TPC. Unlike AMEGO and COSI, which are both under consideration for a launch in the coming decade, the HARPO instrument, of which a prototype has been successfully calibrated on ground \cite{HARPO2}, is currently not under consideration for a launch.

\subsection{Performance Predictions}
\label{sec:instrument-comparison}
Generally, the coming decade looks promising. In the $\sim10\,$--$\,1000\;$keV energy range a number of new detailed 
measurements are foreseen, which should be capable of resolving the current differences in PD reported by different groups. This is illustrated in 
Fig.~\ref{fig:sensitivity} that shows the yearly number of measurements capable of excluding a non-polarized flux as a function of the true polarization 
degree of GRBs for three different instruments, GAP, POLAR and POLAR-2. For this figure the instrument response of POLAR, as used in the POLAR analysis, 
was used as well as that for POLAR-2 in combination with the Fermi-GBM GRB catalog. For GAP, for which the response is not available, the numbers were 
produced by scaling the POLAR numbers based on the performance of GAP and POLAR for respectively detected GRBs, again in combination with the Fermi-GBM GRB 
catalog. It should be noted that for GAP, for which the detailed response is not known, a fixed $M_{100}$ was used, which, given its design, should be close 
to the truth.

\end{paracol}
\begin{figure}
    \widefigure
    \centering
    \includegraphics[width=1.0\textwidth]{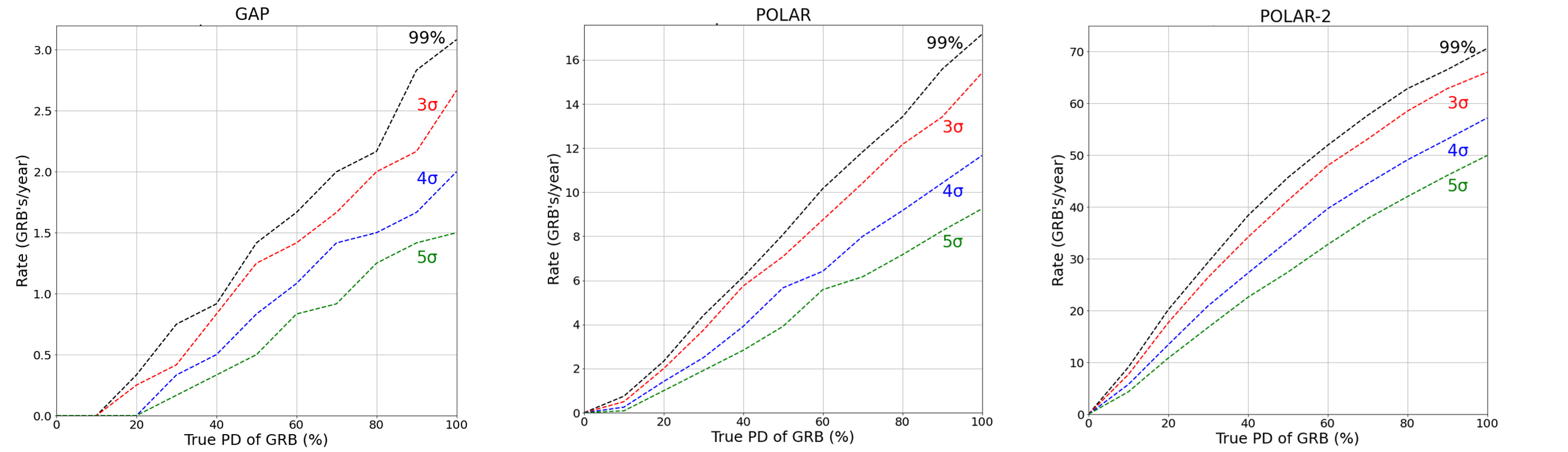}
    \caption{The rate of measurements capable of excluding a non-polarized flux, for different confidence levels, as a function of the true polarization degree (PD) of GRBs for three different instruments, GAP, POLAR and POLAR-2. Although exact numbers are not available it can be assumed that LEAP will be capable of similar rates as POLAR-2, albeit slightly lower.}
    \label{fig:sensitivity}
\end{figure}

\begin{figure}
    \widefigure
    \centering
    \includegraphics[width=1.0\textwidth]{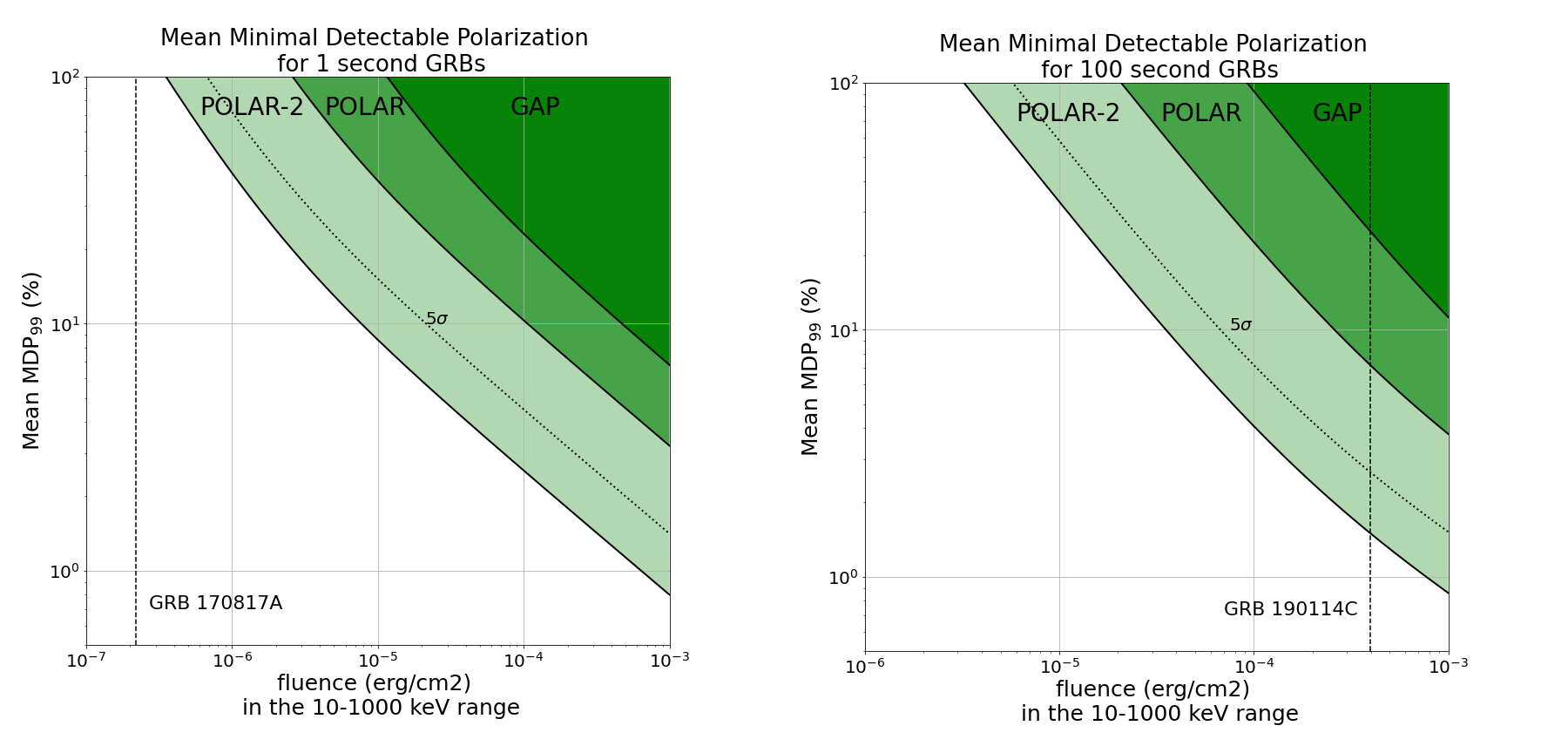}
    \caption{The mean minimal detectable polarization for $99\%$ confidence level (MDP averaged over PA) as a function of the fluence in the 10$\,$--$\,1000\;$keV energy band for GAP, POLAR and POLAR-2 for both short ($1\;$s observed duration) and long ($100\;$s observed duration) 
    GRBs. Short GRBs with a fluence above $10^{-6}\, \mathrm{erg/cm^2} $ occur at a rate of approximately 10 per year on the full sky
    whereas for long GRBs the rate is about 200 per year. For POLAR-2 the MDP for $5\sigma$ confidence is added as well, 
    using a dotted line. The fluences of 2 well known GRBs, the weak and short 170817A and the long and bright 190114C, 
    are added as an illustration. }
    \label{fig:sensitivity2}
\end{figure}
\begin{paracol}{2}
\switchcolumn

It can be seen that with GAP excluding a non-polarized flux was possible for a handful of GRBs per year only in cases where the true PD of the emission 
is relatively high. For POLAR the situation improves and, as was the case, with less than a year of data it was able to claim exclusion of polarization 
levels above $\approx50\%$. It could not, however, effectively probe polarization levels below $30\%$ with a high confidence. With POLAR-2 this region will 
be probed within a few months, while with 1 year of data it will be capable of determining whether GRB emission is polarized to levels as low as $10\%$. 
To illustrate the type of GRBs which can be probed with the different instruments Fig.~\ref{fig:sensitivity2} shows the mean MDP for the 3 different 
instruments as a function of the GRB fluence for both short (1$\;$s observed duration) and long (100$\;$s observed duration) GRBs. As an illustration the fluence of the short and very weak GRB 170817A 
as well as the long and very bright GRB 190114C are added. It should be noted that the energy ranges used for the different instruments differs, the 
energy range of  $50\,$--$\,300\;$keV was used for GAP, $50\,$--$\,500\;$keV for POLAR and $20\,$--$\,500\;$keV for POLAR-2. Although no detailed response is 
available, the performance of LEAP is foreseen to be similar to that of POLAR-2 with a typical effective area $\sim30\%$ smaller than that of POLAR-2. A launch of LEAP would therefore further improve the situation, 
not only regarding the statistics but more importantly regarding the systematics. As for Daksha, not enough details on the instrument are available to make any clear predictions, while the SPHiNX performance would be similar to that of POLAR.

It is evident that the next generation of polarimeters will be capable of almost probing GRBs with fluences as weak as GRB 170817A, a GRB which was hard 
to even detect with both Fermi-GBM and INTEGRAL-SPI but was important due to its association with a gravitational wave signal \cite{Abbott+17-GW170817-GRB170817A}. 
Additionally for very bright GRBs such as 190114C, which was observed at TeV energies \cite{GRB190114C-MAGIC-2019,GRB190114C-Fermi-Swift-2020}, 
highly detailed polarization measurements will become possible, 
indicating that fine time or energy binning  will become an available tool to study such GRBs. It should again be stressed that the mean MDP is simply a figure of 
merit and the estimates given here are not exact, as the details will depend not only on the fluence and length of the GRBs, but also on its energy spectrum, 
incoming angle and position of the polarimeter along its orbit. Additionally systematic errors, which can be significant, are not taken into account in an MDP 
calculation. The predictions should therefore be taken only to give an indication of the advancement in the field as well 
as the possibilities during the coming decade.

Apart from an improvement in the Compton scattering regime, the first polarization measurements of the prompt emission at MeV energies can be 
expected towards the end of this decade. There still remains an additional 
need for energy-dependent polarization measurements. Whereas the eXTP instrument can probe the polarization  at keV energies it is unlikely 
to detect the prompt emission due to its narrow field of view. An instrument such as the LPD would, especially when placed closed to POLAR-2, allow 
to provide an energy range of $2\,$--$\,800\;$keV for many GRBs per year. This would allow to study a potential change in PD in the $10\,$--$\,50\;$keV energy 
range, as proposed in some photospheric emission models \citep{Lundman+18}. In addition, if either COSI or AMEGO will be launched, detailed 
energy resolved studies will become possible for bright GRBs in the 2 keV to 5 MeV energy range, thereby fully probing the prompt emission over several 
orders of magnitude in energy.

\subsection{Improvements in analysis}
\label{sec:improvements-analysis}
From the measurement results published to date it can be seen that increasing the number of measurements alone is likely not enough to provide 
clear conclusions on the polarization of GRB prompt emission. The clearest example of this is the discrepancy between 
the results of POLAR and Astrosat CZTI. Out of the 11 GRBs analysed by the Astrosat CZTI collaboration in \cite{Chattopadhyay+19}, the 6 GRBs 
for which statistically significant measurements (based on the calculation of a Bayes factor required to be above 2) 
are possible the polarization levels for all exceeds $50\%$. For 12 out of the 14 GRB measurements presented by POLAR the PD is found to be 
 below $25\%$ with the two remaining having a low significance. Although the number of measurements is low, the difference in the results is striking. In order 
 to advance the field it is prudent to first understand the cause of these differences in these results as 
 well as other earlier published results.

\subsubsection{Need for public analysis tools and data}

 Polarization analysis is complex and mistakes can easily lead to high levels of PD being measured. As the field is not yet mature 
 and collaborations are small, every analysis has so far almost exclusively been performed using a tool developed for that specific data. This constant reinvention 
 of the wheel not only allows for mistakes, but more importantly results in instrument-specific analysis tools. Such tools are incapable of being applied to other data and their performance is difficult to verify by a referee or other interested scientists. If additionally, the code and 
 the data are not public, as is often the case, and publications lack details on the analysis then it remains nearly impossible to investigate discrepancies 
 with other results.

 What is therefore required, arguably more so than more measurements, is a standardized analysis method which can be adapted to each polarimeter 
 with a public code. Such tools, similar to those widely used in spectrometry  such as \texttt{Xspec} \citep{Arnaud-96} and \texttt{3ML}, would 
 not only allow to understand any potential discrepancies but would also remove the need to reinvent the method by each new collaboration. Furthermore, 
 if additionally instrument data and responses exist publicly, it would remove the requirement to have an in depth understanding of the instrument 
 for being able to perform analysis. This would allow, similar to what happens in spectrometry, for experts in the field of data analysis 
 and statistics to perform the analysis instead of only instrument experts as is now often the case, allowing for more detailed and innovative 
 analyses to be performed.

 A first step towards this was produced as part of the 3ML framework \cite{3ML} for the analysis of the POLAR data. 
 The developed tools aim to provide a framework in which the instrument response and the measurement data are combined to perform the 
 polarization analysis in a transparent way that is usable by anyone. Both for the instrument response and the data format a 
 standardized format is proposed similar to that used in spectrometry and the tool can therefore easily be adapted for other polarimeters. 
 The tool has been used first to analyze GRB 170114A \cite{Burgess-19} in detail using POLAR data, and subsequently to produce the full 
 GRB catalog published by POLAR \cite{Kole+20}. The POLAR data used for this analysis is furthermore 
 public\footnote{https://www.astro.unige.ch/polar/grb-light-curves} allowing further analysis by anyone interested as well as to 
 perform rigorous tests of the validity of the different POLAR results. The public data alone could, for example, already be used by the 
 Astrosat CZTI collaboration using the tools used for the results in \citep{Chattopadhyay+19} to find if their tools provide consistent results 
 are those published in \citep{Kole+20}. Although not perfect, such a study would arguably progress the field further than the analysis 
 of additional Astrosat CZTI or POLAR data by the collaborations themselves.
 
 \subsubsection{Multi-Instrument Analysis}
 
 Thanks to the properties of the 3ML framework data from different instruments can be combined. So far this feature was used 
 only to combine the POLAR data with that from \textit{Fermi}-GBM and \textit{Swift}-BAT. This allowed to improve the spectral fits, as the error 
 on the spectrum adds to the systematic error on the polarization measurement, which in turn led to more precise polarization measurements. The 3ML framework additionally allows to fit physical models directly to the data, 
 rather than fitting the data with empirical models and subsequently comparing the results with a parameterized outcome of a theoretical prediction. 
 Although easier, the latter method has, especially in the field of gamma-ray spectrometry, been found to result in over interpretation 
 of data analysis results and to inconsistent conclusions (see discussion in section \ref{sec:Synchrotron}). Fitting of physical models directly to data is
 especially desirable in the field of polarimetry as it allows to fit these models, potentially unbinned in time and energy, directly both to spectral and polarization data at the same time.
 
 Apart from combining spectral and polarization data in the analysis, in theory, the same can be done using data from two polarimeters 
 in case two different polarimeters observed the same GRB. In fact, several GRBs were observed by both Astrosat and POLAR \cite{GCN170101B, Kole+20}. 
 It would therefore be highly desirable to perform combined analysis of the Astrosat and POLAR data for such GRBs as it would, firstly, allow to 
 study the cause of the likely discrepancy between the results from both instruments. Secondly, it can allow for more detailed measurements of the 
 polarization of these GRBs. 
 
 With upcoming instruments sensitive in different energy ranges, such analysis tools will in the future allow to fit physical models to both spectral 
 and polarization data over a broad range in energy by, for example, combining the data of the LPD, LEAP, POLAR-2 
 and AMEGO or COSI. Whereas with the current level of polarimetry analysis tools the data has to be studied separately, leaving the full potential of the data 
 unexploited.  
 
 As the polarization tool in the 3ML framework discussed here is new and has not been used for the polarization analysis of other instruments, it is to 
 be seen if it will be used by the wider community. However, with the potential of 2 large scale polarimeters in LEAP and POLAR-2 launching in the coming 
 years, as well as polarimeters sensitive at keV and at MeV energies,  there is a clear need for a collaborative effort between the groups to either 
 further develop this tool or construct a completely new one.

\subsection{Improvements in Theoretical Modeling of Prompt GRB Polarization}
\label{sec:improvements-theory}
Pulse-integrated polarization from semi-analytic models of axisymmetric flows with different prompt GRB radiation mechanisms and B-field 
configurations have been presented in many works \citep{Granot-03,Granot-Konigl-03,Lyutikov+03,Lazzati+04,Granot-Taylor-05,Toma+09,Gill+20}. 
The same setup was used to make predictions for the time-dependent polarization for synchrotron emission in some works \citep{Cheng+20,Lan+20,Gill-Granot-21}. 
On the other hand, only a few works have attacked the problem using MC simulations \citep{Ito+14,Lundman+14,Parsotan+20} or radial integration 
of the transfer equations for the Stokes parameters \citep{Beloborodov-11}. Many of these have focused only on photospheric emission.

As the next decade may see the launch of more sensitive instruments to measure GRB polarization with high fidelity, it calls for time- and 
energy-dependent polarization predictions ($\Pi(E,t)$, $\theta_p(E,t)$) for more realistic outflow models, which would also predict the 
time-dependent flux density, $F_E(t)$. 

One of the weaknesses of current theoretical models is the assumption of an axisymmetric flow, which is usually 
made for simplicity and convenience. This restricts the change in PA to only $\Delta\theta_p=90^\circ$, whereas some observations do show, although not so 
convincingly yet, hints of gradual PA swings. To obtain a change in the PA other than $\Delta\theta_p=90^\circ$ or to get a gradually changing PA the condition for 
axisymmetry must be broken, e.g. the magnetic field configuration/orientation and/or the emissivity can change as a function of $(\theta,\phi)$. 

One possibility is that if the different pulses that contribute to the emission arise in `mini-jets' within the outflow 
\citep[e.g.,][]{Shaviv-Dar-95,Lyutikov-Blandford-03,Kumar-Narayan-09,Lazar+09,Narayan-Kumar-09,Zhang-Yan-11}. In this case the different directions of the 
mini-jets or bright patches w.r.t. the LOS \citep[e.g.][]{Granot-Konigl-03,Nakar-Oren-04} would cause the PA to also be different between the pulses even for 
a field that is locally symmetric w.r.t the local radial direction (e.g. $B_\perp$ or $B_\parallel$) as well as for fields that are axisymmetric w.r.t to 
the center of each mini-jet (e.g. a local $B_{\rm tor}$ for each mini-jet). Finally, broadly similar results would follow from an ordered field within 
each mini-jet ($B_{\rm ord}$) which are incoherent between different mini-jets. Time-resolved measurement in such a case would naturally yield 
a time-varying PA. 

Alternatively, as shown by \citet{Granot-Konigl-03} for GRB afterglow polarization, a combination of an ordered field component 
(e.g. $B_{\rm ord}$) and a random field, like $B_\perp$, can give rise to a time-varying PA between different pulses (with a different 
ratio of the two field components) that, e.g., arise from internal shocks. The ordered field component here would be that advected 
from the central engine and the random field component can be argued to be shock-generated. Notice that the ordered field component 
should not be axisymmetric in order for the position angle to smoothly vary.

Realistic theoretical predictions can be obtained by coupling radiation transfer modeling with MHD numerical simulations of relativistic jets after 
they break out of the confining medium. A step towards this direction was taken by \citet{Parsotan+20} who used the MHD code \texttt{FLASH} to first obtain 
the jet's angular structure by injecting variable jets into stellar density profiles of Wolf-Rayet stars at core-collapse. They then used an MC code to 
carry out the radiation transfer of the Stokes parameters and obtain the time-resolved polarization for the photospheric emission 
(see Fig.~\ref{fig:Photspheric-Pol}). In another recent work, \citet{Ito+21} carried out global neutrino-hydrodynamic simulations of a relativistic 
jet launched in a binary NS merger scenario. The photospheric emission and polarization from the short GRB was then calculated using a 
relativistic MC code. While these works focused only on photospheric emission, polarization modeling for other radiation mechanisms performed 
in the same vein is lacking and can prove to be very fruitful.

MC radiation transfer and MHD numerical simulations of relativistic jets can be computationally expensive. They are nevertheless a useful tool that can be used to 
calibrate semi-analytic models by delineating the relevant parameter space expected in GRB jets. Ultimately, when high quality observations are made in this 
decade, fast and computationally inexpensive theoretical models will be required to carry out time-resolved spectro-polarimetric fits in a reasonable amount 
of time. This further stresses the need for a library of models, akin to \texttt{Xspec} \citep{Arnaud-96} that is used routinely for spectral fitting or 
\texttt{boxfit} \citep{vanEerten+12} for GRB afterglow lightcurve modeling, which can be conveniently used by observers. Combining the library of models with 
the 3ML framework for spectro-polarimetric data analysis will become a very powerful tool for GRB science.

In order to test the different model predictions, e.g. from different radiation mechanisms, on an equal footing, a single underlying theoretical 
framework should be devised for the jet structure and dynamics, which allows the same freedom in the different model parameters. Such an approach 
can help to isolate the dominant prompt GRB radiation mechanism when compared with observations.

To conclude, the next decade appears very promising for answering many fundamental questions in GRB physics. With the launch of several dedicated 
instruments capable of performing high-fidelity $\gamma$-ray and X-ray spectro-polarimetry, a larger sample of statistically significant prompt GRB 
polarization measurements will be obtained. Improvements in polarization data analysis using a single underlying framework that allows simultaneous 
fitting of both spectrum and polarization from different instruments will yield unbiased and high-quality results. More realistic theoretical models 
of both time- and energy-dependent polarization based on advanced numerical simulations will allow to better understand the true nature of GRB jets.

\vspace{6pt} 


\authorcontributions{
Writing--original draft preparation--review and editing, R.G., M.K., J.G.
}

\funding{
This research was funded in part by the ISF-NSFC joint research program under grant no. 3296/19 (R.G. and J.G.) and the Swiss National Science Foundation (M.K.)
}

\acknowledgments{
We thank Stefano Covino and Mark McConnell for a thorough read of an earlier version of the manuscript and for their comments and feedback.
}

\conflictsofinterest{
The authors declare no conflict of interest.
} 




\reftitle{References}


\externalbibliography{yes}
\bibliography{refs}



\end{paracol}
\end{document}